\documentclass[useAMS,usenatbib]{mn2e}
\usepackage[dvips]{graphicx}
\usepackage{pdflscape}
\usepackage{amsmath}
\usepackage{amssymb}
\usepackage{multirow} 
\usepackage{rotating}
\usepackage{sidecap}
\title[Spatially Resolved Colors and Stellar Population Properties in Early-Type Galaxies at z $\sim$ 1.5.]{Spatially Resolved Colors and Stellar Population Properties in Early-Type Galaxies at z $\sim$ 1.5.}
\author[A. Gargiulo et al.]{A. Gargiulo$^{1}$\thanks{E-mail:
adriana.gargiulo@brera.inaf.it}, P. Saracco$^{1}$, M. Longhetti$^{1}$, F. La Barbera$^{2}$, S. Tamburri$^{1,3}$\\
$^{1}$INAF-Osservatorio Astronomico di Brera, via Brera 28, 20121 Milano, Italy\\
$^{2}$INAF-Osservatorio Astronomico di Napoli, salita Moiariello 16, 80121 Napoli, Italy\\
$^{3}$Universit\'{a} degli Studi dell’Insubria, via Valleggio 11, 22100 Como, Italy}

\begin{document}

\date{Accepted 2012 July 5.  Received 2012 June 13; in original form 2012 March 26}

\pagerange{\pageref{firstpage}--\pageref{lastpage}} \pubyear{2002}

\maketitle

\label{firstpage}

\begin{abstract}

We present F850LP-F160W color gradients for 11 early-type galaxies (ETGs) at 
1.0$<$z$_{spec}<$1.9 selected from the GOODS South field. 
Significant negative F850LP-F160W color gradients (core redder than the outskirts) 
have been detected in $\sim 70\%$ of our sample within the effective radius R$_{e}$, 
the remaining 30$\%$ having a flat color profile consisten with a null gradient.
Extending the analysis at R$>$R$_{e}$, enclosing the whole galaxy, we have found that the fraction of high-z ETGs with 
negative F850LP-F160W color gradients rises up to 100$\%$.  
For each galaxy,  we investigate the origin of the radial color variation 
with an innovative technique based on the matching of both the spatially resolved color and 
the global spectral energy distribution (SED) to predictions of composite stellar population models.
In fact, we find that 
the age of the stellar populations is 
the only parameter whose radial variation alone can fully account for 
the observed color gradients and global SEDs for half of the galaxies in our sample (6 ETGs), without
the need of radial variation of any other stellar population property. For four out of these six ETGs, a pure metallicity variation can also reproduce 
the detected color gradients. 
Nonetheless, a minor contribution to the observed color gradients from radial variation of
star-formation time scale, abundance of low-to-high mass
stars and dust cannot be completely ruled out.
For the remaining half of the sample, our analysis suggests a more complex scenario whereby 
more properties of the stellar populations need to simultaneously vary, likely with comparable weights, to 
generate the observed color gradients and global SED.
Our results show that, despite  the young mean age of our
galaxies ($<$3-4 Gyr), they already exhibit significant
differences among their stellar content. We have discussed our results within the framework of
the widest accepted scenarios of galaxy formation and conclude that none of them can satisfactorily account for the observed distribution of 
color gradients and for the spatially resolved content of high-z ETGs. Our results suggest that 
the distribution of color gradients may be due to ETGs forming by different mechanisms. 

\end{abstract}


\section{Introduction}

A viable way to gather insight on the
processes that concur to accrete the stellar mass in early-type galaxies (ellipticals plus S0's, hereafter ETGs)
is to analyze the spatial distribution and properties of their stellar 
content which, in principle, can be directly connected to the events experienced by the galaxies. 
Indeed, the different scenarios proposed to explain the formation of ETGs 
give different predictions on their stellar population content.

The $revised$ monolithic model predicts that, in a cold dark matter framework, 
massive ETGs assemble the bulk of their mass at z$>$ 2-3 through the merger of small substructures 
moving in a common potential well. This initial collapse might be regulated by cold gas streams from
the cosmological surroundings, also known as cold accretion~\citep{dekel09}, the latters becoming less important, 
for high-mass galaxies, at lower redshift. The subsequent evolution should be mainly characterized 
by the aging of their stellar populations, with small new episodes of star-formation at z$<$1, related, e.g., 
to the capture of satellites \citep{katz91,kawata01,kobayashi04,merlin06}. 
During the gravitational dissipative collapse, the metal-enriched gas should naturally flows towards the center 
of the galaxy, leading to an ETG with stellar populations
more metal-rich in the center than in the external regions (i.e. negative metallicity gradient).
Moreover, because of the deeper potential well,  the star-formation is expected to last longer
in the central than in the outer regions. This would lead to null or mildly positive age gradients, 
with stellar populations in the center $\sim$ 10 $\%$ younger than those in the outskirts~\citep{kobayashi04}. Nonetheless the effect of metallicity variation on color profile should be the dominant one, thus in this scenario ETGs are generally expected with negative color gradients. \\
The competing formation scheme is the $hierarchical$ scenario. Following the hierarchical assembly of 
cosmic structures, ETGs are supposed to form through gas-rich (``wet'') mergers of disc 
galaxies \citep[e.g.][]{toomre72, delucia06} at high-redshift (z$\sim$4-5). In this phase, a large fraction 
of the stellar mass of a galaxy is assembled   through central intense bursts 
of star formation \citep[e.g.][]{renzini06}. Concurrently, a ``dry'' merger picture has also been advocated,
where bright ETGs would form through the merging of quiescent galaxies \citep[e.g.][]{bell04}. 
Actually, in the last years a new scheme of mass accretion of massive ETGs, 
known as \textit{inside-out growth}, is becoming widely 
accepted. This scenario is motivated by the observational evidence of ETGs at 1.0$<$z$<$2.5 
with effective radii 3-5 times smaller than the mean radius of local ETGs with the same stellar 
mass \citep[e.g][]{daddi05,longhetti07}. 
In this context, supported by a wealth of simulations \citep[e.g.][]{khochfar06, hopkins09, wuyts10, naab09, bezanson09} ETGs are supposed to be formed  at high redshift (z$\sim$4-5) as compact spheroids result 
of gas-rich mergers. Then, at lower redshift, compact ETGs would undergo subsequent 
minor-``dry'' mergers, whose main effect is to add an external low-mass 
density envelope to the compact core ETGs, enlarging the effective radius while leaving the stellar mass nearly
constant. Indeed, this scenario shows some limitations, such as the 
not plausible number of minor-``dry'' mergers necessary to enlarge the ETGs' size which could produce a scatter in the fundamental plane much larger than the observed one, and its failure to explain 
the presence of normal ETGs observed at high-z in number similar to the compact 
ones~\citep[e.g.][]{nipoti09,saracco11, saracco10}. From a theoretical point of view, the ``wet''-merger scenario 
predicts ETGs with significant radial age variations. Indeed, the galaxy remnant of a wet merger should be 
characterized by a central stellar population younger, and more metal-rich, than the outer one, i.e. by
a positive (negative) age (metallicity) gradient~\citep{kobayashi04}. On the contrary, ``dry''-mergers, mixing 
the pre-existing stellar populations of progenitor galaxies, should dilute any radial variation of age and 
metallicity, producing flatter distributions (but see also \citet{dimatteo09}). 
In the \textit{inside-out} scenario minor dry-mergers are actually believed to only $add$ an $outer$ low-density envelope on top of a compact 
core, without mixing the pre-existing stellar-populations but redistributing the star content of the 
satellites in the outer regions of the compact ETG.\\
The picture described so far shows how important can be to spatially resolve the properties of the 
underlying stellar populations of ETGs in order to pinpoint their formation scenario.
The most viable way to carry out information on the radial variation of the stellar content of a galaxy 
is to investigate its radial color variation, being the color 
of a stellar population tightly dependent on its age, star-formation time scale, 
metallicity, and (eventual) presence of dust. \\
ETGs at high redshift (hereafter with ``high redshift'' we mean
$z>$1) represent the best-suited benchmark to investigate and
constrain the mechanisms driving the mass assembly of
spheroids due to the short time elapsed since their formation. 
So far, instrumental limits have allowed only studies of the global
(i.e. integrated) properties of high-z spheroids, preventing, indeed,
any measure of their spatially resolved information. Recently, the capabilities of 
the Hubble Space Telescope (HST) have
made possible to overwhelm part of these limitations for the first time. 
Gargiulo et al. (\citeyear{gargiulo11}, hereafter GSL11) taking advantage of the deep
and high-resolution HST Advanced Camera for Surveys (HST/ACS) images of the GOODS South field 
derived F606W - F850LP color gradients ($\sim$ (UV -U)$_{\it{rest
frame}}$ at z=1.5, $\lambda_{eff}$ of F606W and F850LP filter $\sim$ 5810\,\AA\,and 9010\,\AA\,respectively) for 20 ETGs at 1 $<$ $z_{spec}$ $<$ 2.
In their work they ascertained the feasibility of this analysis up to z$\sim$2 and presented the first
spatially resolved information for ETGs at high-z. Despite the short wavelength baseline covered, $\sim$
50\,$\%$ of the galaxies showed a significant (positive or negative) color gradient.
These results clearly showed that, after 3-4 Gyr (at most) from
their birth, ETGs do not exhibit a unique spatial distribution of their stellar populations,
implying that they followed different mass assembly paths. 
Unfortunately, the available HST data, mainly optical, 
prevented the authors to discriminate the drivers of the observed 
color gradients (e.g. radial variations of age, metallicity, ...) and 
consequently, to constrain the mechanisms responsible for them. Indeed, at $<$z$>$ $\sim$ 1.5, the galaxy 
emission sampled by the F606W and F850LP filters is sensitive to both age  
and dust variation. Moreover, it is dominated by the youngest ($\sim$ 1Gyr) stellar populations, missing
any information on the distribution of the oldest stellar populations. \\
The recent advent of the first HST Wide Field Camera 3 (HST/WFC3) near infrared images 
for part of the GOODS South area (see Section 2 for details) 
has opened new possibilities to study color gradients of high-z ETGs and to
constrain their origin. In this paper we combine the information provided by the WFC3/F160W-band ($\sim$ R-band 
rest-frame at $<$z$>$ $\sim$ 1.5, $\lambda_{eff}$ of the filter $\sim$ 15369\,\AA) and F850LP-band emission to derive the F850LP-F160W color gradients ($\sim$ (U-R)$_{\it{rest
frame}}$ at z=1.5) for a sample of 11 ETGs at $<$\textit{z}$>$\,$\sim$1.5. The bands we 
selected sample emissions dominated by different stellar populations. In particular, 
differently from F606W-F850LP color, the F850LP-F160W color is much more sensitive to age variations 
than dust content. This allows us to extend and to complement the analysis presented by GSL11.\\
Actually, \citet{guo11}  presented F850LP-F160W color gradients for 4 massive 
passively evolving galaxies in the GOODS South area at  1.3 $<$ \textit{z}$_{spec}$ $<$ 2. 
They derived the color profiles by measuring optical-NIR  colors in concentric annuli, 
and found that high-z ETGs have cores redder than the outerskirts. The observed 
radial trend in color gradients is not reproduced by the radial variation of a single 
stellar population parameter (age, metallicity, extinction), although they found that dust 
should partially contribute to generate the observed color distribution. In this paper, we examine a 
sample three times larger than that of \citet{guo11}, deriving F850LP-F160W color profiles from 
the 2-dimensional fit of the light profiles in the two bands. Then, we 
present a new approach to exploit the wealth of available information to constrain the radial 
variation of the underlying stellar populations. In fact, taking advantage of a unique set 
of data and an innovative procedure, we are able to constrain the radial variation of 
stellar population parameters (age, metallicity, dust, star-formation time scale, and initial 
mass function) and their contribution to produce the color gradients we observe in high-z ETGs. 
Finally, we compare our  findings to the prediction of theoretical formation models.\\
Throughout the paper we adopt a standard $\Lambda$CDM cosmology with 
H$_{0}$=70km\,s$^{-1}$\,Mpc$^{-1}$, $\Omega$\,$_{m}$=0.3 and  $\Omega$\,$_{\Lambda}$=0.7. All 
the magnitudes are in the AB system.
\section{The sample}
\label{sec:thesample}
We have derived F850LP-F160W 
color gradients for 11 ETGs at $<$z$>$ $\sim$ 1.5. The sample has been
extracted from the complete sample of 34 ETGs selected on the whole GOODS
South field (GOODS-South v2; Giavalisco et
al. \citeyear{giavalisco04}) and presented in \citet{saracco10}. The morphological classification was performed by the authors
both on the basis of a visual inspection of the
F850LP images and of Sersic index $n_{F850LP}$
(\textit{n}$_{F850LP}$\,$>$ 2).
Starting from this complete sample we have restricted our preliminary analysis to those ETGs with near-IR WFC3 data (16 ETGs out 34). Indeed, as pointed out in the Introduction, two
areas of the GOODS South field have been imaged in the F160W band with the
WFC3 extending the space-based data from the
pre-existing optical domain to the near-infrared one.
The Early Release Science (ERS, propID: 11359, PI: R. W. O'Connell)
imaged an area of $\sim$ 40 arcmin$^{2}$. Briefly, the observations were acquired with a
2-cycles-long exposures for a total exposure time of
$\sim$ 6 ks, reaching a 5$\sigma$ depth in the AB system of F160W =
27.25 (Windhorst et al. \citeyear{windhorst11}). Additionally, the
HUDF09 HST Treasury program (GO 11563, PI: Illingworth) provides the
first ultra-deep near-IR WFC3 observation of the Hubble Ultra Deep
Field (HUDF). The images cover an area of $\sim$ 4.7
arcmin$^{2}$. At the time of our analysis, the first epoch data were
available for a total exposure time of $\sim$ 80 ks and a depth at
5$\sigma$ of 28.8 in the F160W filter (AB magnitude, Oesch et
al. \citeyear{oesch10}). 
Starting from the raw WFC3 images, we have created the mosaic images
with the software MULTIDRIZZLE \citep{koekemoer02} reducing the pixel/scale of the mosaic from the original
value of 0.128 ``/px to 0.06 ``/px. In the final
mosaics we have stacked  only those single exposures not affected by the presence of
persistence. Thus, we have obtained the WFC3/F160W-band mosaics of the ERS and HUDF09 area. Both mosaics are
characterized by a full width at half maximum (FWHM) of $\sim$
0.2''. The exposure times are $\sim$ 80ks and $\sim$ 6 ks for the HUDF09 and ERS, respectively.  S/N $>$ 5 at 3r$_{e}$ are reached 
for the faintest galaxies of our sample in  the ERS area,
the one with lower exposure time. This S/N and the resolution of the WFC3 assure us reliable
estimates of the surface brightness parameters even in the images with
the shorter exposure time (see Sec. 3.1). 

Out of the 16 ETGs with NIR WFC3 data available, five objects have been excluded from the present
analysis because of either potential problems
in the modelling of the near-IR galaxy light distribution (4 ETG, Sec.~3), or of large
uncertainties in the estimate of their structural parameters (1 ETG, Sec.~3.1).
These selections result into a final sample of 11 galaxies.
Each one of these galaxies is provided with spectroscopic redshift (Vanzella et al.
\citeyear{vanzella08} and references therein) and with the wide
photometric coverage of the GOODS survey: four deep HST/Advanced
Camera for Surveys (ACS) images in the F435W, F606W, F775W, F850LP
bands, extensive observations with ESO telescopes both in three
optical U bands and with J, H, K$_{s}$ near-infrared filters and four
$Spitzer$-IRAC images in the 3.6, 4.5, 5.8, 8.0$\mu$m bands.
Morphological parameters (effective radius and Sersic index) in the F850LP filter and physical parameters (age and stellar mass) were already estimated for each galaxy of the sample 
\citep{saracco10}.
Four out of the 11 ETGs were already studied in GSL11, thus F606W-F850LP color gradients are also available (GSL11).

\section{Surface brightness parameters estimation}

Following GSL11, we have estimated the internal color gradient of a galaxy 
as the logarithmic slope of its color profile, given by the difference between the  $\mu_{F850LP}$(r) 
and $\mu_{F160W}$(r) surface brightness profiles.  As for the F850LP and F606W bands, 
we have modelled the F160W surface brightness profile of high-z ETGs with a Sersic law:
\begin{equation}
\mu(R) = \mu_{e} + \frac{2.5b_{n}}{\ln(10)}[(r/r_{e})^{1/n}-1].
\label{mu}
\end{equation}
To estimate the free parameters of the profile, i.e. the effective
radius r$_{e}$ (in units of arcsec), the Sersic index $n$, and 
the normalization term $\mu_{e}$, we have used GALFIT \citep[v2,][]{peng02}. This
software models the galaxy light distribution with a two-dimensional fit.
In the fitting procedure,
the galaxy model is convolved with the PSF, provided by the user, and the software
retains as final solution the parameters of the PSF-convolved model that
minimizes the residuals to the input galaxy image. For each galaxy, we have
constructed different PSF models, each model being obtained by averaging the 
light profiles of unsaturated stars as near as possible to the position of the given 
galaxy on the frame. We have run GALFIT for all different PSFs and have retained the case
that returns the best residual map (i.e. flattest residuals).  For four out of 16 initially selected
galaxies, we found significant substructures in the residual maps, possibly indicating 
an inaccurate modeling of the galaxy light distribution and/or of the PSF.
We have excluded these four galaxies from the analysis (see Sec.~\ref{sec:thesample}).

\subsection{Robustness of structural parameter estimates}

\begin{figure}
	\includegraphics[angle=-90,width=8.0cm]{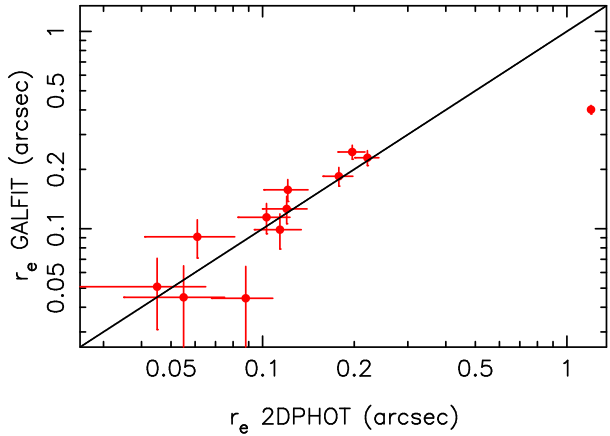} \\
	\caption{Comparison of effective radius estimated with GALFIT and 2DPHOT (see the text). The solid line
marks the one-to-one relation.}
	\label{confre}
\end{figure}

\begin{figure*}
	\includegraphics[angle=-90,width=17.5cm]{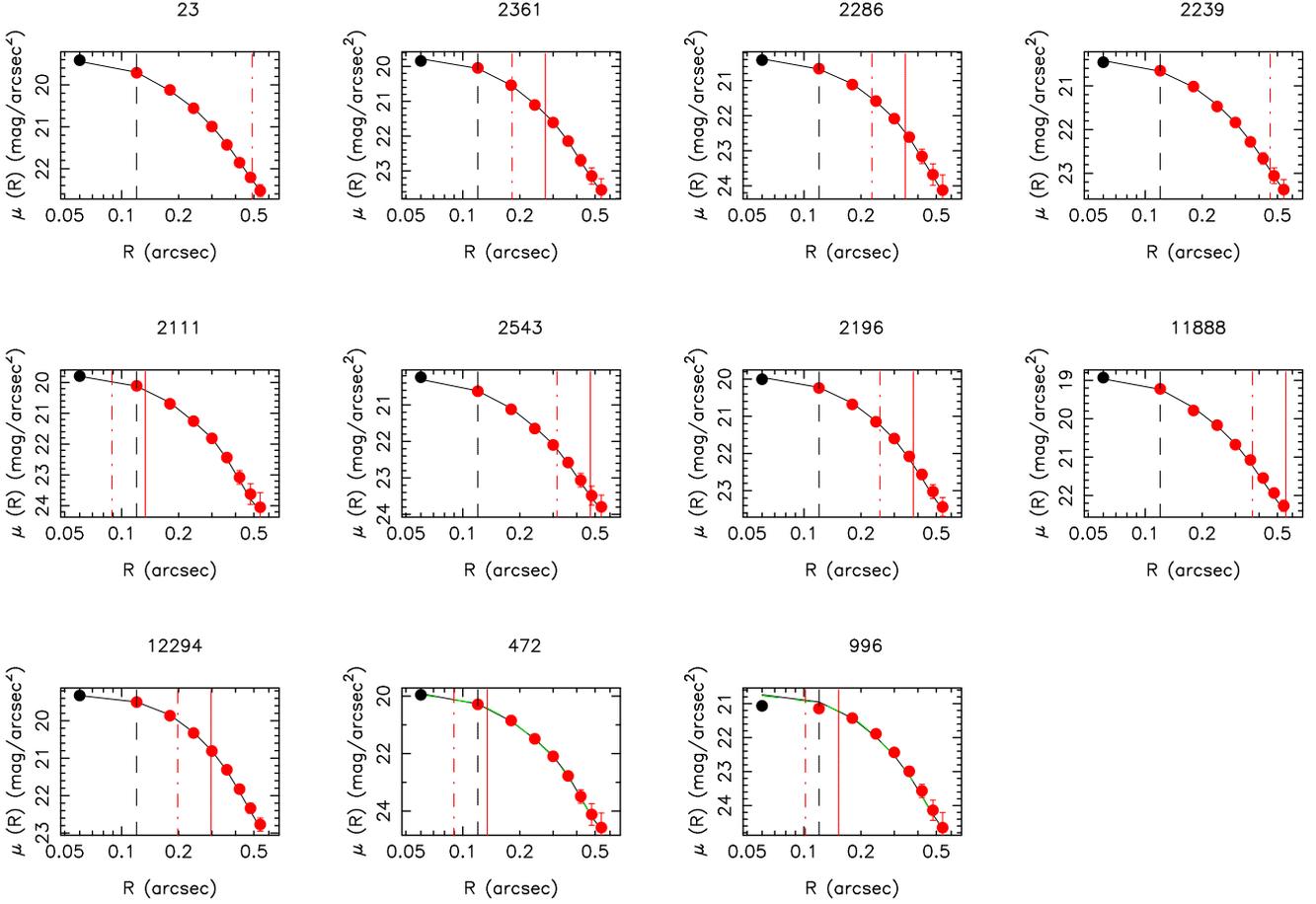} \\
	\caption{Comparison of F160W PSF-convolved  Sersic profiles (black curves) for our sample of 11 ETGs, 
         with the observed surface brightness profiles measured directly on the given images (red
	points). Vertical dashed black lines mark the radius of the FWHM of the PSF, while red dash-dotted and solid lines
	correspond to 2 and 3~r$_{e}$, respectively. In panels related to ETGs 472 and 996, black lines correspond to PSF-convolved  Sersic profiles derived on the mosaic image of $\sim$ 80ks, while green lines, not visible given the almost perfect overlap to the black curves, the same quantity measured on mosaic images of $\sim$ 6ks.}
	\label{checksbp}
\end{figure*}

Given the pixel scale of 0.128"/pixel (0.06''/pixel after drizzling)
and the FWHM of the PSF of $\sim$ 0.2", the  WFC3 images in the F160W passband are close
to the sampling limit.
Moreover,  even with the excellent spatial resolution of WFC3, most of the high-z ETGs
in our sample have sizes comparable to the FWHM.
This makes of paramount importance the accuracy and reliability of the derived PSF used to convolve the galaxy model. To assess the independence of our results from the method we have adopted to derive the PSF and the consequent robustness of the derived structural parameters, 
we have  derived them also with 2DPHOT \citep{labarbera08}, 
a software working by means of different approach than GALFIT. 2DPHOT is a fully automatic tool,
allowing galaxy surface photometry to be performed by fitting galaxy images with two-dimensional,
PSF convolved, (Sersic) models. Differently from GALFIT, 2DPHOT constructs a PSF model 
in a user-independent manner, fitting star images on a given frame 
with a combination of two-dimensional Moffat functions, taking into account PSF 
asymmetries and image (under-)sampling~\footnote{To this effect, at each step of the PSF fitting,
the Moffat functions are convolved with a box kernel, i.e. the pixel of the given image.}. 
In order to reproduce the spikes of the HST PSF, we have modified the 2DPHOT PSF fitting algorithm 
by smoothing the (average) residual map from the fitted stars (with a 3x3 pixels median smoothing),
and adding the smoothed residuals to the Moffat-based PSF. The 2DPHOT initial parameters for Sersic 
fitting are also computed through a user-independent approach (in contrast to GALFIT), by comparing the 
galaxy image with a discrete grid of PSF-convolved Sersic models, reducing the problem 
of spurious convergences in the (non-linear) optimization procedure.
In general, we find a very good agreement between 2DPHOT and GALFIT structural parameters.
As an example, in Fig \ref{confre}, we compare the effective radii derived with the two softwares.
Errors in r$_{e}$ has been set equal to 0.02 arcsec, the typical uncertainty of this measure as we have derived through montecarlo simulations \citep[e.g.][]{longhetti07}.
No systematic difference is found between r$_{e,GALFIT}$ and r$_{e,2DPHOT}$, despite of the different approach and methodology adopted. Only 
for one galaxy, with r$_{e,GALFIT} \sim 0.4''$, the r$_e$ estimate differs significantly, with 
r$_{e,2DPHOT}$ being $\sim$ 3 times larger than  r$_{e,GALFIT}$. Given this discrepancy, we have excluded 
this object from our sample thus leaving with 11 galaxies in our analysis (see Sec.~\ref{sec:thesample}). To be consistent with GSL11, where we estimated 
structural parameters in the F850LP and F606W passbands with GALFIT, we have decided to adopt GALFIT estimates of 
surface brightness parameters also for the WFC3/F160W filter.\\
To further verify the reliability of surface brightness parameters, in Fig.~\ref{checksbp} 
we compare the best-fitting PSF-convolved Sersic profiles  (black curves), with the observed 
surface brightness profiles (red points), measured in concentric circular coronas of fixed width 
on the F160W images. Errors on surface brightness were derived with SEXTRACTOR \citep{bertin96}
and take into account the correlated noise introduced by the drizzling technique \citep{casertano00}. 
The vertical dashed black lines mark the radius of the FWHM of the PSF (0.1 arcsec), while red dash-dotted and continuous lines
correspond to 2~and 3~r$_{e}$, respectively. 
The figure shows an excellent agreement between observed and fitted profiles, from the 
very central region out to a radius of $\sim$ 3r$_{e}$, and in many cases well beyond,
both for galaxies in the deepest images (ID 472, 996) and those in the shallowest 
fields. This confirms the accuracy of PSF modeling and Sersic fitting.
For the two galaxies on the HUDF area (ID 472 and 996), the green lines show 
the PSF-convolved Sersic profiles estimated on a mosaic image created to match the
exposure time of the shallowest images available ($\sim$ 6 ks; see Sec.~\ref{sec:thesample}). 
The excellent agreement between the black ($\sim$ 80~ks) and green ($\sim$ 6~ks) solid curves (in fact overlapping) 
in the Figure indicates the homogeneity of our measures and proves the
feasibility of deriving reliable structural parameters even with 2-orbit-long images.\\
In Table 1 we report our effective radii, r$_{e}$ (in units of arcsec), and R$_{e}$ (in kpc), 
as well as the Sersic index and total magnitude for each of the 11 ETGs in our 
sample. 
\begin{table*}
\caption{Our sample of galaxies. $Column$ $1$: id number $Column$ $2$:
  spectroscopic redshift; $Column$ $3$: total magnitude from GALFIT. Errors on magnitude are the formal GALFIT errors;
  $Column$ $4$: effective radius in arcsec. A typical error on the estimates of r$_{e}$ is 0.02\,arcsec corresponding to 0.17kpc at z=1.5; $Column$ $5$: effective radius in kpc; $Column$ $6$: Sersic
  index; $Column$ $7$: F850LP-F160W colour gradient. At the
  median redshift z = 1.5 1\,arcsec corresponds to $\sim$ 8.5\,kpc.}
  \footnotesize
\begin{tabular}{ccccccc}
\hline
Object & z & F160W$_{tot}$ & r$_{e,160}$ & R$_{e,160}$ & n$_{160}$ & $\nabla_{F850LP-F160W}$ \\
       &   & mag           & arcsec         & kpc         &           &  mag/dex\\
\hline
23     & 1.04 & 20.80$\pm$0.01 & 0.2 & 1.98 & 3.47 & -0.3$\pm$0.2\\
11888  & 1.04 & 20.44$\pm$0.01 & 0.18 & 1.50 & 5.39 & -0.4$\pm$0.1\\
12294  & 1.21 & 20.93$\pm$0.01 & 0.10 & 0.82 & 2.18 & -0.72$\pm$0.05\\
996    & 1.39 & 22.57$\pm$0.01 & 0.05 & 0.43 & 5.52 & -1.0$\pm$0.1\\
2239   & 1.41 & 21.74$\pm$0.01 & 0.23 & 0.38 & 2.85 & -0.5$\pm$0.3\\
2286   & 1.60 & 22.20$\pm$0.01 & 0.11 & 0.97 & 2.02 & -0.1$\pm$0.2\\
2361   & 1.61 & 21.60$\pm$0.01 & 0.09 & 0.77 & 3.35 & -0.1$\pm$0.1\\
2196   & 1.61 & 21.65$\pm$0.01 & 0.12 & 1.06 & 2.81 & -0.2$\pm$0.2\\
2543   & 1.61 & 21.93$\pm$0.02 & 0.15 & 1.33 & 5.62 & -0.8$\pm$0.4\\
2111   & 1.61 & 21.85$\pm$0.01 & 0.04 & 0.38 & 4.73 & -0.54$\pm$0.07\\
472    & 1.92 & 22.12$\pm$0.01 & 0.04 & 0.38 & 3.23 & -0.6$\pm$0.1\\
\hline
 \label{sp}
 \end{tabular}
 \normalsize
\end{table*}
\section{F850LP-F160W color gradients of high-z ETGs}
\begin{figure*}
	\includegraphics[width=12.0cm,angle=-90]{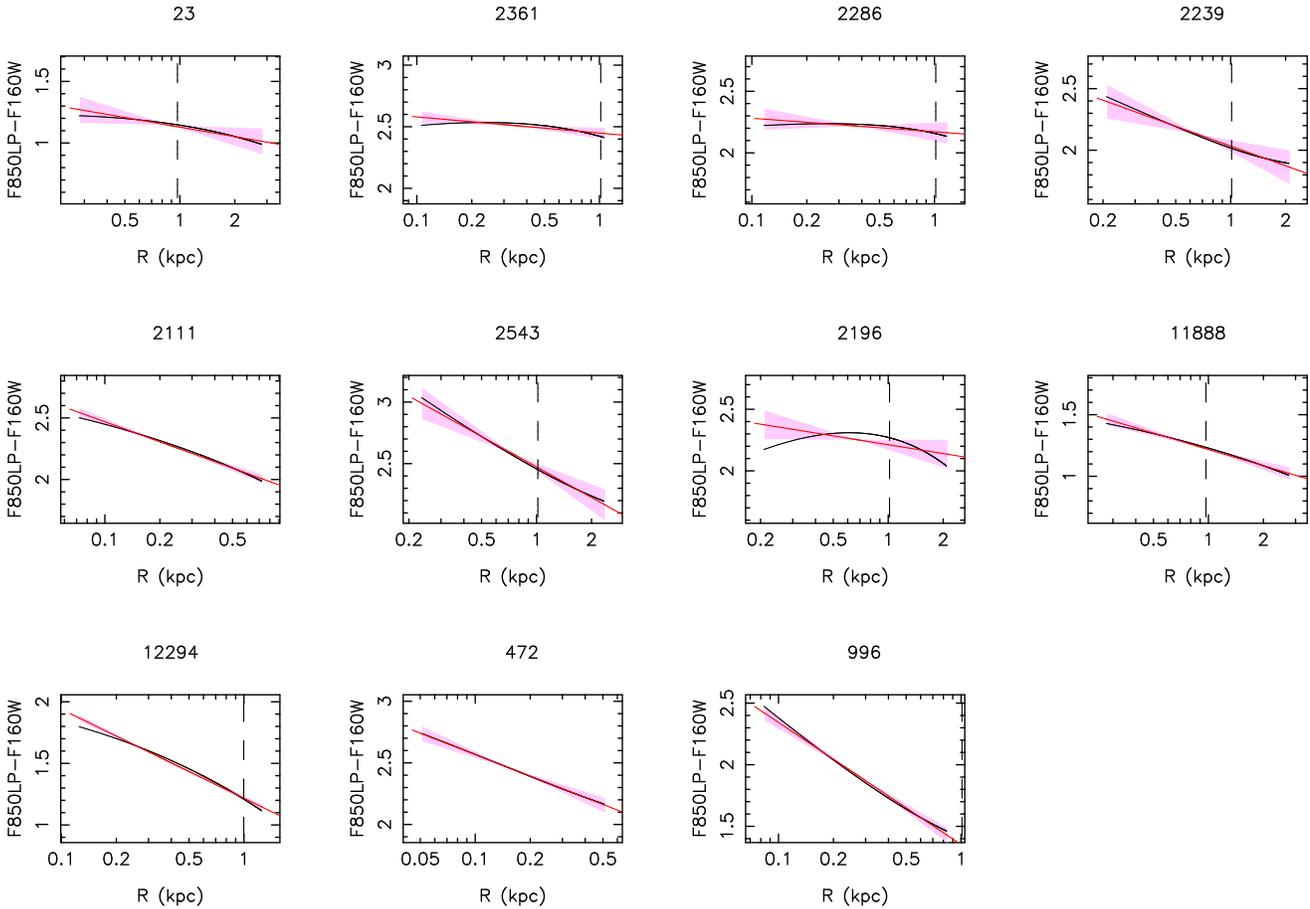} \\
	\caption{The colour gradients for the galaxies of our
	sample. Black lines represent the deconvolved colour profiles
	between 0.1R$_{e}$ and R$_{e}$, and the red lines are the best
	fitted lines to the models. The dashed vertical lines correspond 
	to the radius of the FWHM of the PSF. Colored
	area indicate the 1$\sigma$\, errors on colour profile
	slopes.}
	\label{cg}
\end{figure*}

We recall that color gradients are generally defined as the logarithmic slope of the color
profile:
\begin{equation}
\nabla_{F850LP-F160W} = \frac{\Delta
(\mu_{F850LP}-\mu_{F160W})(R)}{\Delta \log R}.
\label{sb}
\end{equation}
To obtain the measure of this quantity, we have re-estimated the surface brightness parameters in the 
F850LP band, keeping fixed in the fit of each galaxy
the ellipticity and position angle to the values 
derived for the F160W filter. This ensures that both the $\mu_{F850LP}$(r) 
and $\mu_{F160W}$(r) profiles trace the variation of light density in the same radial
direction. Following the prescription commonly adopted for nearby ETGs and
for consistency with GSL11, we have fitted the slope of the color profile between
0.1r$_{e,F850LP}$ and 1r$_{e,F850LP}$. The best-fitting line is obtained
by an orthogonal least-squares fit, less sensitive to 
outliers than the direct fit. Fig. \ref{cg}
reports the results. Black curves indicates the color profiles, while
red lines are the fitted slopes. Dashed lines mark the radius of the
FWHM of the PSF, while colored area set 1$\sigma$ errors. 

\begin{figure*}
	\includegraphics[width=4.1cm,angle=-90]{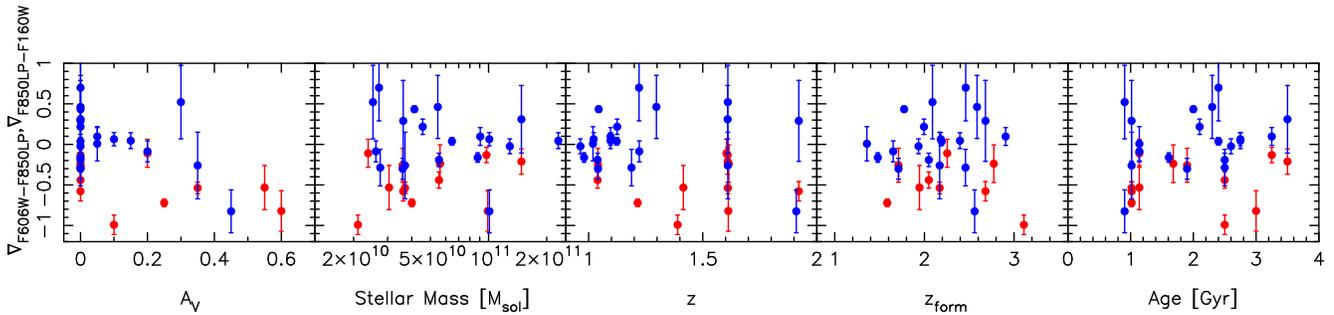} \\
	\caption{The colour gradients F606W-F850LP  and F850LP-F160W (blue and red points, respectively) 
	versus, from left to right, the dust extinction A$_{V}$, the stellar mass, the redshift, the formation redshift and the global age.}
	\label{cgav}
\end{figure*}

The values of the gradients and their
relative errors are reported in Table 1. 
The error on color gradients is an upper limit to the true error. We  estimated it by associating,
at each point of the color profile, the error on color, $\delta col$,
measured directly from the real images in a thin circular corona of radius
$r_e$ (for galaxies with $r_e>$FWHM/2), or FWHM/2 (for galaxies with $r_e<$FWHM/2).
The  $\delta col$ was estimated after reporting the F160W and
F850LP images to the same pixel scale and PSF (see Sec. 5.1).
Our results show that
the high-z ETGs in our sample have negative or null F850LP-F160W color
gradients. Indeed, 3 galaxies have color gradients
comparable at 1$\sigma$ with zero.
On the contrary, the remaining galaxies show
significant negative values which reveal the presence of stellar
populations redder in the center  than those at 1R$_{e}$. The
differences observed in the radial variation of the F850LP-F160W color within
the galaxies of our sample show that after 3-4 Gyr at maximum from their birth, the  stellar content distribution of high-z ETGs was not homogeneous.
This non homogeneous distribution of the stellar content
within the galaxies of our sample corroborates the idea that ETGs have assembled their mass through different mechanisms.\\
To gain insight into the origin of the observed distribution, we firstly have 
looked for the presence of any correlation of observed color gradients with global properties of ETGs. 
Fig. \ref{cgav} reports the F606W-F850LP (blue points) and F850LP-F160W (red points) color gradients versus the dust extinction A$_{V}$ (first panel), the total stellar mass (second panel), the redshift (third panel), the formation redshift (fourth panel; see \citet{saracco11} for the definition and estimate of formation redshift) 
and global age (last panel) of the galaxy. No correlation was detected.  Indeed, the absence of a correlation could be due to the fact that we plot color variation within 1R$_{e}$ versus quantities estimated on the entire galaxy.
We have verified that the result does not change when the color gradients are estimated throughout the whole galaxy. 
 Thus, the various mechanisms responsible of the different stellar content observed in our high-z ETGs, at the extend of our sample, produce color gradients which are independent from redshift, formation redshift, stellar mass, global age and dust extinction. 
This result confirms that the observed distribution of color
gradients is primarily due to differences in the stellar content
of high-z ETGs. Different results were reached by \citet{guo11} who found a correlation between 
dust and color gradients in the direction of steeper gradient for galaxies with higher 
dust-obscuration. 
\citet{guo11} fixed the dust extinction assuming a Salpeter IMF  in the SED fitting, while we have adopted a Chabrier IMF. Furthermore they derived their color gradients fitting the color profiles between 
$\sim$ 1r$_{e}$ up to 5-8r$_{e}$. 
To properly compare our results with those of Guo et al., we have re-estimated the color gradients in our sample in regions similar to those selected in their paper, and computed A$_{V}$ assuming a Salpeter IMF (see Table 2). Even in this case we do not find any correlation between dust extinction and color gradients. Actually, two galaxies of our sample were analyzed also in Guo's paper (ID 472 and 996 our paper, 23555 and 22704 in their). 
Our estimates of the effective radius of these two galaxies are $\sim$ 13$\%$ smaller than their ones, thus consistent within the typical errors on this quantity. Moreover,
for both galaxies our estimates of A$_{V}$ are consistent with those reported by \citet{guo11}. Since \citet{guo11} did not report the gradients values,
we try to qualitatively compare our color profiles with their ones. We would like to stress here that, for these galaxies, the typical error on color at 7r$_{e}$ is $\sim$ 0.15 mag. For the galaxy 472 (see our Fig.\ref{analysis} and magenta line in the lower panel of 
 their fig. 5) the agreement between the two color profiles seems very good. 
For the galaxy 996 we have found a global negative trend in color
values, but the amplitude of the color gradient we detect is less steep than the one measured by Guo, due to the external region (5r$_{e}<$r$<7r_{e}$) where their profile falls suddenly
down, while our one follows a quite costant variation (see our Fig.\ref{analysis} and green line 
in the lower panel of their fig. 5).
The reason for this difference can be due 
to the presence of another galaxy at very low projected distance (see our Section 5.1 and their Fig. 1). In the F850LP band, this source is not detected due to its faint magnitude, thus the extreme blue color observed in the external region by Guo et al. could be due to an underestimate of the flux  in the solely F160W band. 
On the contrary, in the fitting procedure of the surface brightness profile, we have
simultaneously modeled both galaxies and assured from the residual map that the contribution of the other galaxy was properly taken into account. 
If this explanation is correct, the consequent flattening of the color 
gradient value of this galaxy would strongly weaken the correlation between color gradients and extinction also in their sample and the eventual residual correlation should be totally due to the small statistics of their sample.
\section{Spatial distribution of stellar populations in high-z ETGs and radial variation of their
properties.}
In order to constrain the possible physical mechanisms responsible of the mass accretion within the ETGs, 
we have designed a new method able to investigate 
the spatial distribution of the underlying stellar populations in high-z ETGs which is directly 
correlated to the mass assembly processes. Briefly, we have derived the color maps of our galaxies 
and, on their basis, we have defined for each ETG of our sample an internal and an external region.
Then, we have modelled the \textit{global} stellar content of each ETG as formed by two main stellar components,
one dominating the internal regions, and the other the external ones.
With synthetic models of composite stellar populations we have constrained the stellar properties of 
the two components in order to simultaneously reproduce 
the observed color gradient(s) (in one or two colors)
and the whole global SED of the galaxy from 0.3$\mu$m to 8$\mu$m.

\subsection{Color maps of high-z ETGs: a direct look at the spatial distribution
of their stellar content.}

\begin{figure*}
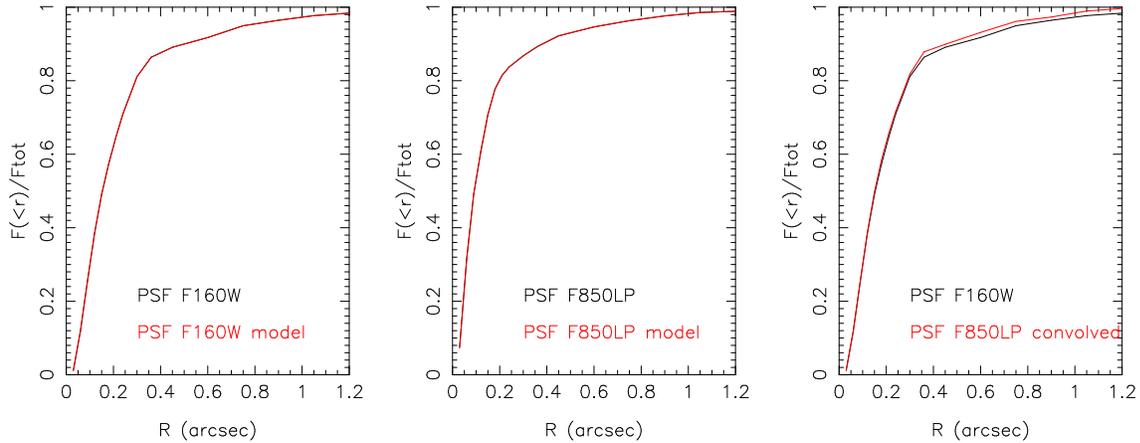

  \begin{tabular}{ccc}
	\includegraphics[width=5.8cm,angle=-90]{figure5.ps} &
	\includegraphics[width=5.8cm,angle=-90]{figure5_1.ps} &
	\includegraphics[width=5.8cm,angle=-90]{figure5_2.ps}\\
  \end{tabular}
  \caption{$Left$ $panel$: the fractional encircled energy for the
  real PSF in the F160W image (black line) and for the model (red
  line). $Central$ $panel$: the same for the F850LP image. $Right$
  $panel$: the fractional encircled energy for the real PSF in the
  F160W image and for the PSF of the F850LP image convolved with the
  kernel function $K$(r).}
  \label{psfmatch}
\end{figure*}

\begin{figure*}
  \begin{tabular}{cccc}
	\includegraphics[width=4.1cm,angle=0]{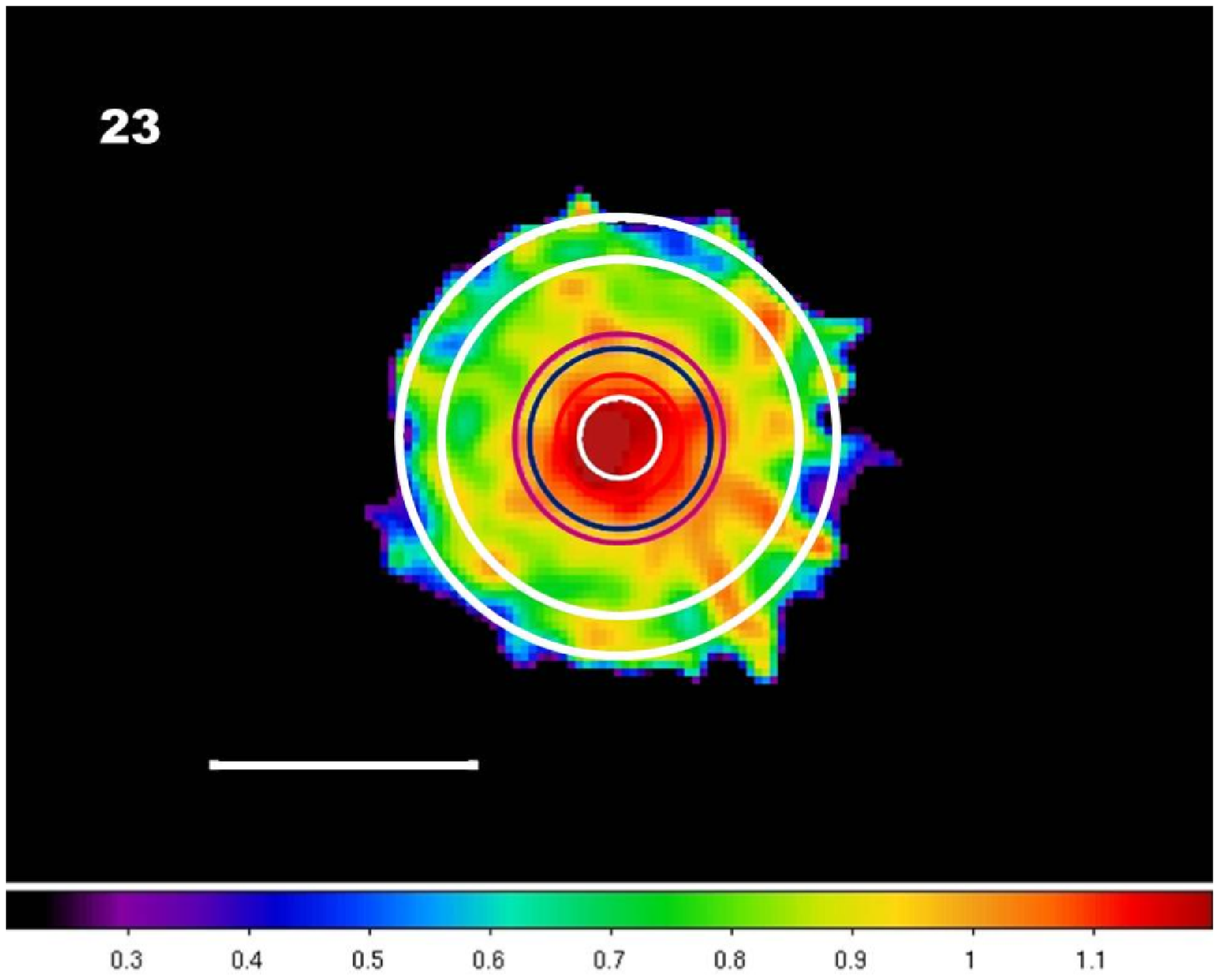} &
	\includegraphics[width=4.1cm,angle=0]{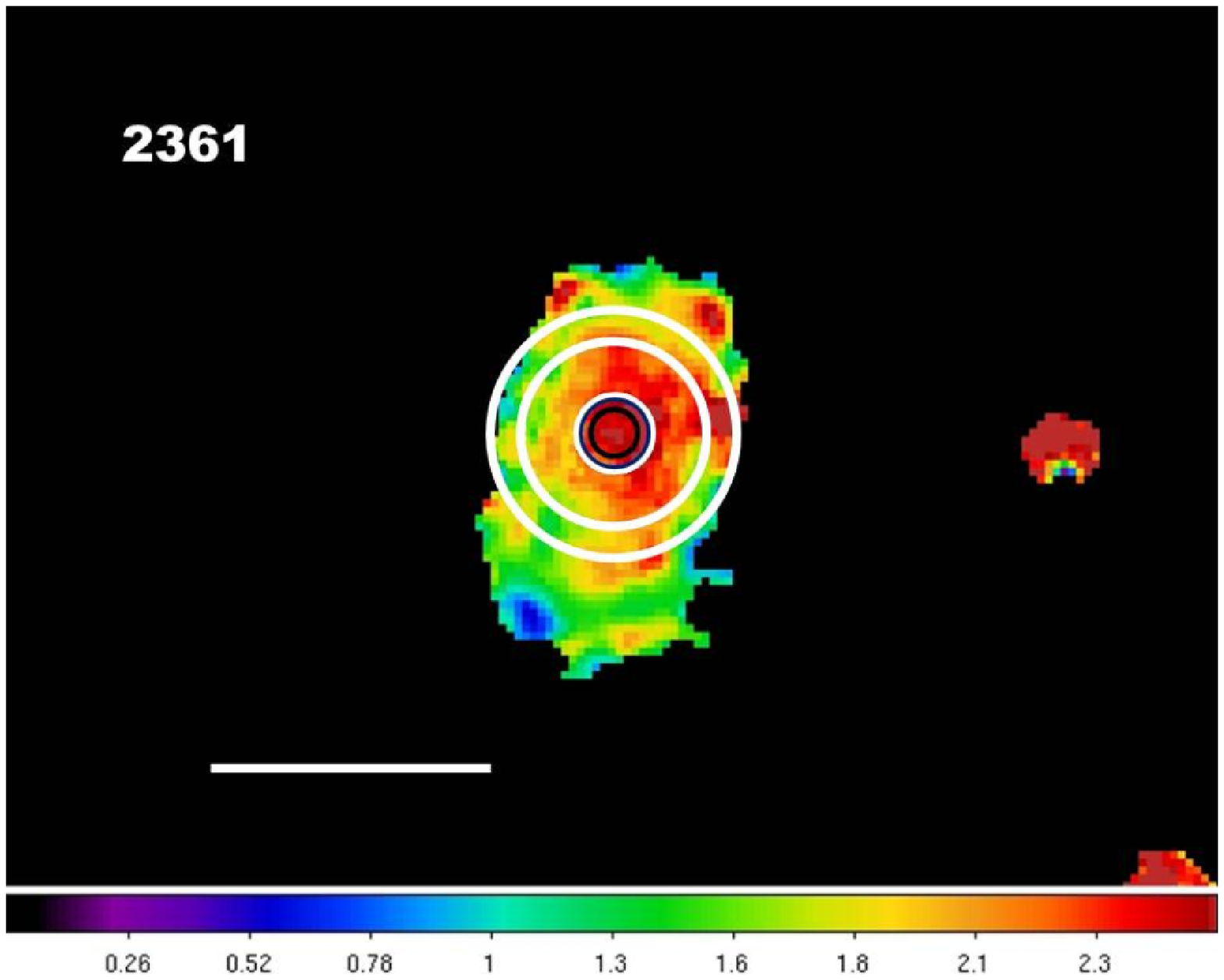} &
	\includegraphics[width=4.1cm,angle=0]{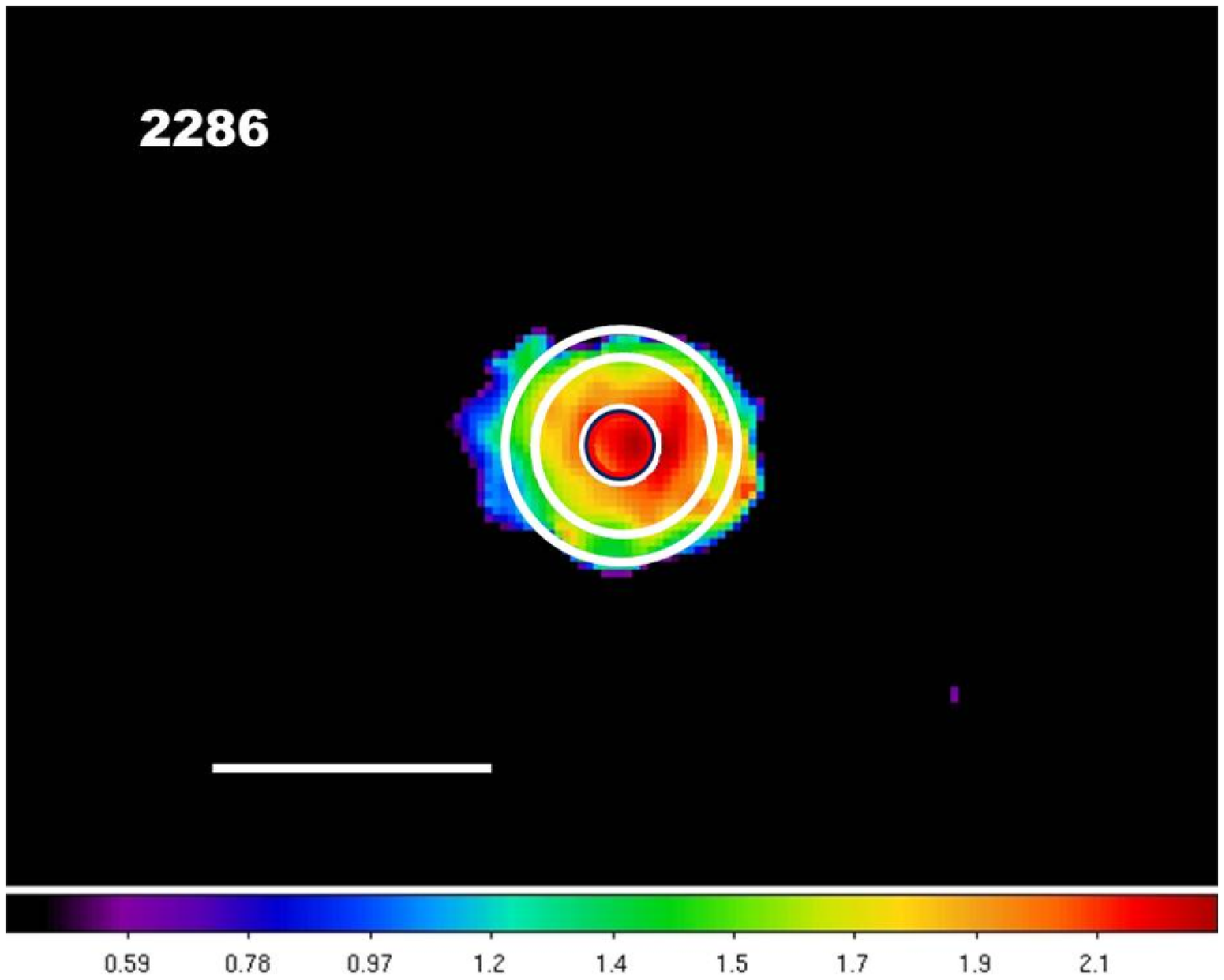} &
	\includegraphics[width=4.1cm,angle=0]{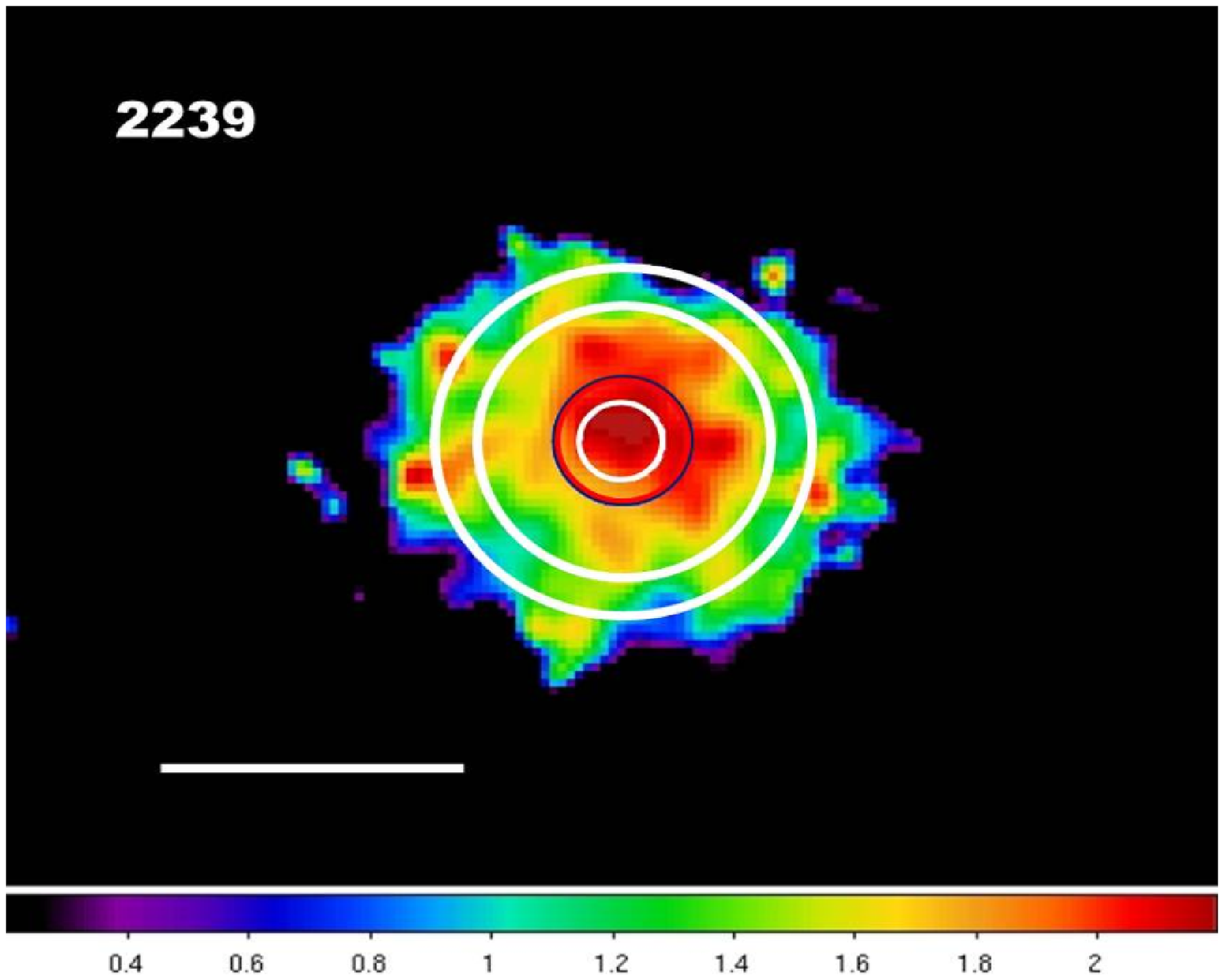} \\
	\includegraphics[width=4.1cm,angle=0]{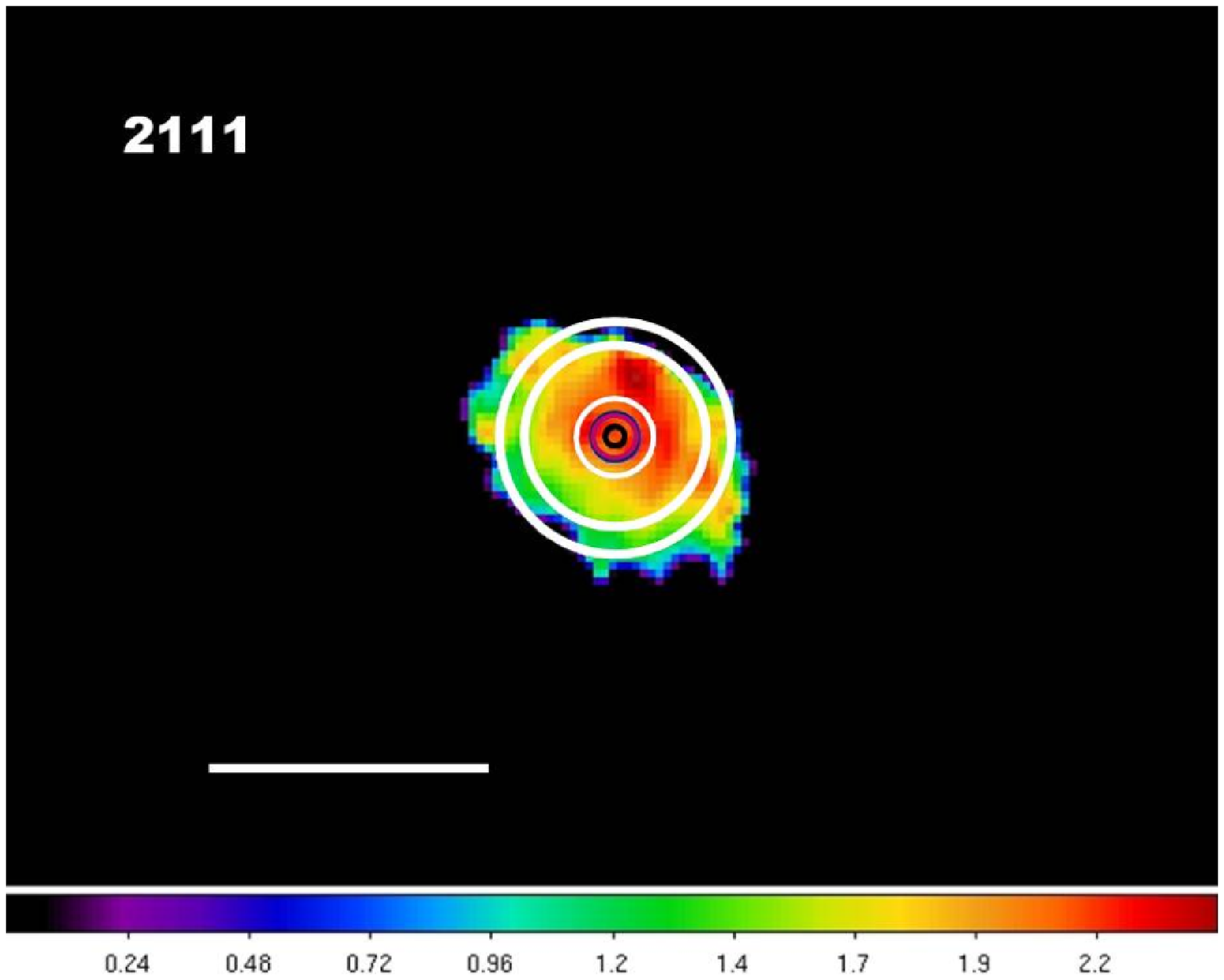} &
	\includegraphics[width=4.1cm,angle=0]{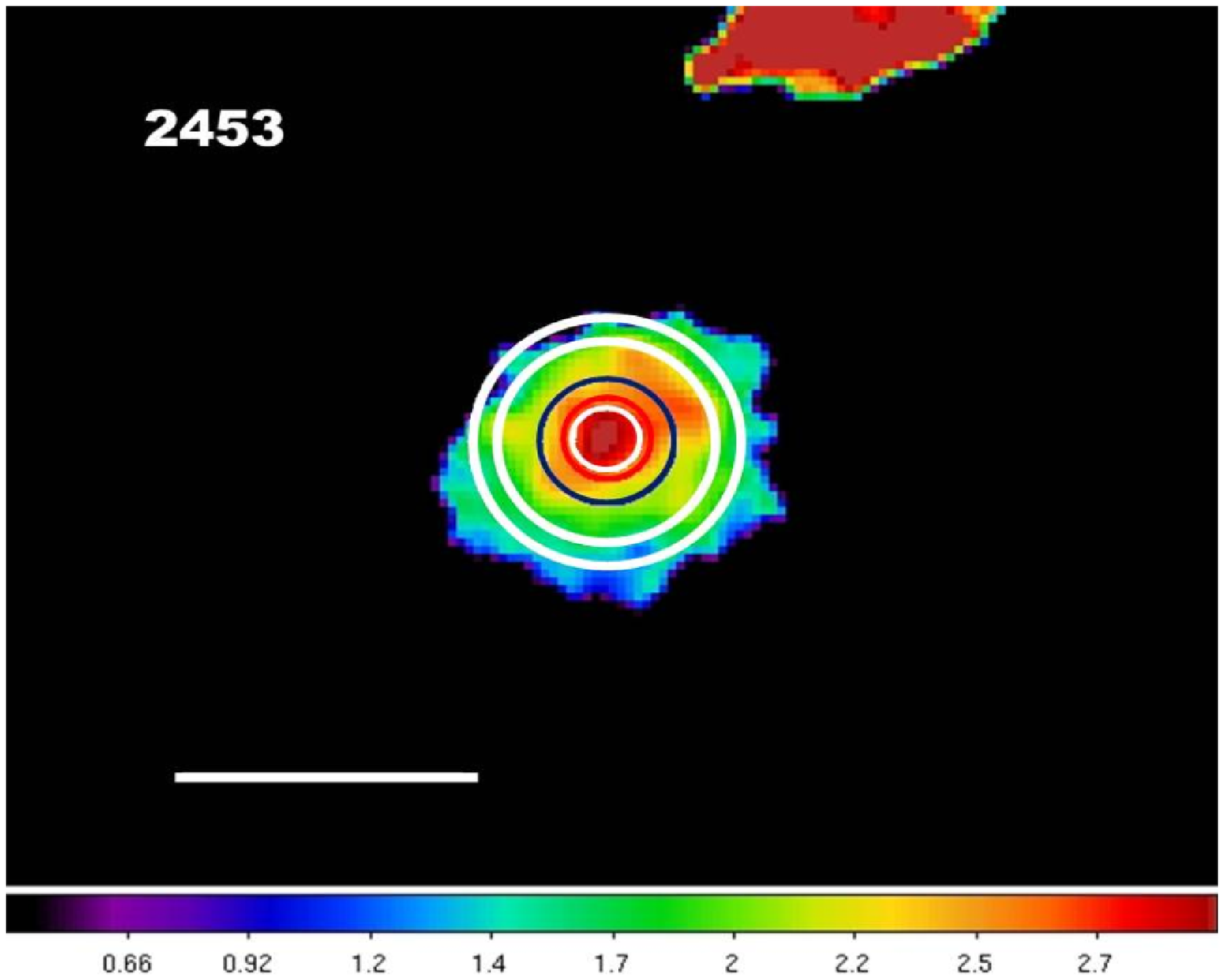} &
	\includegraphics[width=4.1cm,angle=0]{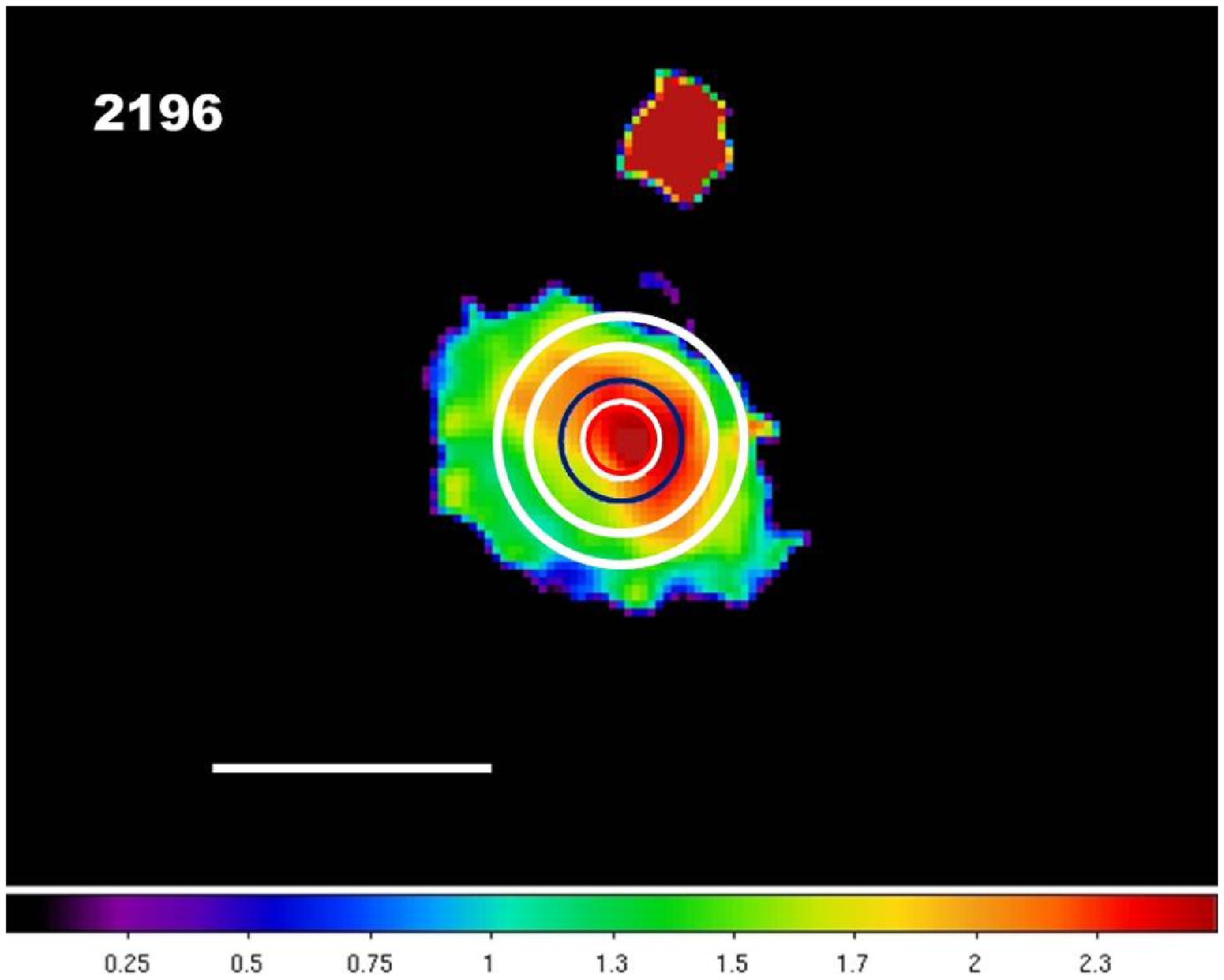} &
	\includegraphics[width=4.1cm,angle=0]{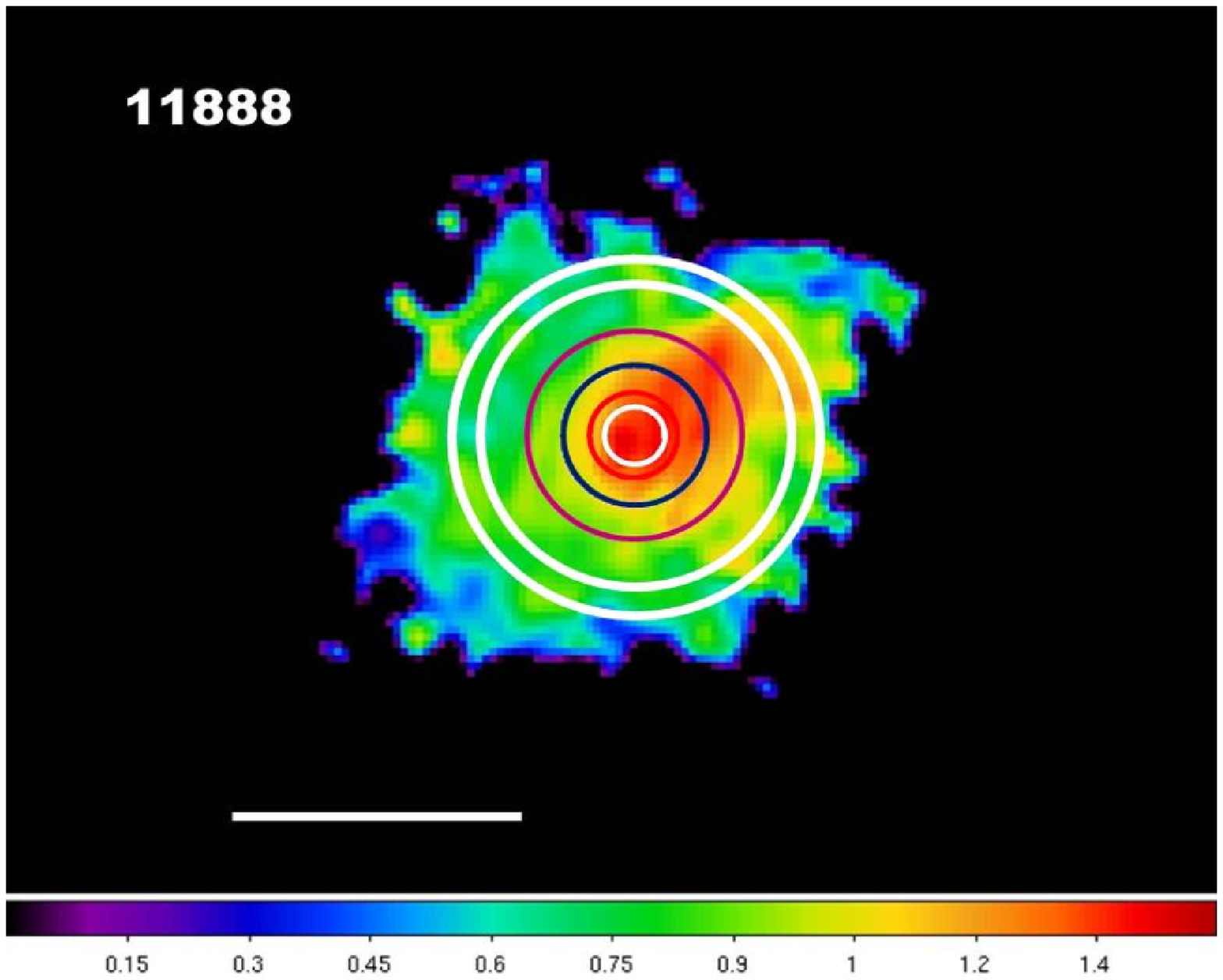} \\
	\includegraphics[width=4.1cm,angle=0]{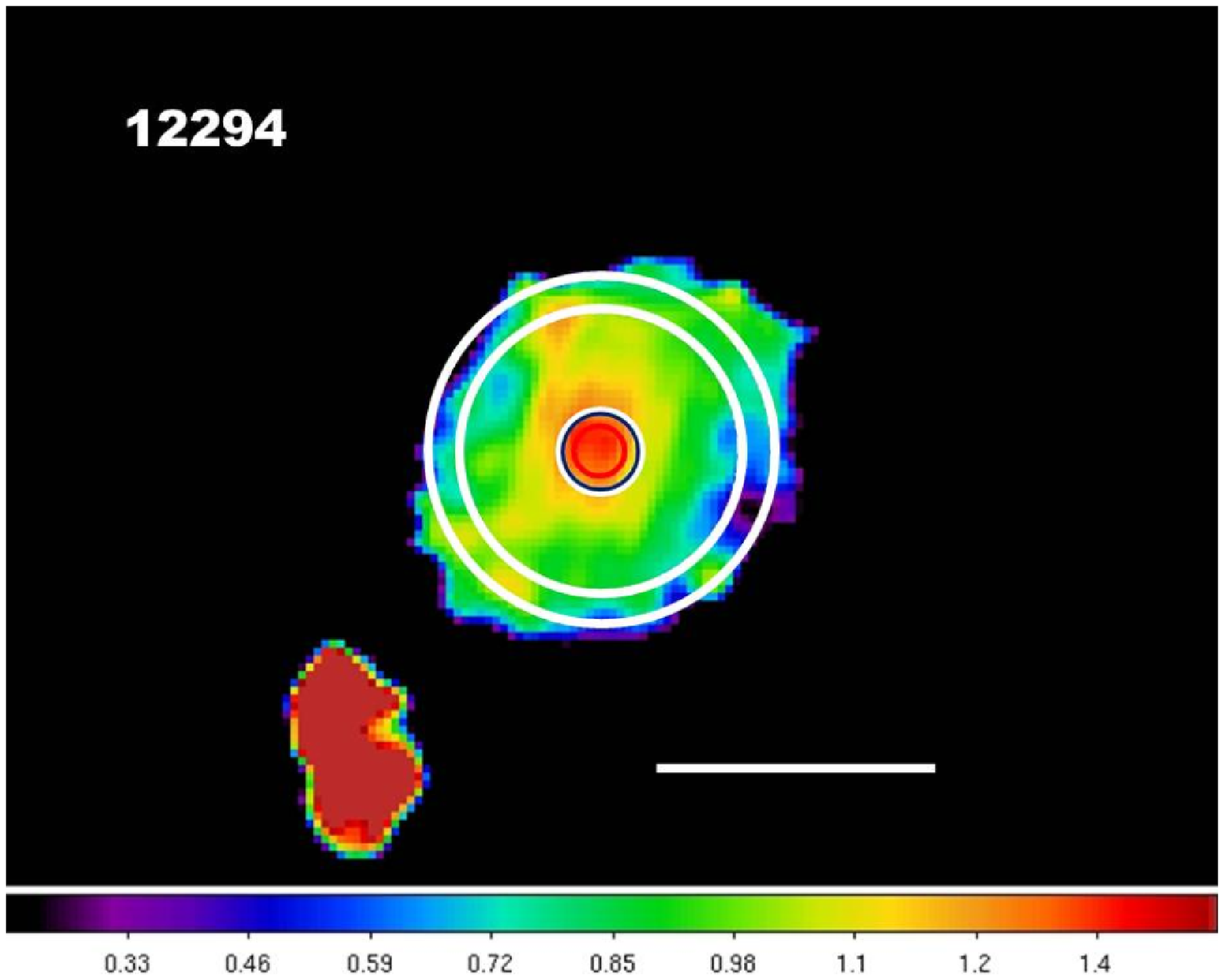} &
	\includegraphics[width=4.1cm,angle=0]{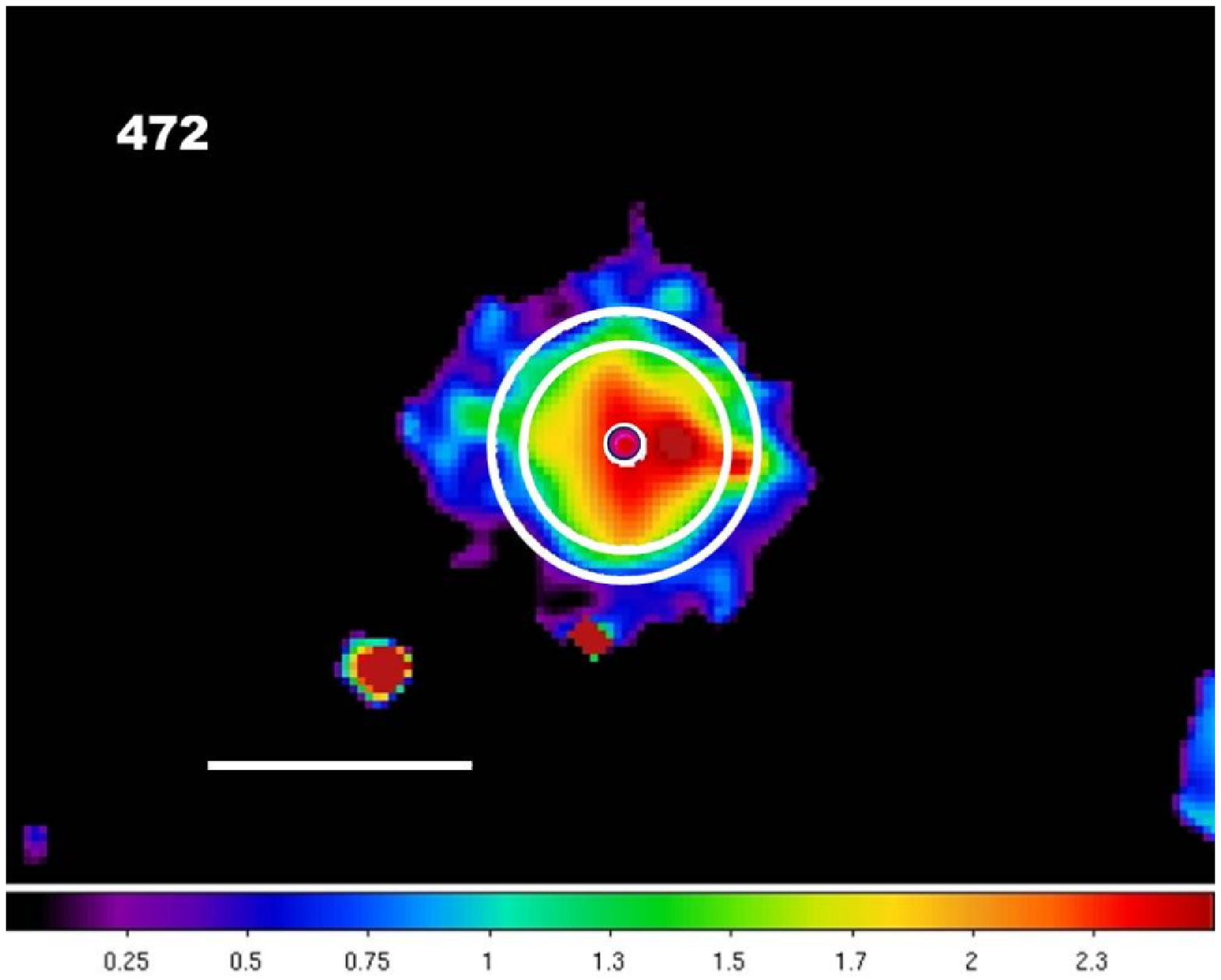} &
	\includegraphics[width=4.1cm,angle=0]{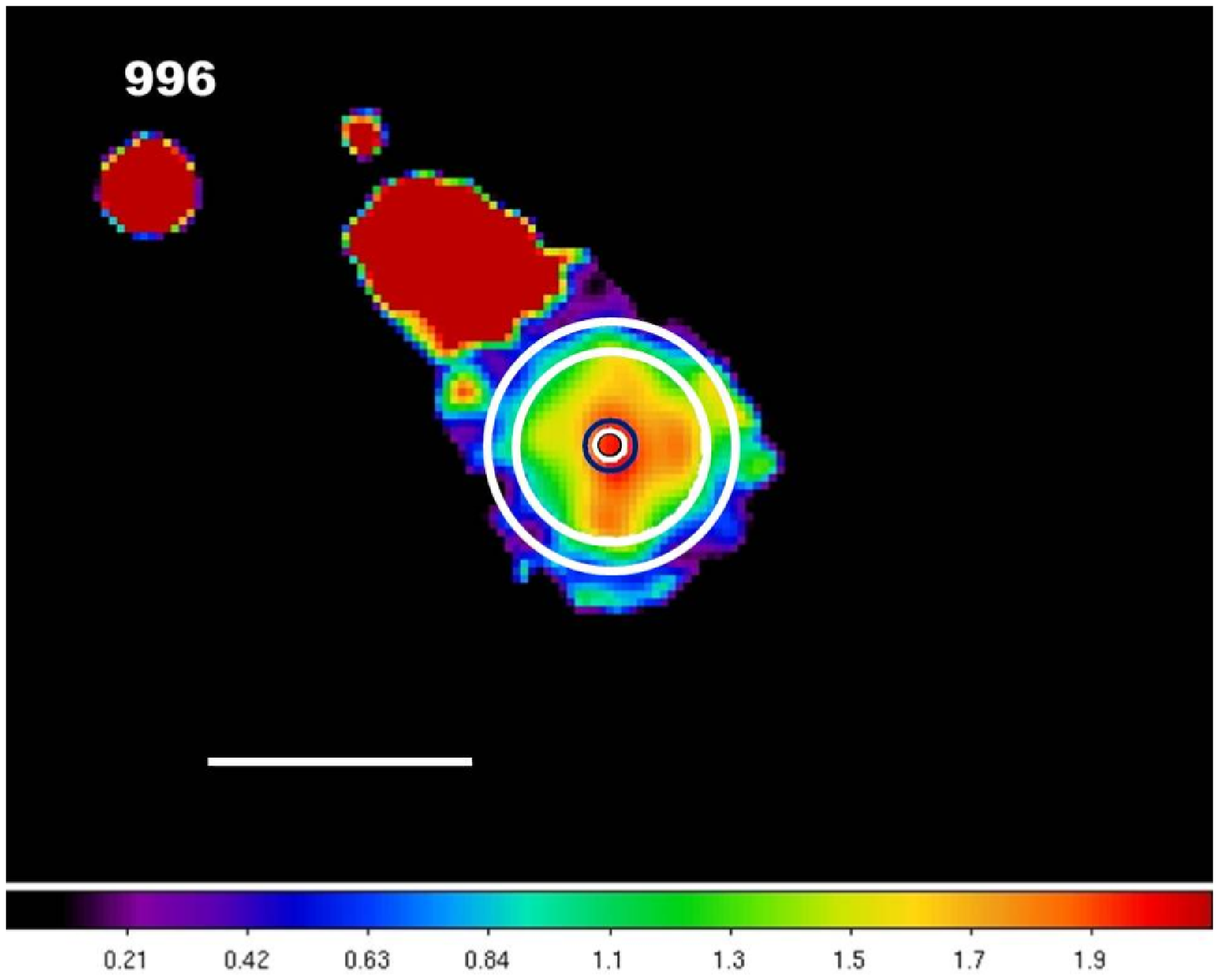} & \\
  \end{tabular}
  \caption{F850LP-F160W colour maps for the galaxies of our sample. The maps are color coded, indicating, from blue to red, region of the galaxy characterized by a redder emission. For each galaxy, the inner white circle constrain the internal region, while the two external ones define the circular corona which identify the external region. Magenta, blue and red circles set the effective radius in the F606W (when available), F850LP and F160W band, respectively. The white line corresponds to 1 arcsec.}
  \label{colormap}
\end{figure*}

To have a direct look on the punctual 2-dimensional distribution of F850LP-F160W color in high-z ETGs, we have derived their color maps. 
To this aim, we have ri-reduced the WFC3 images to a pixel scale of 0.03 ``/px to match the one of F850LP-mosaic images, and used the IRAF tasks
geomap/geotran to align the two mosaics.  Then,
we have degraded the F850LP images (FWHM $\sim$
0.12'') to the same PSF of the F160W
images (FWHM $\sim$ 0.2''). To this aim, we needed to 
derive the kernel function $K$(r) which regulates the
transformation between the PSFs in the two bands, i.e.
which holds:
\begin{equation}
PSF_{F850LP}(r) \star K(r) = PSF_{F160W}(r)
\label{kernel}
\end{equation}
where PSF$_{F850LP}$(r) and PSF$_{F160W}$(r) are the point spread
functions in the F850LP and F160W band, respectively, and where the
symbol $\star$ denotes a convolution.  For each galaxy, we have
modeled its PSF$_{F850LP}$(r) and PSF$_{F160W}$(r) functions and
through Eq.\ref{kernel} we have recovered its kernel function. 
The PSF$_{F850LP}$(r) and PSF$_{F160W}$(r) functions of a given
galaxy have been derived by fitting the light profiles of the unsaturated
stars we have adopted as PSF in GALFIT and that we have already tested to be a
good approximation of the $true$ PSF for that galaxy.  
In Figure \ref{psfmatch}
we report a representative example of the fractional encircled energy
for the real PSF (black line) of a star in the F160W (left panel) and
F850LP (central panel) band and the corresponding fitted model (red
line). The agreement between the two curves is extremely good in both
cases, confirming the quality of our fit.  
In the last panel of Fig. \ref{psfmatch} we report the encircled energy
distribution of the PSF$_{F160W}$(r) (black line) and of the PSF$_{F850LP}$(r)
model convolved with the kernel (red line). The match between the two
functions is perfect and differences $<$ 2$\%$ appear at R$>$ 4'',
when present. In particular, it is to note the optimal agreement in
the central region that, as said before, is the one mainly affecting
the galaxy light profile. 
This has allowed us to
degrade the F850LP images to the same PSF of F160W
images and to derive the color maps presented in Fig. \ref{colormap}.
The images are color coded with the reddest color indicating the
region of the galaxies with highest F850LP-F160W values. Red, blue and magenta
circles indicate the r$_{e}$ in the F160W, F850LP and F606W band,
respectively. As pointed out before, we have at disposal information
on F606W band only in 4 galaxies out of 11. In some cases the
r$_{e}$ in the F160W band is represented with a black circle being the
red one not visible on the color map. The white lines correspond to 1
arcsec. 
In Fig. \ref{colormap} we have marked all the pixels 4.5$\sigma$ 
above the mean background in the F160W images as belonging to
the galaxy. In spite of this high threshold, the color maps extend to regions well beyond 2r$_{e,F850LP}$ and at radius greater than 4r$_{e,F850LP}$ for 5 cases (12294, 996, 2361, 2111, 472).
The color map of the whole galaxy
show that differently from the gradients estimated up to 1r$_{e,F850LP}$, $all$
high-z ETGs of our sample have a negative color gradient.
Moreover, they point out that although the general trend
is the same, the F850LP-F160W color varies from the internal to the external
regions with different gradients showing that
both the spatial distribution and the properties of the stellar
content are not homogeneous in high-z ETGs.
\subsection{Radial variation of the stellar populations properties of high-z ETGs.}
To identify the stellar population parameters whose variation can account for the observed color gradients, 
we have used the following procedure. On the basis of the color maps which show a quite radial variation of color within the ETGs of our sample, we have schematized the whole stellar content of each of them with a two-components model assuming that a
component dominates the central regions and the other dominates the
external regions. By means of the color maps
we have
selected the two areas (white circles in Fig. \ref{colormap}) 
in order to maximize both the difference in the mean color of the two regions and the radial distance among them. At fixed gradient, maximizing the distance of the two regions means maximizing their color difference, and this will facilitate in detecting   
differences between the
stellar population parameters of the two components and hence in detecting possible radial variations. 

We have assumed as initial guess for the stellar parameters of the two components the values derived from the fit
of the global SED [Age$_{global}$, Z$_{global}$, A$_{V,global}$,
$\tau_{global}$]. 
Briefly, for each galaxy, the global stellar
population parameters were derived through the fit of the spectral
energy distribution defined by 14 photometric points and with  known
spectroscopic redshift. The fit was performed with the software
Hyperzmass \citep{bolzonella00} using the composite stellar population models of Charlot $\&$ Bruzual (hereafter CB07) with an exponentially declining star-formation history ($\propto$ \,$\exp^{-t/\tau}$),
and a Chabrier initial mass function \citep{chabrier03}
The best solution was selected on a grid of model with
varying age, star-formation time scale $\tau$ [0.1, 0.3, 0.4, 0.6] Gyr, and solar
metallicity. Extinction A$_{V}$ was left as free parameter in the 0.0 - 0.6 range and extinction curve of \citet{calzetti00} was adopted.
For more details on SED fitting see \citet{saracco10}. For
each galaxy of our sample, the set of the four best-fit parameters are reported
in Table \ref{globalSED}.
\begin{table*}
\caption{Stellar population parameters derived from the fit of global the SED
assuming a Chabrier IMF and a Salpeter IMF. 
  $Column$ $1$: id number; $Column$ $2$:
  Age obtained with a Chabrier IMF$|$Age obtained with a Salpeter IMF; $Column$ $3$: $\tau$ obtained with a Chabrier IMF$|\tau$ obtained with a Salpeter IMF;
  $Column$ $4$: A$_{V}$ obtained with a Chabrier IMF$|A_{V}$ obtained with a Salpeter IMF; $Column$ $5$:
  $\log$ M$_{\star}$ obtained with a Chabrier IMF$|\log$M$_{\star}$ obtained with a Salpeter IMF. In each fit we assume solar metallicity.}
  \footnotesize
\begin{tabular}{ccccc}
\hline
Object          & Age$_{Cha}|$Age$_{Sal}$&  $\tau_{Cha}|\tau_{Sal}$  & A$_{V,Cha}|$A$_{V,Sal}$&  $\log$ M$_{\star,Cha}|$ $\log$ M$_{\star,Sal}$                                             \\ 
                &  Gyr                    &  Gyr                          &   mag                   &  M$_{\odot}$                                  \\
\hline
23              & 1.90$|$2.1                & 0.1$|$0.1                       & 0.0$|$0.0                 &  10.5$|$10.9                                       \\
11888           & 2.5$|$2.4                 & 0.3$|$0.3                       & 0.0$|$0.0                 &  10.7$|$11.0                                       \\
12294           & 1.0$|$1.0                 & 0.1$|$0.1                       & 0.25$|$0.25               &  10.6$|$10.9                                       \\
996             & 2.5$|$1.7                 & 0.3$|$0.2                       & 0.10$|$0.30               &  10.3$|$10.5                                       \\
2239            & 1.1$|$1.1                 & 0.1$|$0.1                       & 0.55$|$0.60               &  10.5$|$10.8                                       \\
2286            & 1.1$|$1.1                 & 0.1$|$0.1                       & 0.20$|$0.25               &  10.4$|$10.7                                       \\
2361            & 3.25$|$3.25               & 0.3$|$0.3                       & 0.0$|$0.0                 &  11.0$|$11.3                                       \\
2148            & 3.5$|$3.25                & 0.4$|$0.4                       & 0.0$|$0.05                &  11.2$|$11.4                                       \\
2196            & 1.7$|$1.9                 & 0.1$|$0.2                       & 0.0$|$0.05                &  10.7$|$11.10                                      \\
2543            & 3.0$|$3.25                & 0.4$|$0.5                       & 0.6$|$0.6                 &  11.0$|$11.3                                       \\
2111            & 1.0$|$1.0                 & 0.1$|$0.1                       & 0.35$|$0.60               &  10.6$|$11.3                                        \\
472             & 1.0$|$1.0                 & 0.1$|$0.1                       & 0.0$|$0.0                 &  10.6$|$10.9                                       \\
\hline
 \label{globalSED}
 \end{tabular}
 \normalsize
\end{table*}

In order to reproduce the color gradient we observe, 
it is necessary that one or more parameters of the two components vary from the internal 
to the external region. Unfortunately, the effect of age, metallicity, dust and $\tau$ 
variation on the galaxy emission is degenerate in the spectral region we are observing, 
preventing, in fact, to identify the contribution of each parameter to the total emission.
For this reason, we have performed our analysis by investigating the
effect of
each single parameter in turn, without considering the possibility of
either a correlated or
an anticorrelated variation of age and metallicity within galaxies.
Thus, we have fixed in both components three out of the
four parameters in the fitting. For example to investigate radial age variation as possible driver of color 
variations, we fixed the metallicity, the dust extinction and the star-formation time scale to 
the value obtained by the fit of the global SED: Z$_{global}$, A$_{V,global}$, $\tau_{global}$. 
We have looked for the value
of age to be associated to the internal and to the external populations, Age$_{in}$
and Age$_{out}$ respectively, which best reproduce the observed color
gradients. In particular we have chosen the value of Age$_{in}$ so that
minimizes the quantity 
\begin{equation}
	\begin{split}
& |(F850LP-F160W)_{mod,in} - (F850LP-F160W)_{obs,in}| + \\
& +|(F606W-F850LP)_{mod,in} - (F606W-F850LP)_{obs,in}|
	\end{split}
	\label{eqmin1}
\end{equation}
where (F850LP-F160W)$_{mod,in}$ and (F606W-F850LP)$_{mod,in}$ are the F850LP-F160W and the F606W-F850LP colors predicted by the model defined by the stellar population
parameters (Age$_{in}$, Z$_{global}$, A$_{V,global}$,
$\tau_{global}$), and (F850LP-F160W)$_{obs,in}$ and (F606W-F850LP)$_{obs,in}$  are
the mean value of the colors observed in the internal region. Similar procedure has been adopted to identify  Age$_{out}$.
In the cases for which F606W-F850LP color gradients are not available we clearly ignore the second lines in Eq.\ref{eqmin1}.
To fix the contribution to the total stellar mass of the two components so selected, we fit the global SED sampled by 14 photometric points with a 
linear combination of the SEDs of the two populations.

We have repeated this analysis also for variation of
metallicity Z, keeping fixed [Age$_{global}$, A$_{V,global}$,
$\tau_{global}$], and star-formation $\tau$ with [Age$_{global}$,
Z$_{global}$, A$_{V,global}$] fixed.  We have looked for the values of
metallicity to be associated to the internal and external population,
Z$_{in}$ and Z$_{out}$ respectively, on a grid of sub-solar and
super-solar values: 0.2Z$_{\odot}$, 0.4Z$_{\odot}$, Z$_{\odot}$ and
2Z$_{\odot}$, while for the star-formation time scale we have
considered the range 0.1 -0.6Gyr. 

\subsubsection{An example}

\begin{figure*}
    \includegraphics[width=16.0cm,angle=0]{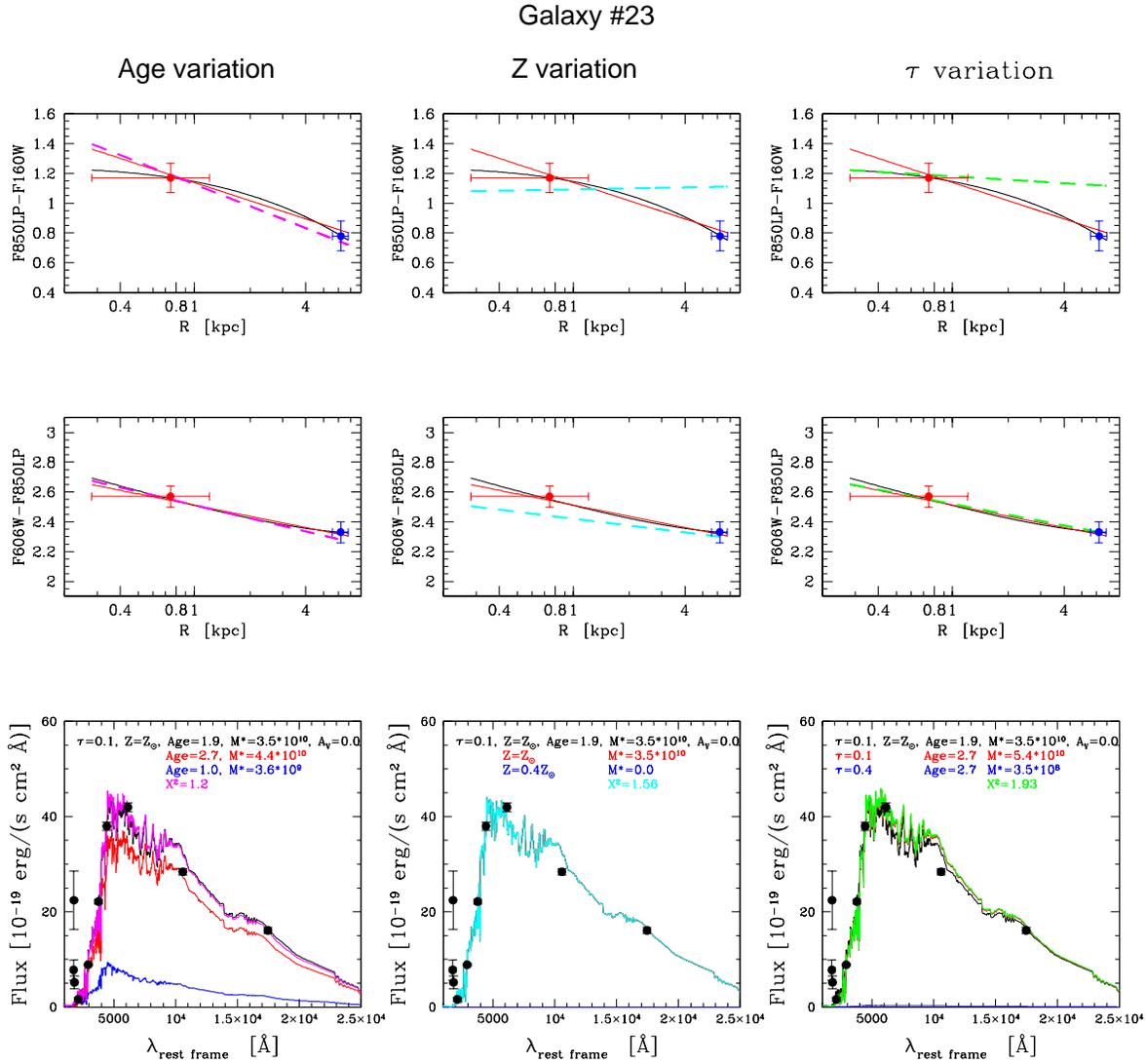}
  \caption{\textit{First column}: analysis of radial age variation as possible driver of observed color gradients. 
Top and middle panels show the F850LP-F160W and F606W-F850LP color profiles, respectively (black lines), whit the relative fit (red lines). The bottom panel shows the observed global SED (black dots) as well as the stellar population parameters (black text) derived from its fit (black line). Among all the stellar populations with varying age but star-formation time scale, metallicity and dust extinction equal to the value obtained from the fit of the global SED, the red text identifies the one that best simultaneously reproduces the colors observed in the internal region (red dots in the top and middle panel). Similarly, blue text reports the value of the age of the stellar population that best reproduce the external colors (blue dots in the upper panels). The pure linear variation of the age of the galaxy stellar populations from the internal value to the external one, will produce the F850LP-F160W and F606W-F850LP color gradients represented with the dashed magenta lines. The contribution to the total stellar mass of the two populations is fixed in order to best fit the whole SED and are reported in the bottom panel (red and blue lines). \textit{Second and third column}: the same of first column but for metallicity and star-formation time scale variation.}
  \label{analysis}
\end{figure*}

In Fig. \ref{analysis} we report an example of the analysis
described above for a galaxy of our sample.
We select a case which presents some peculiarities to guide the interpretation of the analysis in this more complex case and left to the reader the interpretation of the analysis of the other galaxies not discussed here but presented in electronic form. 
The first column reports the analysis relative to the effect on galaxy colors of 
pure age variation from the inner to the outer regions. The second
and the third column show the same but for metallicity and $\tau$ radial variation,
respectively. In each column we report, from top to bottom, the
F850LP-F160W color gradient, the F606W-F850LP color gradient, when available, and the global
SED. In the panels related to the color gradients, the black lines are
the color profiles.  Red lines are the fit to the color
profiles up to the external regions derived following the same method describe in Section 4.  Red points are the mean color values of the internal
regions measured through color profiles, i.e. (F850LP-F160W)$_{obs,in}$ and (F606W-F850LP)$_{obs,in}$ in Eq. \ref{eqmin1}.
Blue points are the same for the external regions. 
The horizontal error bars
indicate the extension of the two areas we select (white circles in
color maps) and over which we computed the mean colors.

In the bottom panel we report the observed SED (black dots), the best fitting template (black line) and the relative best fitting parameters (black text). Red and blue lines are the contribution of the internal and
external population, respectively, to the total emission, and hence to
the total stellar mass. The amount of their contribution was fixed such that 
their linear combination (magenta, cyan or green lines for age, metallicity and $\tau$ variations, 
respectively) minimizes the residual with the observed SED. In each panel we report the $\chi^{2}$ value of this fit.
In red we present the value of the varying parameter which 
minimize the Eq.\ref{eqmin1} as well as their contribute in mass, and in blue the same for the external region.\\
The solely linear radial variation of this parameter from the internal value (red text), to the external value (blue text) will produce a F850LP-F160W and a F606W-F850LP color gradient indicated in the 
top and middle panels by the dashed lines (magenta, cyan and green in the case of age,
metallicity and $\tau$ variation)
Fig. \ref{analysis} shows that the color gradient we observe in this
galaxy can be due to a variation of the age of the stellar populations
from the inner to the outer regions. Indeed, while the global fit
returns us a mean age of 1.9 Gyr (black text in the bottom left panel), our analysis shows that a population
of 2.7 Gyr dominating the central region (red text in the bottom left panel) and a younger population of
1.0 Gyr dominating the external regions (blue text in the bottom left panel), can perfectly reproduce
within the errors (magenta dashed line in top and middle panels) the internal and external observed F606W-F850LP and F850LP-F160W colors, and hence the color gradients. Further, the analysis of the SED suggests that in
this galaxy the older stellar population contributes to the stellar
mass for more than 90$\%$ (red line, bottom left panel). On the contrary, neither a pure metallicity
variation, nor a variation of the star-formation time scale alone can
reproduce the observed color gradients. In the case of metallicity (middle column), the
internal colors are well reproduced by a population with solar
metallicity (red text in the bottom central panel), i.e. the same of the global fit, while the external color
cannot be simultaneously reproduced by a single value of Z. \\
For what concerns the analysis of the $\tau$ variation this galaxy requires particular attention.
The fit of the global SED suggests a mean $\tau$ for this galaxy of 0.1Gyr, that is the lower value in the grid of the 
star-formation time scales considered. To cope with this, we assume a population of 2.7 Gyr with $\tau$ = 0.1 Gyr as 
representative of the inner part, as we already found that this parameter well fits the inner colors.
For the external regions, fixing the age to 2.4 Gyr, the value of $\tau$ that best reproduces both external colors is 0.4 Gyr. 
However, despite it is the best fitting value, it is not able
to reproduce within 1$\sigma$ the observed F850LP-F160W external color.
\begin{landscape} 
\begin{table}
\caption{Color gradients produced by the variation of a single stellar population parameters  from the internal to the external regions. $\textit{Column 1}$: top line: galaxy id number; bottom line: compactness C defined as R$_{e,z=0}$/R$_{e}$ where R$_{e}$ is the effective radius of the galaxy and R$_{e,z=0}$ is the radius that a galaxy of equal stellar mass would have at z=0 as derived by the size mass relation of \citet{shen03}. ETGs with effective radius more than one sigma smaller than those predicted by the local size-mass relation, i.e. our compact galaxies, turn out to have C$\geq$2. ; $\textit{Column 2}$: Radius up to which we extend the analysis (R$_{ext}$, in R$_{e,F850LP}$ unit); $\textit{Column 3}$: F850LP-F160W measured color gradient$|$F606W-F850LP measured color gradient derived up to R$_{ext}$ from the fit of the color profile; $\textit{Column 4}$: top line: F850LP-F160W color gradient$|$F606W-F850LP color gradient produced by a solely radial age variation from Age$_{in}$ in the internal region to Age$_{out}$ in the external region (see bottom line); bottom line: Age$_{in}$: age of the internal population - Age$_{out}$: age of the external population - $\nabla_{age}$: age gradient derived as $d$log Age/$d$log r; $\textit{Column 5}$: the same of column 4 for metallicity radial variation; metallicity gradient is derived as $d$log Z/$d$log r;  $\textit{Column 6}$: the same of column 4 for star-formation time scale radial variation;  $\textit{Column 7}$: the same of column 4 for IMF's slope radial variation. In the case of star-formation time scale and IMF's slope radial variation the $\nabla_{\tau}$ and $\nabla_{\alpha}$ are omitted. In bold case are reported the variations able to reproduce both the color gradients and observed global SED.}
\footnotesize
\begin{tabular}{ccccccc}
\hline
ID & R$_{ext}$ & $\nabla_{F850-F160,mis}|\nabla_{F606-F850,mis}$ & $\nabla_{F850-F160,age}|\nabla_{F606-F850,age}$  & $\nabla_{F850-F160,Z}|\nabla_{F606-F850,Z}$ &  $\nabla_{F850-F160,\tau}|\nabla_{F606-F850,\tau}$ & $\nabla_{F850-F160,\alpha}|\nabla_{F606-F850,\alpha}$\\ 
   &           &                             &   Age$_{in}$ - Age$_{out}$ - $\nabla_{age}$  &   Z$_{in}$ - Z$_{out}$ - $\nabla_{Z}$ & $\tau_{in}$ - $\tau_{out}$ &  $\alpha_{in}$ - $\alpha_{out}$  \\
\hline
  & & mag/dex$|$mag/dex & mag/dex$|$mag/dex & mag/dex$|$mag/dex & mag/dex$|$mag/dex & mag/dex$|$mag/dex \\
  &                  &                   & [Gyr] - [Gyr] - Gyr/dex &      & [Gyr] - [Gyr] & \\
\hline
 23  & 2.4  & -0.40$|$-0.24  & \textbf{-0.49$|$-0.29}     &  0.02$|$-0.15                         &  -0.08$|$-0.24  &  -0.06$|$-0.05 \\
  1.1   &      &                & \textbf{2.7 - 1.0 - -0.45}  &  $Z_{\odot}$ - 0.4$Z_{\odot}$ - -0.42  &   0.1 - 0.4      &  1.5 - 3.0 \\ 
\hline
 11888  & 2.5  & -0.51$|$-0.29  & -0.19$|$-0.31     &  -0.16$|$-0.44                     &  -0.10$|$-0.31  &  -0.02$|$-0.08 \\
  1.6   &      &                   & 3.2 - 2.1 - -0.19  &  2$Z_{\odot}$ - 0.4$Z_{\odot}$ - -0.72  &   0.1 - 0.4      &  3.5 - 2.0 \\ 
\hline
 12294  & 4.5  & -1.07$|$---  & -1.27$|$---     &  -0.46$|$---                     &  -0.81$|$---  &  -1.20$|$--- \\
  2.3   &      &                   & 1.3 - 0.3 - -0.73  &  2$Z_{\odot}$ - 0.2$Z_{\odot}$ - -1.15  &   0.1 - 0.6      &  1.5 - 3.0 \\ 
\hline
 996  & 4.9  & -0.59$|$---  & \textbf{-0.66$|$---}     &  -0.24$|$---                     &  -0.34$|$---  &  -0.64$|$--- \\
  2.6   &      &                   & \textbf{3.0 - 1.1 - -0.38}  &  $Z_{\odot}$ - 0.2$Z_{\odot}$ - -0.6  &   0.1 - 0.6      &  2.0 - 3.5 \\ 
\hline
 2239  & 2.8  & -0.28$|$---  & \textbf{-0.22$|$---}     &  \textbf{-0.24$|$---}                     &  \textbf{-0.16$|$---}  &  \textbf{-0.13$|$---} \\
  1.3   &      &                   & \textbf{1.9 - 1.1 - -0.28}  &  \textbf{2$Z_{\odot}$ - $Z_{\odot}$ - -0.35}  &   \textbf{0.1 - 0.3}      &  \textbf{1.5 - 3.0} \\ 
\hline
 2286  & 3.2  & -0.42$|$---  & \textbf{-0.55$|$---}     &  \textbf{-0.72$|$---}                     &  \textbf{-0.50$|$---}  &  0.20$|$--- \\
   1.9  &      &                   & \textbf{1.8 - 0.8 - -0.51}  &  \textbf{2$Z_{\odot}$ - 0.2$Z_{\odot}$ - -1.45}  &   \textbf{0.1 - 0.4}      &  1.5 - 3.5 \\ 
\hline
 2361  & 4.3  & -0.54$|$---  & -0.51$|$---     &  -0.83$|$---                     &  -0.30$|$---  &  0.06$|$--- \\
   4.4    &      &                   & 3.5 - 1.8 - -0.36  &  2$Z_{\odot}$ - 0.2$Z_{\odot}$ - -1.25  &   0.1 - 0.6      &  1.5 - 3.0 \\ 
\hline
 2196  & 2.0  & -0.70$|$---  & \textbf{-0.86$|$---}     &  \textbf{-0.82$|$---}                     &  -0.52$|$---  &  0.07$|$--- \\
  2.0   &      &                   & \textbf{3.0 - 0.8 - -0.84}  &  \textbf{2$Z_{\odot}$ - 0.2$Z_{\odot}$ - -1.47}  &   0.1 - 0.6      &  1.5 - 3.5 \\ 
\hline
 2543  & 2.0  & -0.73$|$---  & \textbf{-0.68$|$---}     &  \textbf{-0.67$|$---}                     &  -0.42$|$---  &  \textbf{-0.41$|$---} \\
  2.6   &      &                   & \textbf{3.5 - 1.8 - -0.33}  &  \textbf{2$Z_{\odot}$ - 0.2$Z_{\odot}$ - -1.19}  &   0.1 - 0.6      &  \textbf{3.5 - 1.5} \\ 
\hline
 2111  & 4.9  & -0.76$|$0.05  & 0.17$|$0.3     &  0$|$0                    &  0$|$0  &  -0.01$|$0.02 \\
   3.8   &      &                   & 1.0 - 1.1 - +0.05  &  $Z_{\odot}$ - $Z_{\odot}$ - 0  &   0.1 - 0.1      &  3.0 - 3.5 \\ 
\hline
 472  & 8.6  & -0.31$|$0.83  & 0.22$|$0.93     &  0.21$|$0.71                    &  0$|$0  &  -0.08$|$0.30 \\
  5.4    &      &                   & 0.9 - 1.4 - 0.18  &  $Z_{\odot}$ - 2$Z_{\odot}$ - 0.28  &   0.1 - 0.1      &  1.5 - 3.5 \\ 
\hline
\hline
 \label{finalcount}
 \end{tabular}
 \normalsize
\end{table}
\end{landscape} 
The fact that we have changed the global age value does not influence our analysis.
Indeed, we make the starting assumption to associate to the two
components the stellar population parameters of the global fit since
it can be a good representation of the whole galaxy. However what we are
interested in is the $variation$ of these parameters as possible driver
of the observed color gradients, and not their absolute values.
In Table\,\ref{finalcount} we report the summary of the results obtained for the other galaxies of the sample, while the plots related to their analysis similar to Fig.\ref{analysis} and with the same conventions are available in electronic form.
\subsubsection{Dust distribution and color gradient}
Actually, as pointed out before, one of the causes of the radial
variation of color we observe in our ETGs can be a non homogeneous
distribution of the dust within the galaxy. Fig. \ref{cgav} shows that
for half galaxy of our sample the fit of the global SED returns a
value of A$_{V}$ = 0. Thus, for these galaxies (ID 23, 11888, 2361,
2196, 472) seems not plausible a scenario whereby the main driver of the
color gradients is a radial distribution of dust. For the remaining galaxies, the observed global SED was best fitted
assuming the presence of a non null quantity of dust. As in the case
of age, metallicity and $\tau$ we test if, starting from
a flat color distribution, a pure radial variation of the dust extinction (more dust in the center than in the outskirt)
can reproduce the observed color gradients. 

In particular, we have made the working hypothesis
that the ETGs of the sample have no dust in the outskirt. Thus, the colors we measure in the external regions (i.e. (F850LP-F160W)$_{obs,out}$) are the intrinsic colors of the galaxies.  
In these hypothesis, the reddest colors observed in the center of the galaxies (i.e. (F850LP-F160W)$_{obs,in}$) are the direct effect of the extinction of the dust. The necessary amount of dust extinction to redden the intrinsic colors to those observed in the center is: 
\begin{equation}
 \begin{split}
&A_{V} = \\
&\frac{[(F850LP-F160W)_{obs,in}-(F850LP-F160W)_{obs,out}] \times R_{V}}{K(F850LP)  - K(F160W)}
 \end{split}
\label{dust}
\end{equation}
where K(F850LP)A$_{V}$/R$_{V}$ and K(F160W)A$_{V}$/R$_{V}$ are the extinction
in both bands due to the dust. We have derived K(F850LP) and K(F160W) following \citet{calzetti00} and have
assumed R$_{V}$ = 4.05 $\pm$ 0.80.  For the galaxy 2111, Eq. \ref{dust} leads to a dust extinction A$_{V}$ in the center  $\sim$ 1. This value is a lower limit. Indeed, if dust was not localized only in
the center as we have assumed, but also in the periphery, the amount
of dust necessary to generate the observed color gradient should be
even higher.  The fit of the global SED returns a global value
of A$_{V}$ = 0.35, much lower than the one we obtain. Moreover, the
amount of dust predicted, that necessarily affects also the emission in
the F606W and F850LP bands, produce F606W-F850LP color in the central region of the galaxy redder than the observed one.
Thus, for this galaxy, dust does not seem the main driver of the observed color
gradients.

For galaxies 2286, 2239, and 2543 the
analysis indicates that, in our hypothesis, their color gradients can
be obtained assuming a dust extinction in the central regions of
 A$_{V}$ = 0.55, 0.29 and 0.83, respectively. Table \ref{globalSED} shows that for the galaxy 2239 the
dust extinction we need to reproduce the color gradient is lower than
the one predicted by the fit of the global SED. On the contrary,
galaxies 2286, 2543 require a dust extinction $\sim$ 1.5-2 times higher than the one
we obtained from the global SED fitting. These values confirm that for these
galaxies, if not the main driver, a significant contribution of dust
in generating the observed color cannot be excluded.

Finally, for the galaxy 996 and 12294 we have
obtained A$_{V}$ = 1.08 and 1.53 
respectively, results which disagree with the best-fit values of 0.10
and 0.25.  In these cases the dust extinction we need to reproduce the
observed color gradients is 6-10 times higher than the one obtained
from the global SED fitting, suggesting that, although it is not
possible to exclude it at all, the radial variation of dust extinction 
is certainly not the main driver of the color variation we observe.

\subsubsection{Radial variation of initial mass function as possible driver of
color gradients} 

Among the main hypothesis to explain the color gradients we observe,
there is also a change in the initial mass function (IMF) from the inner to
the outer regions of the galaxy. 

\begin{figure}
  \begin{center}
	\includegraphics[width=7.0cm,angle=0]{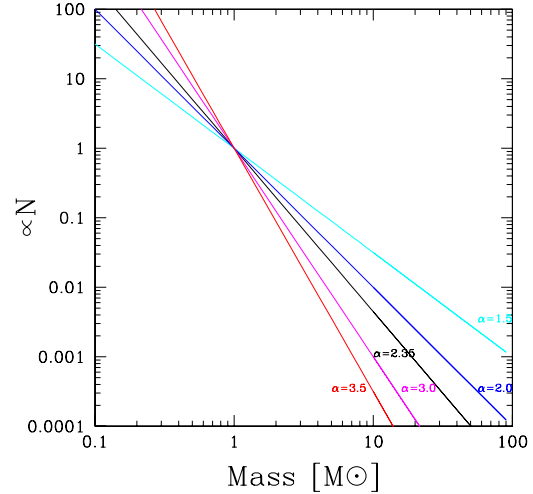} \\
	\caption{The relative amount of low-to-high mass stars predicted by 5 different IMFs with
analytic form dN/dM $\propto$ M$^{-\alpha}$ and $\alpha$ = 1.5, 2.0, 2.35, 3.0, 3.5.}
	\label{imf}
  \end{center}
\end{figure}

In spite of the fact that a radial variation of IMF could be a
reliable way to produce color gradient, it has never been
explored due to the lack of suitable stellar population models
For this reason a grid of stellar population
models has been kindly produced $ad$ $hoc$ by Stephan Charlot for this
purpose: reproducing the spectral energy distribution of galaxies with
different proportion of low-to-high mass stars. In particular we have assumed
an IMF analytically described by a power law form:
\begin{equation}
dN/dM \propto M^{-\alpha}
\end{equation}
with 5 different values of $\alpha$: 1.5, 2.0, 2.35, 3.0, 3.5. The
value 2.35 corresponds to a pure Salpeter IMF \citep{salpeter55}. Fig.
\ref{imf} shows the relative different amount of low-to-high mass
stars produced by the 5 IMFs.

\begin{figure*}
  \begin{tabular}{cc}
	\includegraphics[width=7.0cm,angle=0]{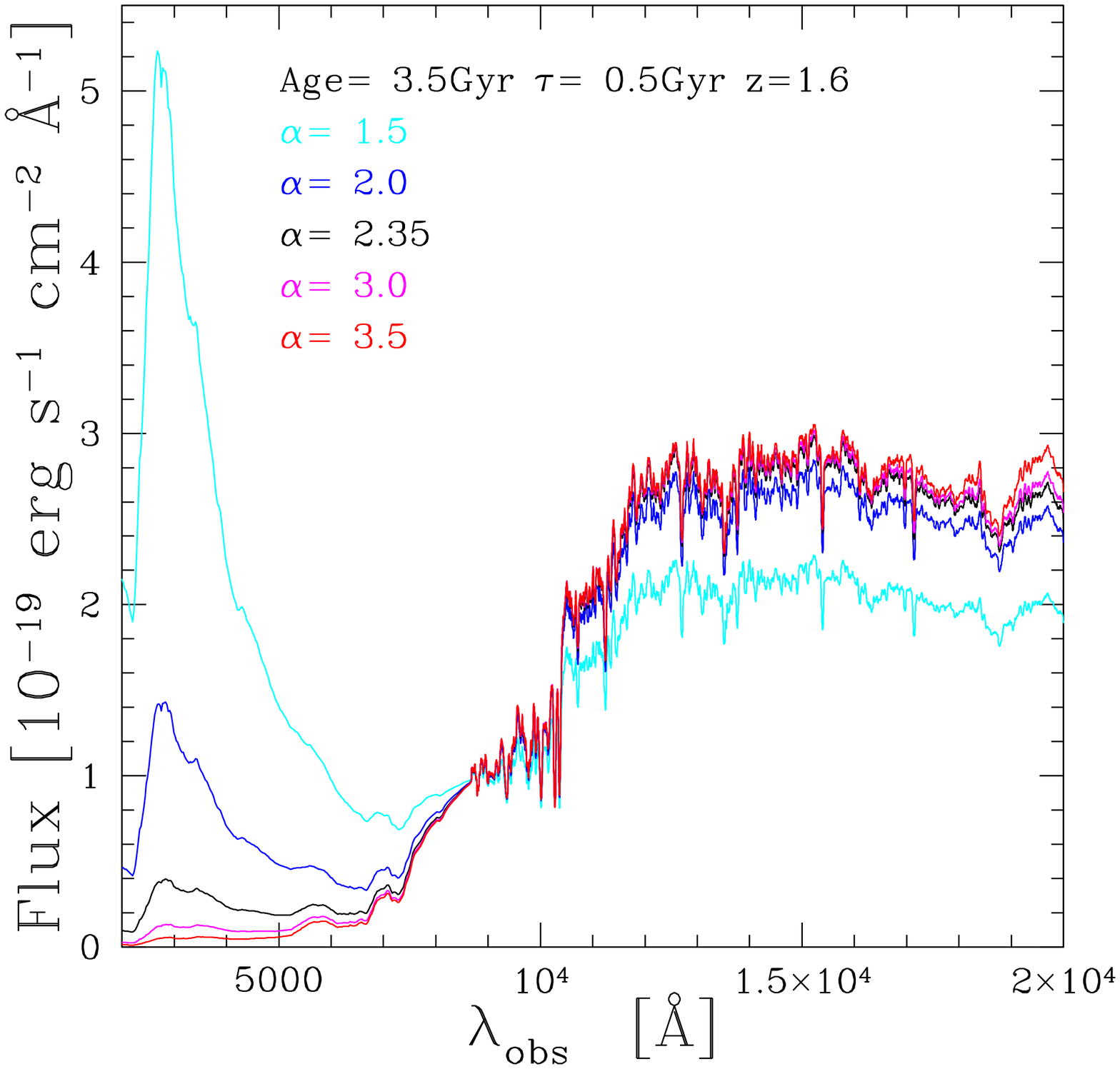} & 
	\includegraphics[width=7.0cm,angle=0]{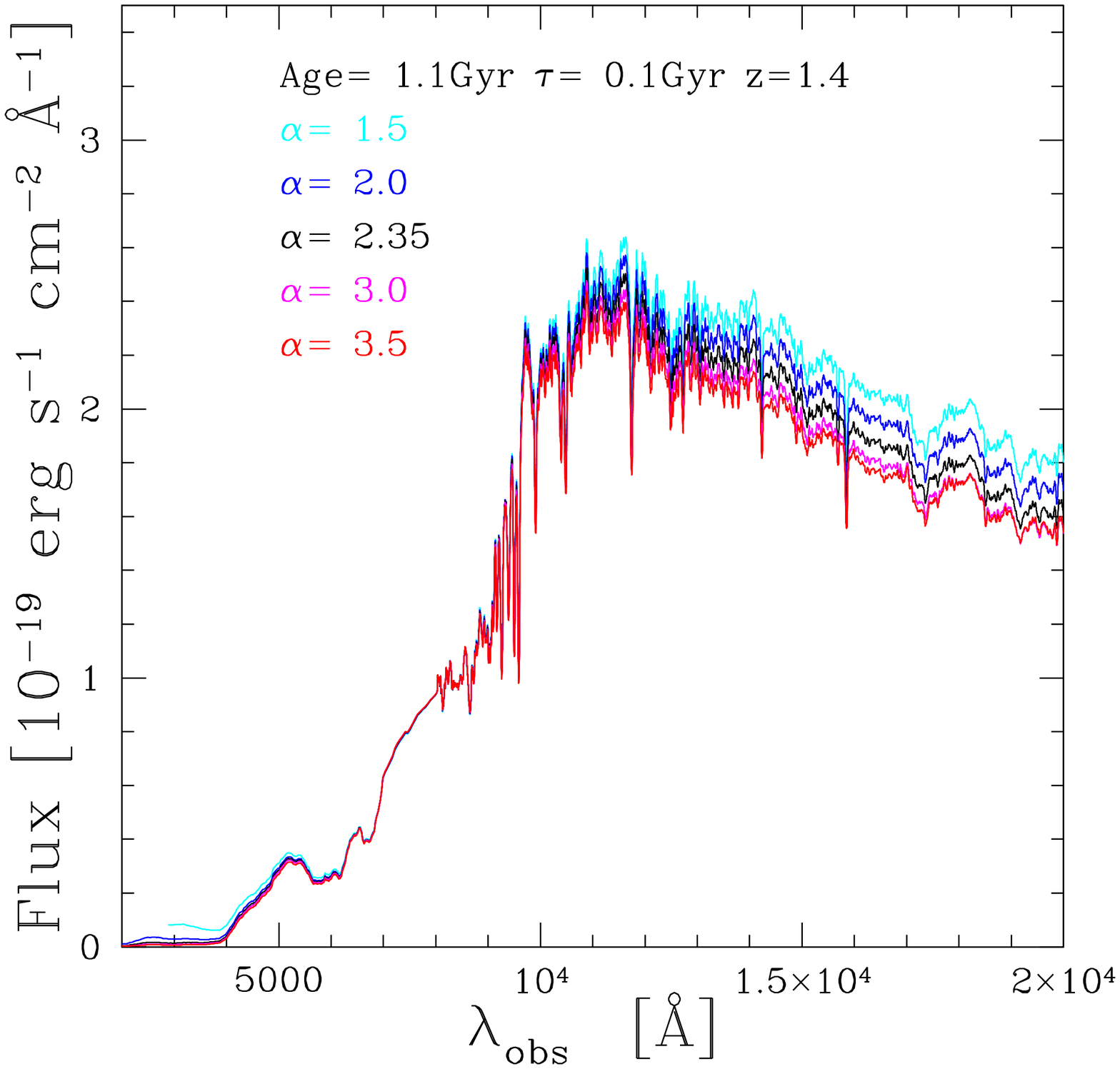}\\
  \end{tabular}
	\caption{\textit{Left Panel}: global SED of a template galaxy with Age= 3.5 Gyr and $\tau$= 0.5 Gyr at 
varying slope $\alpha$ of the IMF described by the functional form dN/dM $\propto$ M$^{-\alpha}$. \textit{Right Panel}:
the same of left panel but for a different ratio Age/$\tau$.}
	\label{sedimf}
\end{figure*}

The evolution of a star with time is strictly
related to its mass. Thus, at fixed age, metallicity, star-formation
time scales, and dust extinction, the spectral energy emission of a
galaxy will change with the IMF due to the different amount of
low-to-high mass stars. Left panel of Fig. \ref{sedimf} shows how the template of the global SED
of a possible galaxy of our sample (age = 3.5 Gyr, $\tau$ = 0.5 Gyr,
z=1.6) would change with different IMFs. To observe the relative differences, we have
normalized the observed SEDs in the wavelength range [8250 \AA - 8750 \AA ]. The
colors of the different SED follow the same code of
Fig. \ref{imf}. Since we normalized all the curves approximately in
correspondence of the F850LP filter, when looking at the emission around
16000 \AA \, this plot returns immediately how the F850LP-F160W color changes with
the IMF. In particular, for this template, IMF with an higher abundance
of low mass stars (red line) returns redder color than the one with
lower number of low-mass stars.
Actually, the color of a star changes through all its life cycle, as it
moves on the HR diagram according to its mass. So, the contribution of low and high mass stars
to the total emission in a fixed band changes at the variation of the ratio age/$\tau$. Right panel of Fig. \ref{sedimf} shows the same plot on the left
but for a galaxy with a different ratio Age/$\tau$. This simulated galaxy has
age = 1.1 Gyr and $\tau$ = 0.1 Gyr. Even in this case, we normalize
the curves in the wavelength region [8250 \AA - 8750 \AA ].  This plot
shows how in this case, the near-IR domain is dominated by the light coming from the high-mass stars contrary to the previous case. 
These plots clearly emphasize two main aspects: colors are affected by the IMF,
hence the radial color variation we detect can be effectively due to 
radial variation of IMF, and the shape of the SED depends on the ratio age/$\tau$ and IMF.\\
To investigate the hypothesis of an IMF radial variation as the main
driver of color gradient, we have adopted the same
method used to study age, metallicity and $\tau$ radial variations. For each galaxy we have fixed the stellar
population parameters of the internal and external populations to those
derived from the global fit and have looked for the value of the IMF's
slope able to reproduce the observed color gradients.
Since our goal is to investigate the radial variation of low-to-high
mass stars abundance, we have re-fitted the observed global SED assuming a Salpeter
IMF instead of the Chabrier. In Table \ref{globalSED} the result of the
SED fitting with a Salpeter IMF are reported. We have assumed the
internal and external populations defined by these new global
parameters (A$_{global,Sal}$, Z$_{global,Sal}$, A$_{V,global,Sal}$,
$\tau_{global,Sal}$, $\alpha$=2.35)  and we have looked for the value of $\alpha$ to be associated to the
internal and external population, $\alpha$$_{in}$ and $\alpha$$_{out}$
so that the two populations defined by the set of parameters
(A$_{global,Sal}$, Z$_{global,Sal}$, A$_{V,global,Sal}$,
$\tau$$_{global,Sal}$, $\alpha$$_{in}$) and (A$_{global,Sal}$,
Z$_{global,Sal}$, A$_{V,global,Sal}$, $\tau$$_{global,Sal}$,
$\alpha$$_{out}$) best reproduce the observed color gradients and the global SED. 

\begin{figure}
\begin{center}
   \includegraphics[width=5.0cm,angle=0]{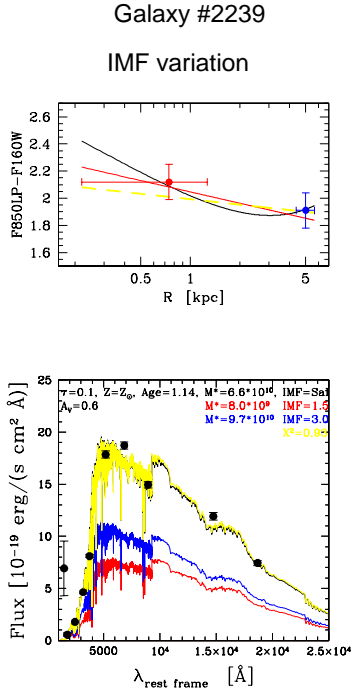} 
\caption{Analysis of the radial variation of IMF's slope as possible driver of observed color gradients. 
Top  panel shows the F850LP-F160W color profile (black line) whit the relative fit (red line). The bottom panel shows the observed global SED (black dots) as well as the stellar population parameters (black text) derived from its fit (black line) assuming a Salpeter IMF. Among all the stellar populations with varying IMF's slope but age, star-formation time scale, metallicity and dust extinction equal to the value obtained from the fit of the global SED, the red text identifies the one that best reproduces the color observed in the internal region (red dot in the top panel). Similarly, blue text reports the value of the slope of the IMF that best reproduce the external color (blue dot in the upper panel). The pure linear variation of the IMF's slope from the internal value to the external one, will produce the F850LP-F160W color gradient represented with the dashed yellow line. The contribution to the total stellar mass of the two populations is fixed in order to best fit the whole SED and are reported in the bottom panel (red and blue lines).}
  \label{imfvar}
\end{center}
\end{figure}

In all but two cases a variation of the
abundance of low-to-high mass stars can be excluded as the main
driver of the observed color gradients. On the contrary Fig. \ref{imfvar} 
shows that for the galaxy
2239 an external population accounting for the great part
of the mass ($>$ 90$\%$) formed with an IMF with $\alpha$ = 3.0,
i.e. with an higher number of low mass stars than a Salpeter, and a
small contribution in mass of an internal population with an IMF
defined by an $\alpha$ = 1.5 can simultaneously reproduce the F850LP-F160W
observed color gradient we have derived and the observed global SED.  At the same way, the
F850LP-F160W color gradient and the global SED of the galaxy 2543 are well reproduced
by a very massive population dominating the internal region characterized by an IMF with $\alpha$ = 3.5 and by a small
contribution of an external population described by an IMF with $\alpha$ =
1.5. As noted before, the different ratio age/$\tau$ of these two galaxies reflects in inverted radial trend of IMF's slope to reproduce negative color gradients. 
Thus, for these two galaxies a variation of the slope of the
IMF,and hence a variation in the abundance of low-to-high mass stars
can be a possible driver of the radial color variations we detect.
In Table\,\ref{finalcount}, the results of the analysis for all the galaxies of the sample are shown, while in electronic form we present their plots.

\section{Summary of the results}

We have found that $\sim\,70\%$ (8 out of 11) of the ETGs in our sample show negative color gradients at more than  $\sim$ 2$\sigma$ 
($>$4$\sigma$ in 5 ETGs) in the 0.1R$_{e}$-1R$_{e}$ range, the effective radius range usually adopted to study color gradients in local ETGs. 
The remaining $30\%$ show a gradient consistent with a flat color profile. 
Extending the analysis at R$>$R$_{e}$, enclosing the whole galaxy, we have found that the fraction of high-z ETGs with 
negative F850LP-F160W color gradients rises up to 100$\%$. 
In fact, we have generally found a steepening of color gradients extending the fit to regions out of R$_{e}$ (2R$_{e}<$R$<8R_{e}$)  \\
For 6 galaxies of our sample (ID 23, 2286, 2239, 2543,
2196, 996) a solely radial variation of the age of the stellar populations
can simultaneously reproduce at less than 1$\sigma$ the observed color gradients throughout the whole galaxy and the global SED. For
four of these galaxies (ID 2286, 2239, 2196, 2543), also radial
metallicity variation can reproduce the color gradient. 
On the contrary, a radial variation of the star-formation time scale, dust content, 
and abundance of low-to-high mass stars are able to reproduce both color gradients and global SED
in only few of these 6 galaxies. Moreover, we pointed out that a radial variation of the slope of the IMF
is not able to reproduce the observed radial color variation and spectral emission in any of the four galaxies for which both F606W-F850LP and F850LP-F160W
color gradients are available.\\
These results assign to age and metallicity a central role in generating the observed color gradients. As pointed out before, 
a contribution of other parameters to the color variations we detected cannot be ruled out.\\
For the remaining five galaxies (ID 2111, 11888, 12294, 472, 2361) the
variation of a single stellar population parameter is not able to
reproduce the observed color gradients and global SED 
consistently with the findings of \citet{guo11} on their sample. 
This suggests that for these galaxies a simultaneous variation of several parameters has to be invoked to
reproduce the observed color variations 
and global SED.
For these galaxies, an approach that investigates the simultaneous variation 
of more than one parameter should be considered.

\section{Discussion and conclusions}

The results of our analysis establish the heterogeneity of the stellar
content in high-z ETGs.  For $\sim$ 50$\%$ of the galaxies of our sample (6 ETGs) we have found that 
a pure radial age variation, with the oldest stars located at the center of the
galaxy, is able to reproduce the observed radial color profile and global SED. The age gradients we have detected
for these galaxies span a range from -0.84 to -0.28 dex per radial decade. For 4 out of these 6 ETGs the color gradients we have observed can be also accounted for by solely metallicity gradients whose strength varies from -0.35 to -1.45 dex per radial decade.\\ 
Due to the lack of similar measurements on other sample of high-z ETGs we have compared the age/metallicity gradient values we have found with those observed in the local ETGs. \\
Studies of ETGs at lower redshift affirm that color gradients are mainly  due to radial metallicity variations. 
Assuming the age of the stellar populations constant throughout the galaxy, as we did in the case of metallicity gradient analysis, color gradients in local ETGs 
turn out to be reproduced by a mean metallicity gradients ranging from -0.16 $\div$ -0.3 \citep{peletier90, idiart03, tamura03}. A further confirmation of the metallicity as main driver of local gradients comes out from studies at intermediate redshift which show that the color gradients evolution is better accounted for by the passive evolution of metallicity gradients \citep{saglia00, hinkley01}. In fact, our results seem to be inconsistent with these findings. In our sample of high-z ETGs we have found that only 4 galaxies out 11 have color gradients well reproduced by pure radial metallicity variation. Moreover, the metallicity gradients we have detected in high-z ETGs ([-0.3$\div$-1.45]) are systematically steeper than those typically observed in local ETGs ([-0.16$\div$-0.3]), even if \citet{ogando05} found that this range become wider ([0.0 $\div$ -1.0]) for nearby massive (M$_{\star}>10^{10}$M$_{\odot}$) ETGs see also \citep[see also][]{spolaor09}. The steeper metallicity gradients that we have detected derive by a radial metallicity variation from supersolar (2Z$_{\odot}$) values in the inner regions to subsolar values in the external regions  (0.2Z$_{\odot}$). Although such extreme values of Z have been observed also in few local ETGs \citep{mehlert03,rickes08}, our results show that metallicity gradients in high-z ETGs of our sample are only marginally comparable with the typical metallicity gradients detected in local ETGs. This result seems to point in favor of a possible evolution of the metallicity gradient in the last 9Gyr.

\begin{figure*}
	\begin{center}
	\includegraphics[width=19.0cm,angle=0]{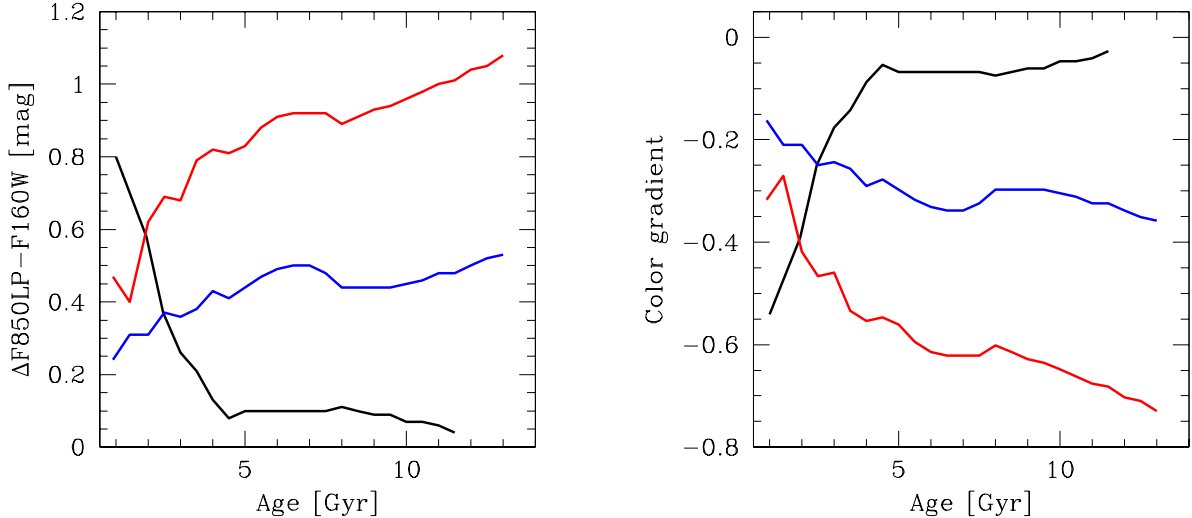} 
	\caption{$Left$ $panel$: Black line shows the difference observed in the F850LP-F160W color of two stellar populations with age differentiating of 2Gyr, as a function of the age of the youngest stellar population. Red line shows the variation in the F850LP-F160W color observed in two populations with metallicities 2Z$_{\odot}$ and 0.2Z$_{\odot}$, while blue line for two populations with metallicities Z$_{\odot}$ and 0.2Z$_{\odot}$. $Right$ $panel$ Color gradients relative to the color variation in the left panel in the hypothesis that they occur between 0.1R$_{e}$ and 3R$_{e}$.}
	\label{evolcol}
	\end{center}
\end{figure*}

Concurrently, studies on cluster ETGs at low and intermediate redshift show that pure age variations in  their stellar populations are not able to account for their color gradients \citep{saglia00}. In contrast to local results, we have found that for $\sim$ 50$\%$ of the galaxies of our sample a radial variation of stellar populations age alone can reproduce the observed color gredients and global SED. In fact, recent studies investigating the simultaneous radial variation of both age and metallicity confirm metallicity gradients ($\sim$-0.4 dex per radial decade) as the main driver of observed color gradients in local ETGs but found also a small contribution to color variation of positive age gradient ($\sim$0.1 dex per radial decade) \citep{labarbera09,wu05}. The age/metallicty degeneracy affecting optical colors does not allow us to consider the simultaneous radial variation of both parameters and hence to detect a possible positive age gradient, whose presence in local ETGs, actually, is still matter of debate. \\
To spread light on this issue, high-z ETGs constitute the ideal place to investigate the presence of an age gradient. Indeed, at fixed radial variation of age $\Delta$age, its effect on color profile is much more enhanced when stellar populations are younger, hence in high-z ETGs. 
This effect is clearly shown in left panel of Fig. \ref{evolcol}, where we report the differences observed in the F850LP-F160W color of two stellar populations with age which differ of 2Gyr ($\tau$ = 0.1Gyr, black solid curve), as a function of the age of the youngest stellar population. 
The same $\Delta$age produces a difference in the F850LP-F160W color of the two populations that is $\sim$ 10 times higher if observed in high-z ETGs (Age $<$ $\sim$ 4Gyr) respect those observed in local ETGs (Age $\sim$ 10Gyr). A typical radial variation of 2 Gyr, as the one we measure in the ETGs of our sample, will produce in a local ETG a variation in the F850LP-F160W color of $\sim$ 0.05 mag, thus at the very limit of the detection. On the contrary, the same age variation will result in a color variation of $\sim$ [0.3-0.5] mag for high-z ETGs. 
Red and blue lines report the same of black line, but for pure metallicity variations. In particular, red line show the variation in the F850LP-F160W color observed in two populations with metallicities 2Z$_{\odot}$ and 0.2Z$_{\odot}$, while blue line in two populations with metallicities Z$_{\odot}$ and 0.2Z$_{\odot}$. In the right panel of Fig. \ref{evolcol} the color gradient that the above age/metallicity variations would produce when occurring between 0.1R$_{e}$ and 3R$_{e}$. Differently from age variation, the effect of a metallicity variation on color, and hence on gradient, increases with the age of the galaxy of a factor $\sim$ 2 from high-z ETGs to local ETGs. 
These plots emphasize how challenging is the detection of an age gradient in local ETGs due to its almost negligible effect on color profile. On the contrary, in high-z ETGs age and metallicity variations produce comparable effect on color profile, thus setting the ideal condition for their detection.
This comparison with local samples is only meant to have an indication on how the results obtained at high-z, both in terms of age/Z gradients and in terms of internal and external age/Z values, relate with the local values. On the other hand, the unknown 
evolution experienced by ETGs in $\sim$9 Gyr from z$\sim$1.5 to z=0 can affect the stellar properties and distribution
(e.g. minor mergers triggering secondary burst of star formation) making complex any comparison with local universe. 
In fact, to properly face on high-z and local values samples of ETGs selected in an homogeneous way, a dataset able to trace the evolution of the same rest-frame color gradient over 9 Gyr, and similar procedures for both the color gradients estimates and the relative analysis should be considered. In a forthcoming paper, taking into account all these factors, we will try to address the origin of the color gradients following their evolution from z$\sim$2 to z=0.

For the remaining five ETGs of our sample, a pure radial variation of a single stellar population parameter is not able to reproduce the observed color gradients and global SEDs. Differently form the previous cases, where color gradients could be reproduced by a pure age or metallicity gradient as well as by a simultaneous radial variation of more than one parameter, these galaxies $need$ a more complex scenario whereby more than one property of the stellar
populations have to vary from the center to the periphery to generate the observed color gradients.

Thus, our analysis clearly indicates that the properties of the stellar population and their 
distribution within high-z ETGs do not follow an homogeneous and
common scheme. 

In the following we try to investigate if the theoretical expectations of the widely accepted scenarios
of galaxy formation can explain the observed color distributions.

In the monolithic revised scenario, color gradients are supposed to be mainly due 
to metallicity gradient, being the contribution of age null or mild (and positive).
Our findings of
none correlation between color/metallicity gradients and total mass  suggest that the monolithic revised scenario 
is not the favored mechanisms with which ETGs assembled their mass, 
although the 
narrow mass range covered and the assumption of the stellar mass as a proxy of the total stellar mass 
can affect this conclusion.

\begin{figure}
	\includegraphics[width=7.5cm,angle=0]{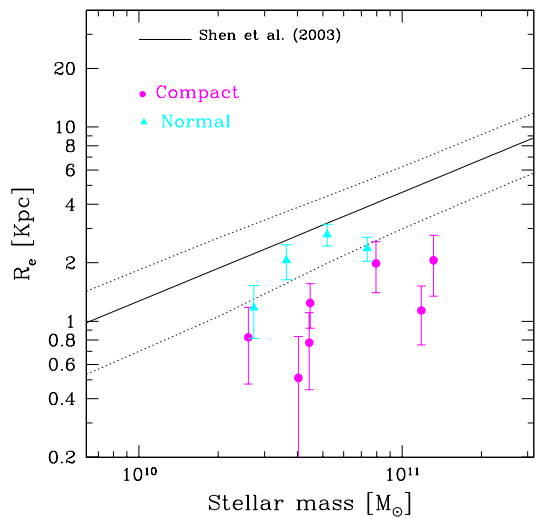}
	\caption{SM relation for local ETGs (solid line, Shen et al. 2003) and for
our sample (solid symbols). The dashed lines are the scatter lines at 1$\sigma$.
We shifted the Shen et al.’s relation by a factor ∼1.2 towards lower masses
to take into account the systematic shift observed in the mass estimations
using our models or those adopted by Shen et al. (2003). The circles are
compact galaxies, that is, galaxies having the effective radius more than 1$\sigma$
smaller than those predicted by a local relation for that mass. On the contrary,
galaxies having the effective radius comparable at 1$\sigma$ with those expected
by Shen et al.’s relations are classiﬁed as normal (triangle symbols).}
	\label{smr}
\end{figure}

\begin{figure}
	\includegraphics[width=5.8cm,angle=-90]{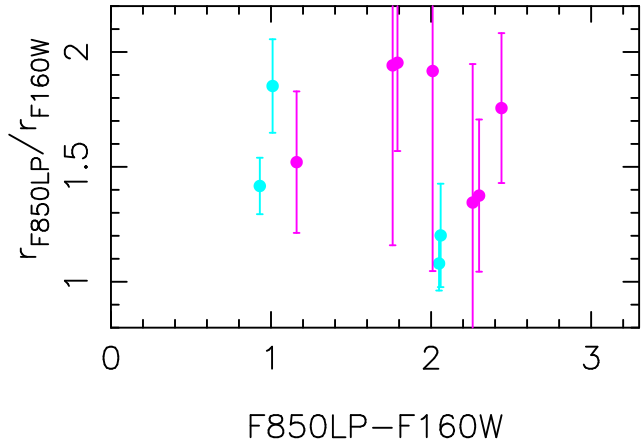} \\
	\caption{The r$_{e,F850LP}$/r$_{e,F160W}$ ratio as proxy of F850LP-F160W color gradient vs the F850LP-F160W global color.}
	\label{cgvscolor}
\end{figure}

Theoretical predictions of the \textit{inside-out-growth} scenario point in favor of compact ETGs with cores \citep{wuyts10} redder than the outskirt regions. This negative color gradient seems to be due to a combined effect of negative metallicity and positive age gradients, with a non negligible effect of dust \citep{wuyts10}. To compare our results with the theoretical prediction of this model, we  define compact galaxies those ETGs with effective radius more than 1$\sigma$ smaller than those predicted by the local size-mass relation (SMR) for that mass. In Table 3 we report the compactness C for the galaxies of our sample defined as the ratio R$_{e,z=0}$/R$_{e}$ where R$_{e}$ is the effective radius of the galaxy and R$_{e,z=0}$ is the radius that a galaxy of equal stellar mass would have at z=0 as derived by the SMR of \citet{shen03}. In our sample, compact galaxies, as we defined them, turn out to have C$\geq$2. In Fig \ref{smr} we report the SMR in the F850LP band for our sample (solid symbols) and for local galaxies \citep[solid line,][]{shen03}. The dashed lines represent the scatter at 1$\sigma$. It turns out that seven out of 11 galaxies (solid points) of our sample are compact (magenta points) and 4 are normal ETGs (cyan triangles). In Table 3, we report the classification in normal (N) and compact (C) for our galaxies.
Compact ETGs in our sample (as well as normal ETGs) show redder cores supporting the \textit{inside-out-growth} scenario even if we cannot firmly establish the origin and the nature of this gradient. Actually, we cannot test the presence of a positive age gradient, since due to age/metallicity degeneracy we do not treat the case of simultaneous radial variation of age and metallicity. On the other hand, our results show that only two out of seven compact ETGs have color gradients well reproduced by a metallicity gradient.\\
Nonetheless, the simulations predicts that the negative color gradients produced by the interplay of age, metallicity and dust should result to  correlate with the integrated rest-frame optical color. Following \citet{wuyts10} in Fig. \ref{cgvscolor} we 
show the ratio between the F850LP-band effective radius and F160W-band effective radius as a proxy of the color gradients  versus the F850LP-F160W color. The points follow the same convention of Fig. \ref{smr}.
We do not find any correlation neither in the whole sample, nor in the compact selection.
The absence of such correlations cast some doubts on the validity of this scenario as a viable process to assemble the stellar mass in compact ETGs.\\
Despite the fact that both the widely accepted formation scenarios do not seem to be able to reproduce the stellar 
content of high-z ETGs, we have to test the hypothesis whether ETGs can be assembled through a common 
formation process and the distribution we observed in color gradients can be
the final product of subsequent merger events.  We have to bear in mind that our galaxies are 
all younger than 3.5 Gyr. This severely constrains the time each galaxy has at disposal to experience a merger event. 
Table 1 of \citet{boylan08} shows the dynamical friction merging time in Gyr, $\tau_{merger}$, for a host halo
with virial mass M$_{host}$ = 10$^{12}$M$_{\odot}$, measured from numerical simulations. They assumed different 
initial orbital angular momentum, initial orbital energy, and ratio of initial satellite mass to initial host 
halo mass and only for two cases they found 3.5 Gyr is enough to complete the merger, while in all the other 
cases $\tau_{merger}$ is greater than 4.3 Gyr.  Thus, the different stellar content of high-z ETGs does
not seem to be due to the effect of subsequent merger events,
but primarily to the formation process. Actually, the continuum distribution of ETGs in the size-mass plane, both at high redshift and in the local Universe, together with the systematic direction of the color gradient (negative or null) of high-z ETGs, point toward a common formation process responsible of this continuity. The possible
different initial conditions, such as the different time scale
of collapsing gas cloud, can be responsible of the observed structural and dynamical differences as we previously suggested \citep{saracco11, saracco12}.

\section*{Acknowledgments}
This work is based on observations made with the ESO telescopes at the
Paranal
Observatory and with the NASA/ESA Hubble Space Telescope, obtained from the
data archive at the Space Telescope Science Institute which is operated by
the Association of Universities for Research in Astronomy.
This work has received financial support from ASI-INAF (contract
I/009/010/0) and from PRIN-INAF (1.05.01.09.05).
\nocite{}
\bibliographystyle{mn2e}
\bibliography{paper_gargiulo_referee}
\newpage

\begin{figure*}
 \begin{tabular}{c}
	MATERIAL PRESENTED IN ELECTRONIC FORM\\
    \includegraphics[width=16.0cm,angle=0]{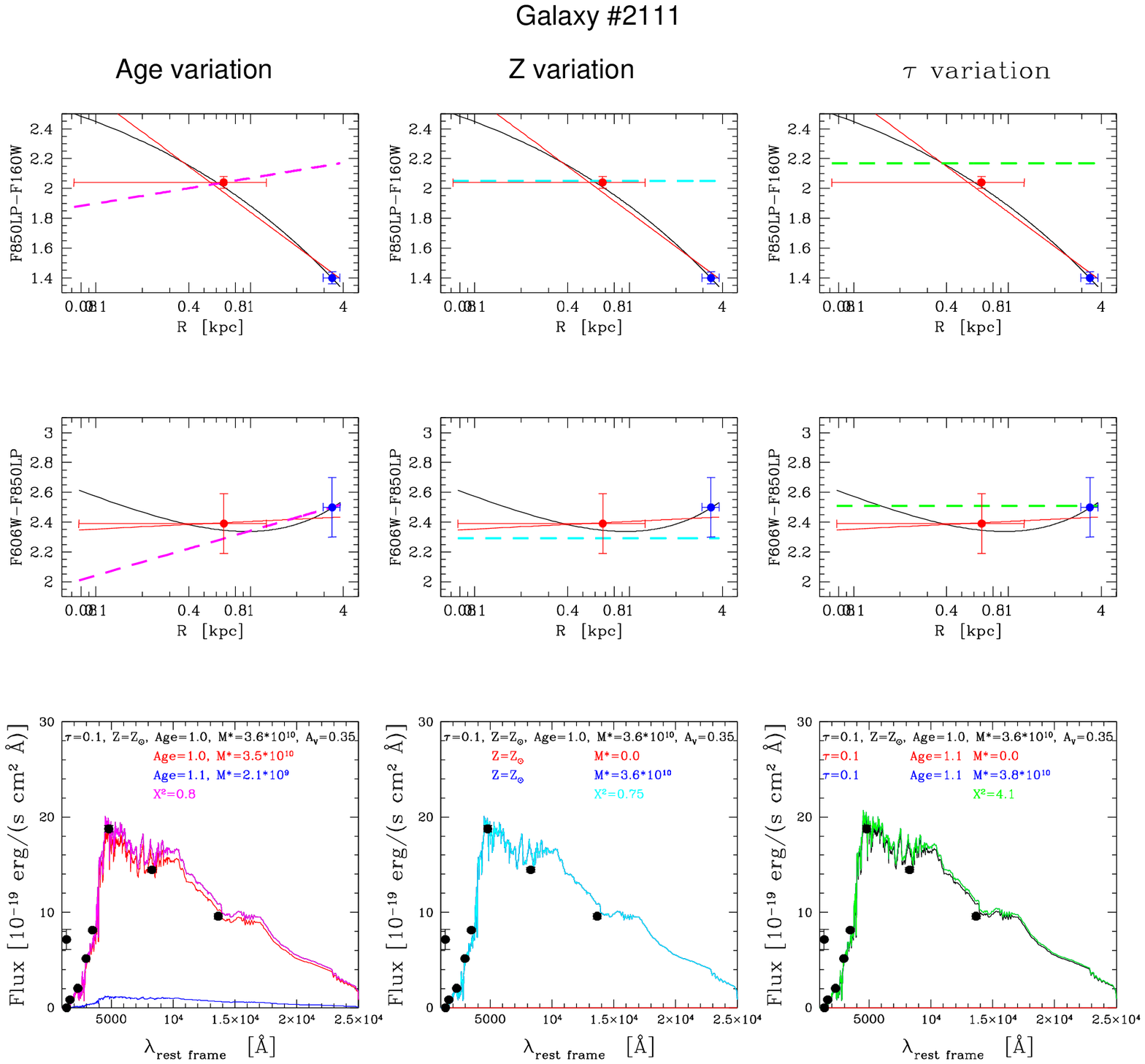}\\  
 \end{tabular}
  \begin{tabular}{c}
   \textbf{Figure 7. (continued)}
  \end{tabular}
\end{figure*}
\begin{figure*}
    \includegraphics[width=16.0cm,angle=0]{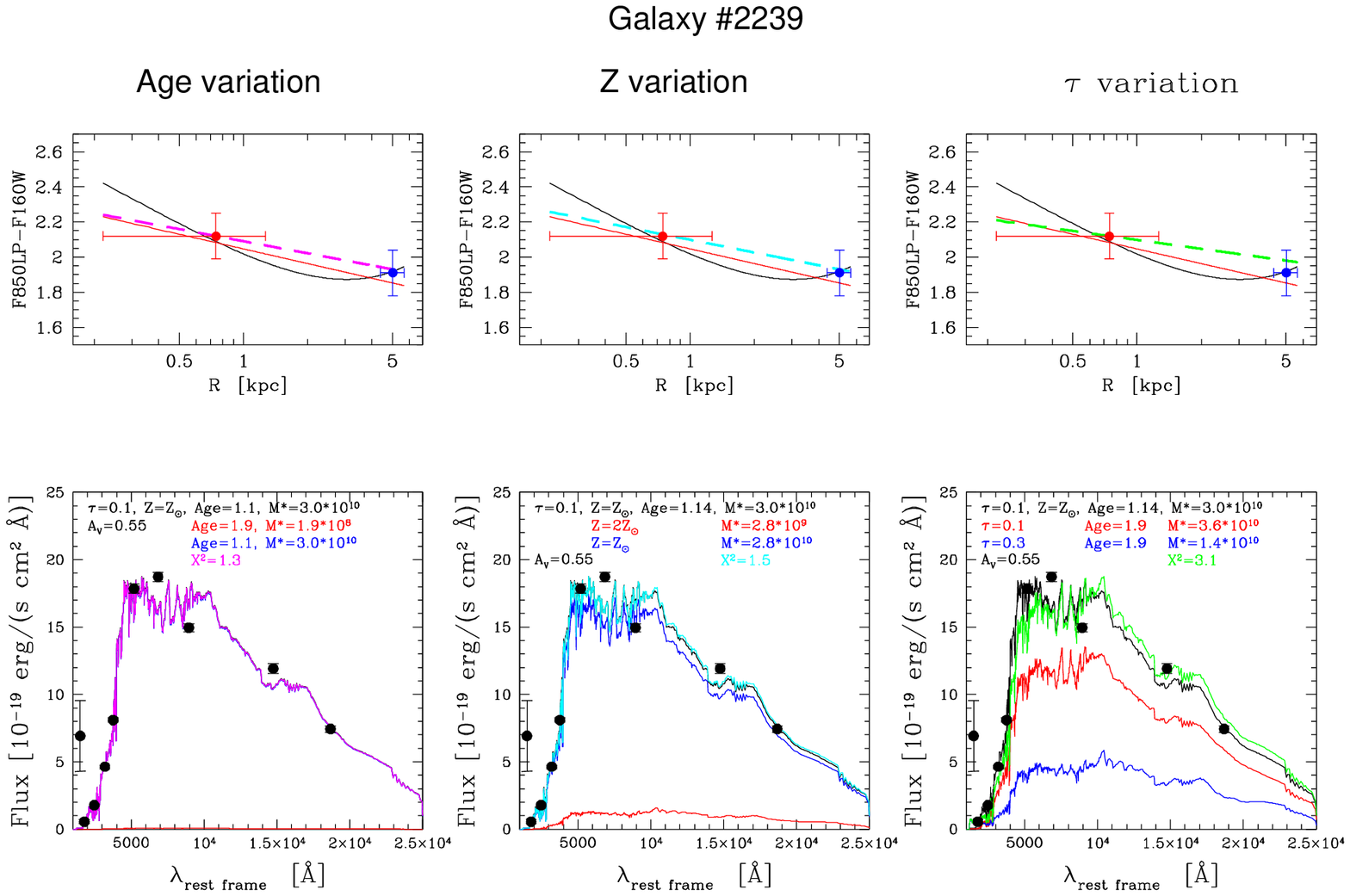}
\end{figure*}
\begin{figure*}
    \includegraphics[width=16.0cm,angle=0]{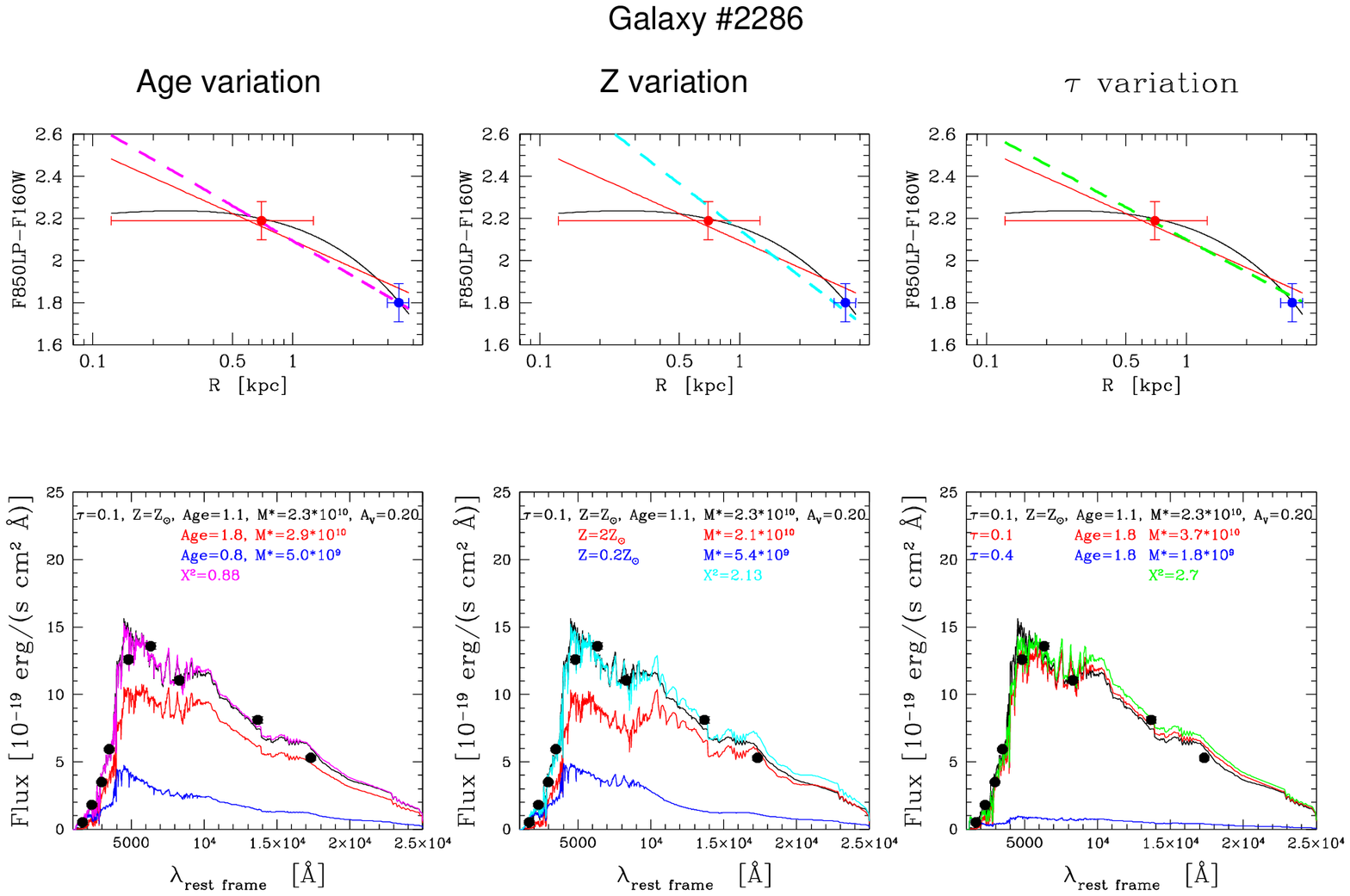}
  \begin{tabular}{c}
   \textbf{Figure 7. (continued)}
  \end{tabular}
\end{figure*}
\begin{figure*}
    \includegraphics[width=16.0cm,angle=0]{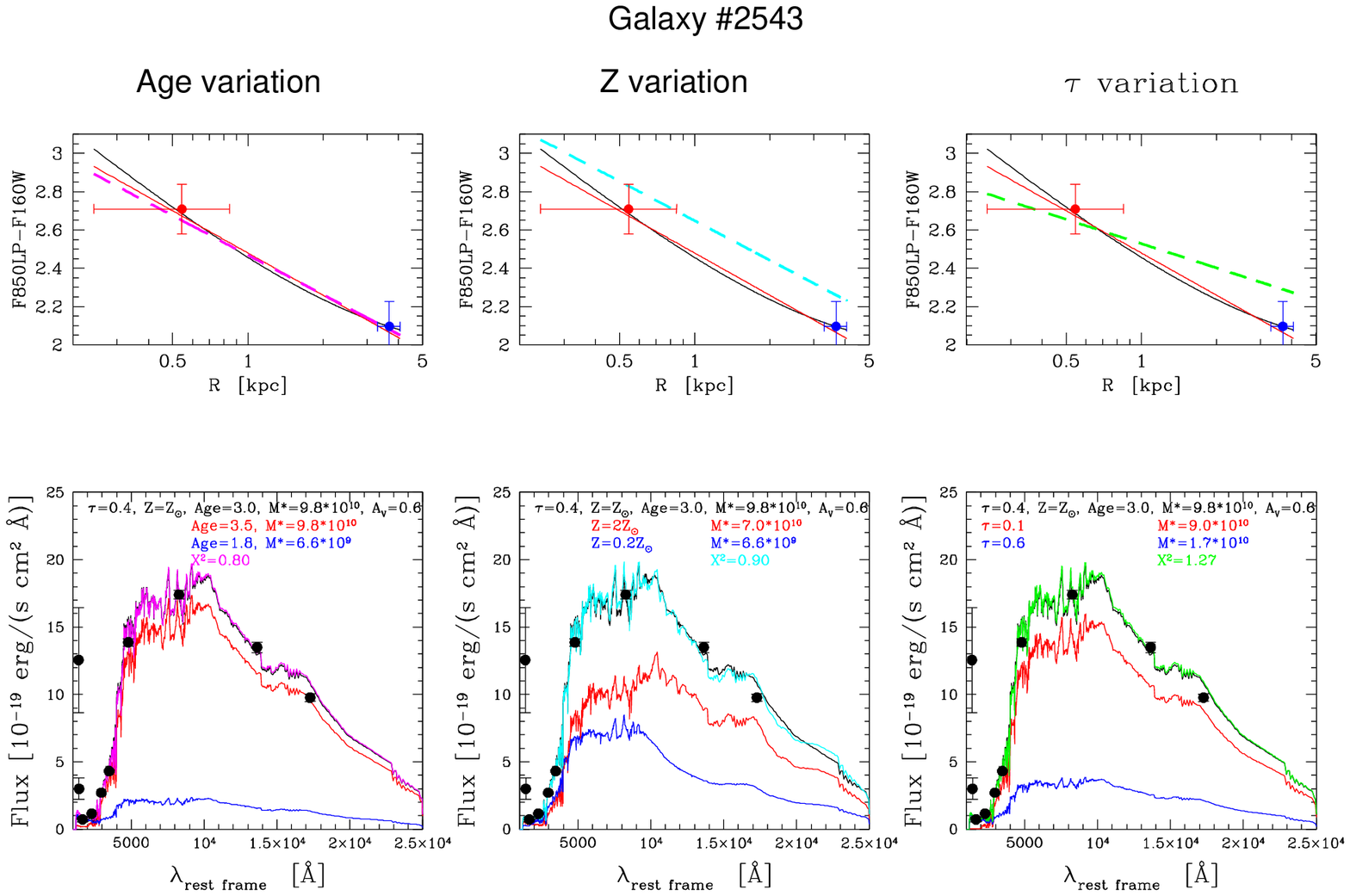}
\end{figure*}
\begin{figure*}
    \includegraphics[width=16.0cm,angle=0]{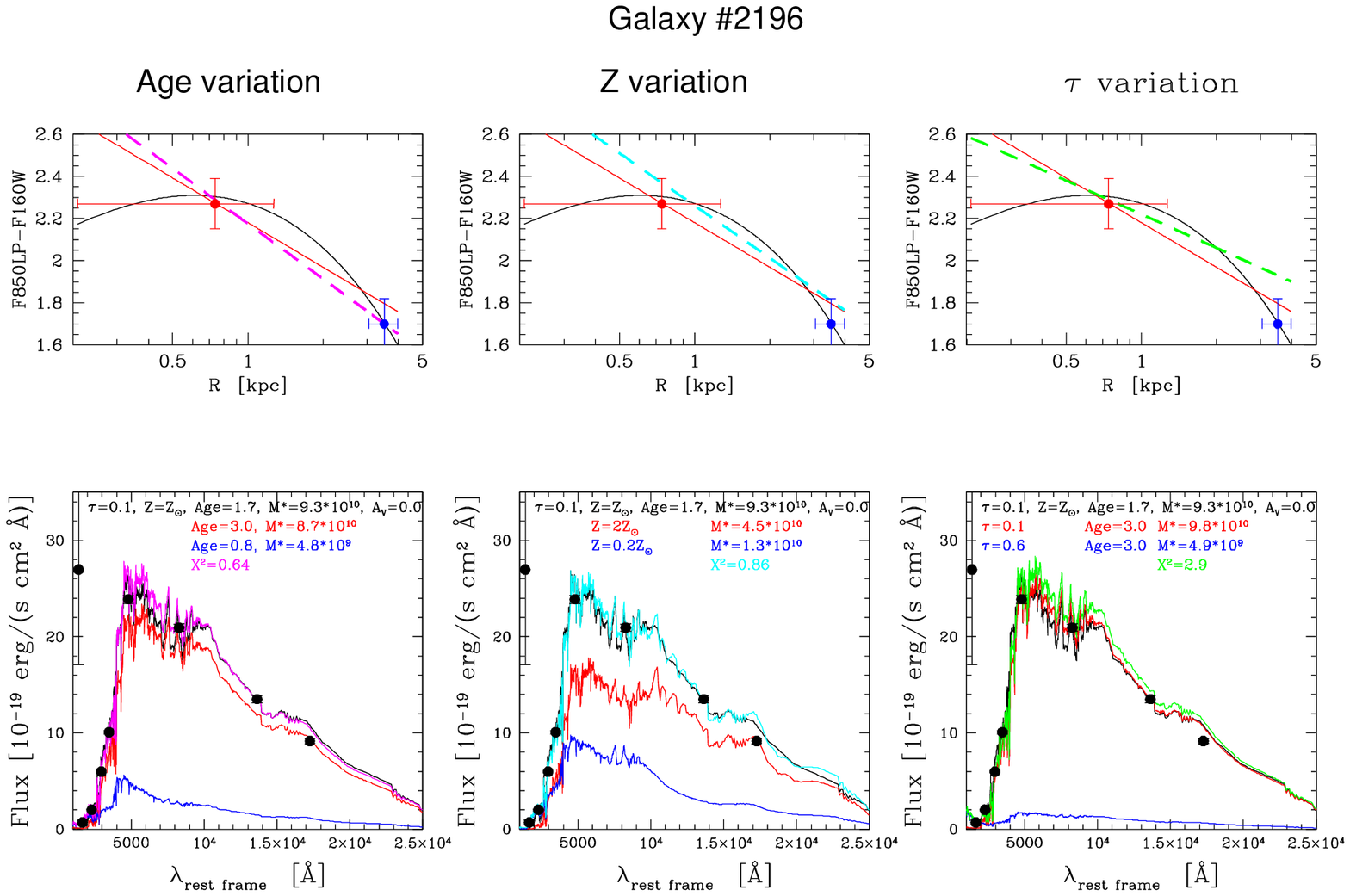}
  \begin{tabular}{c}
   \textbf{Figure 7. (continued)}
  \end{tabular}
\end{figure*}
\begin{figure*}
    \includegraphics[width=16.0cm,angle=0]{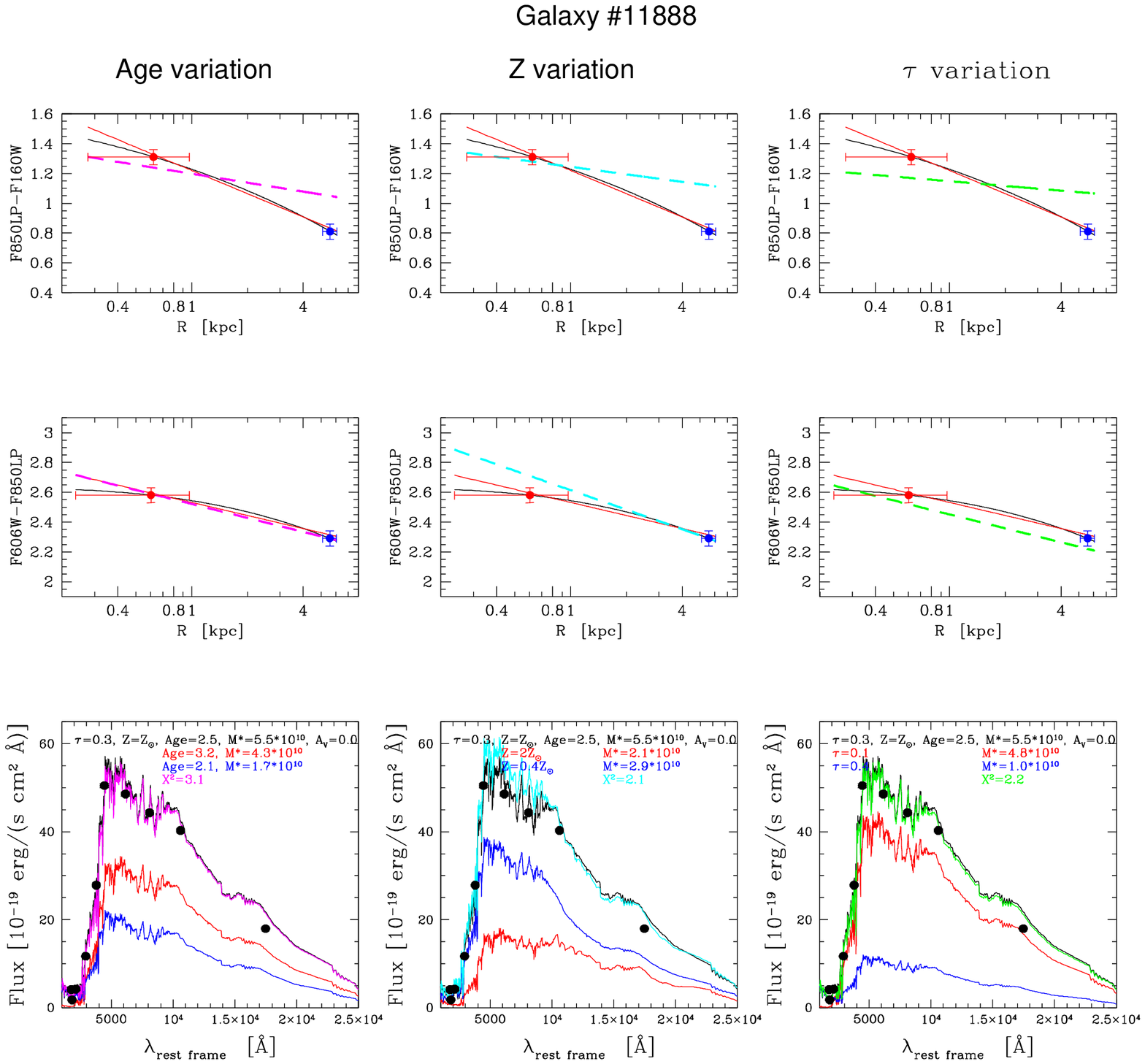}
  \begin{tabular}{c}
   \textbf{Figure 7. (continued)}
  \end{tabular}
\end{figure*}
\begin{figure*}
    \includegraphics[width=16.0cm,angle=0]{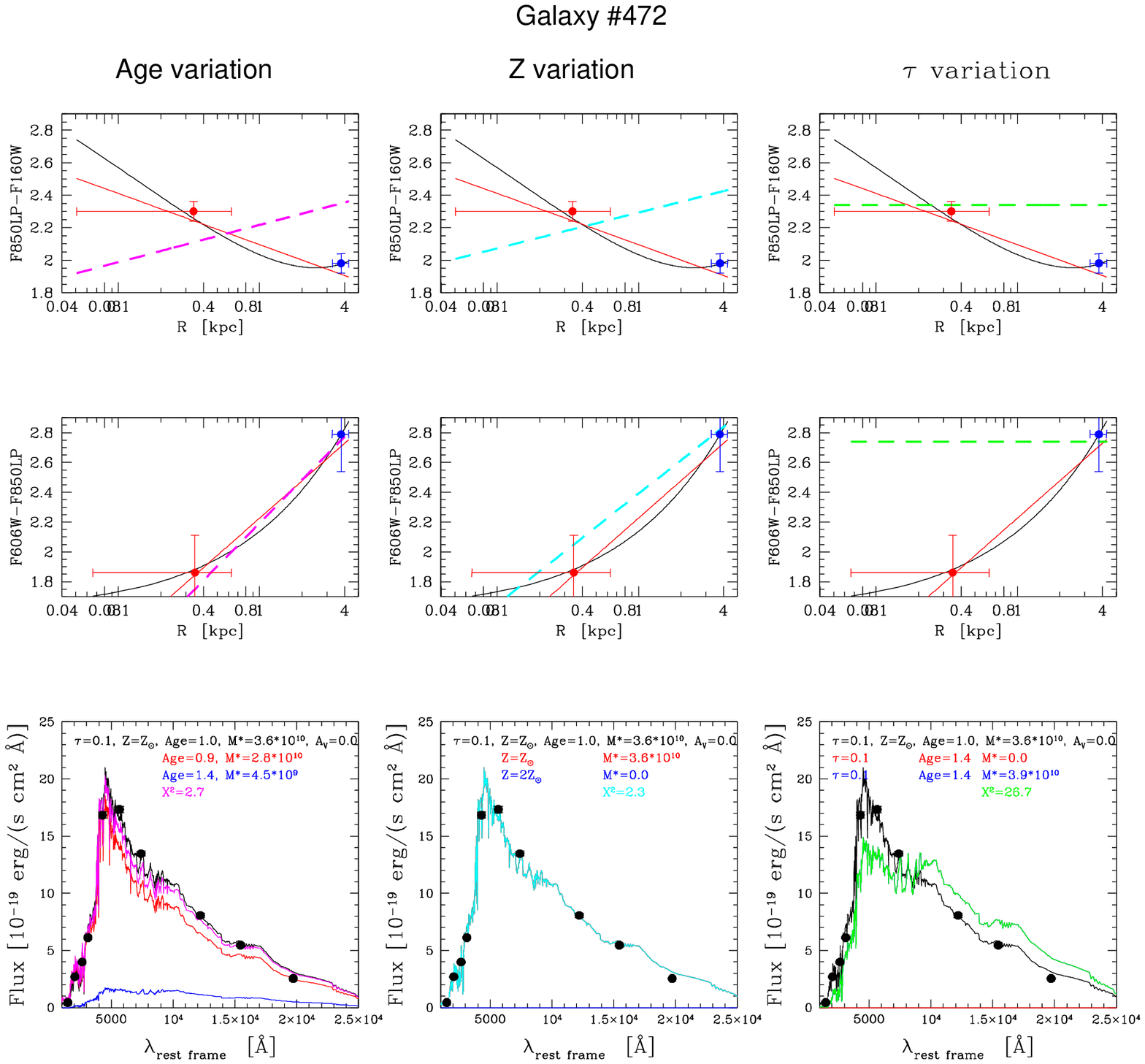}
  \begin{tabular}{c}
   \textbf{Figure 7. (continued)}
  \end{tabular}
\end{figure*}
\begin{figure*}
    \includegraphics[width=16.0cm,angle=0]{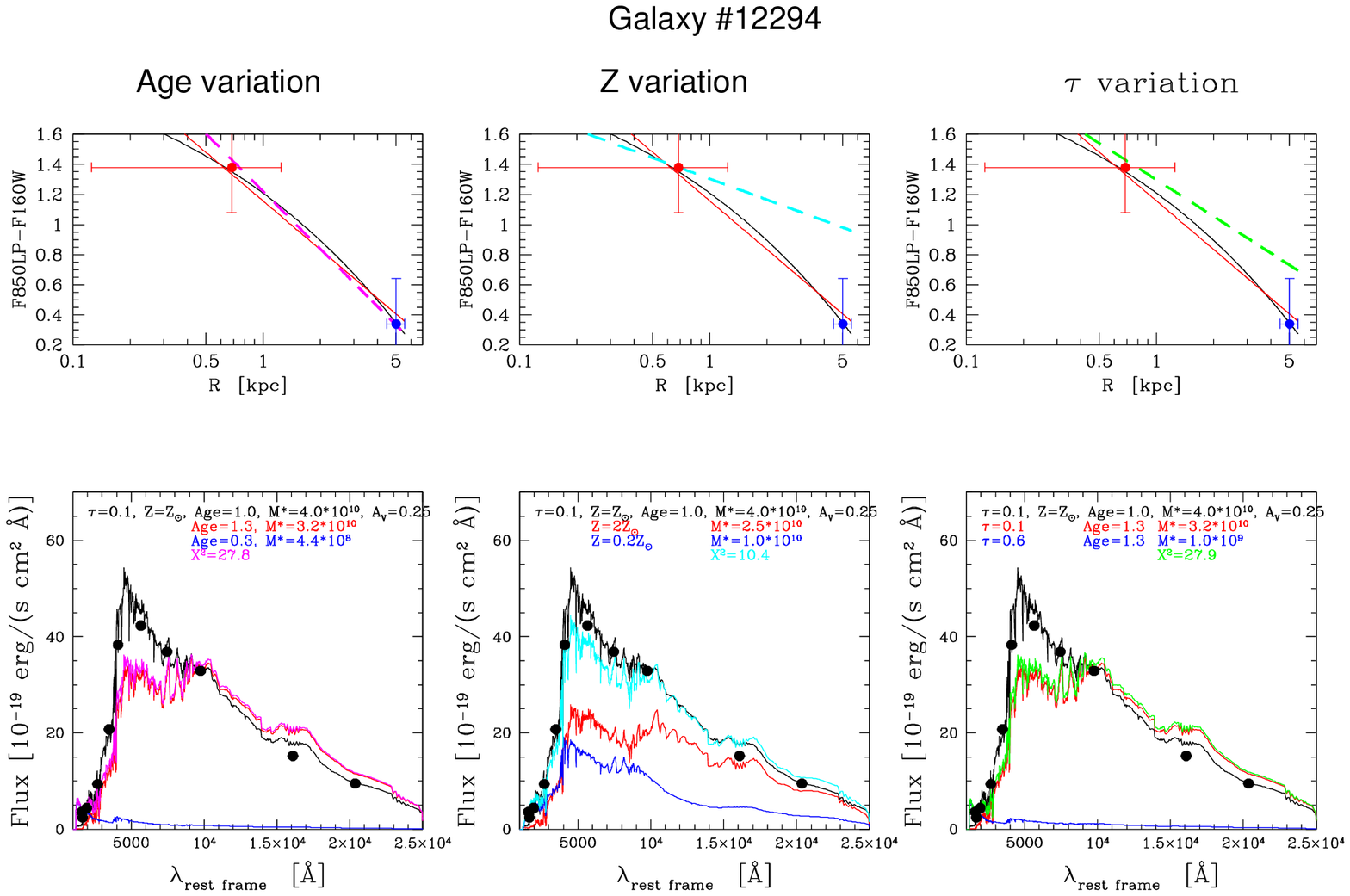}
\end{figure*}
\begin{figure*}
    \includegraphics[width=16.0cm,angle=0]{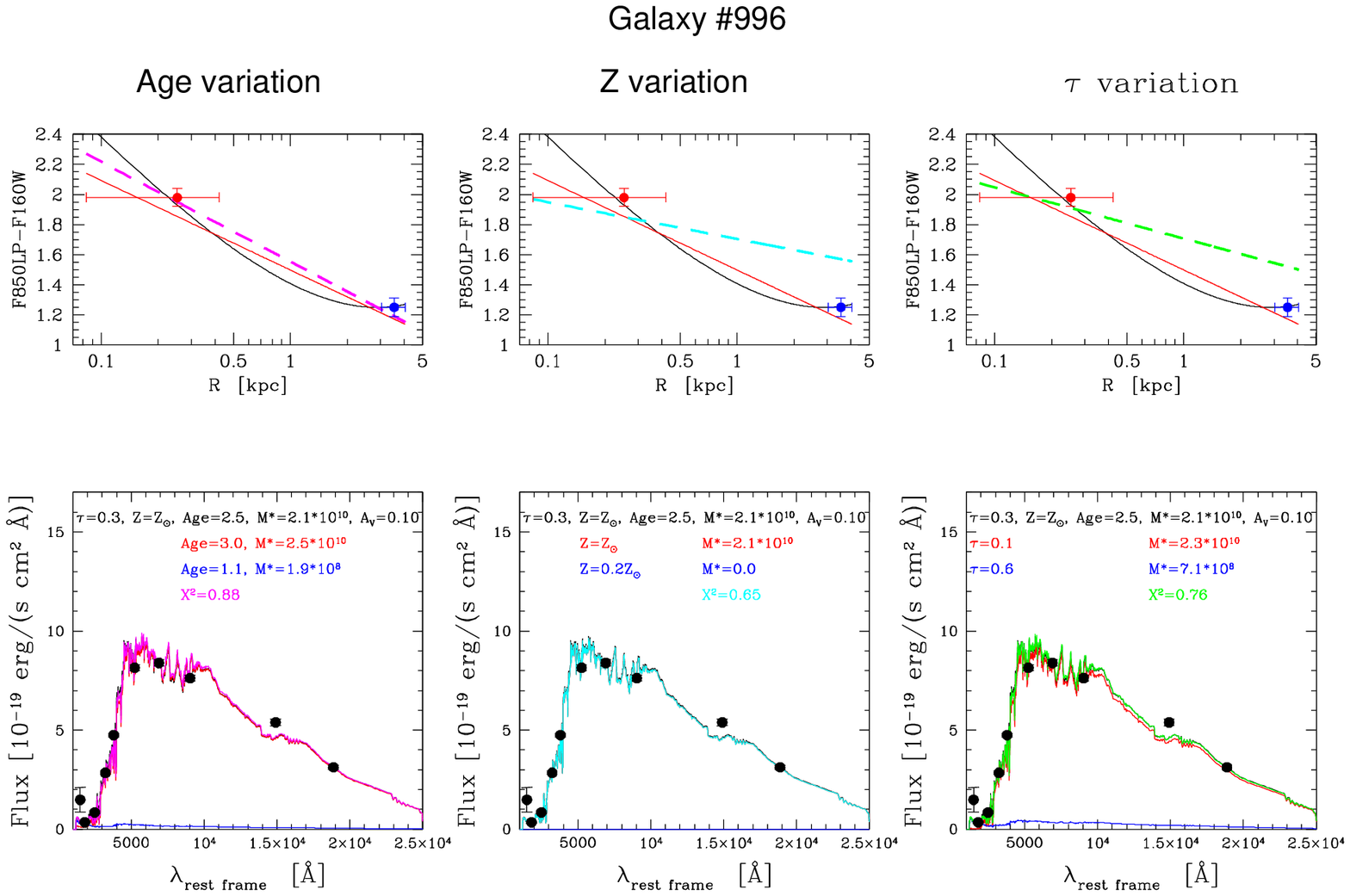}
  \begin{tabular}{c}
   \textbf{Figure 7. (continued)}
  \end{tabular}
\end{figure*}
\begin{figure*}
    \includegraphics[width=16.0cm,angle=0]{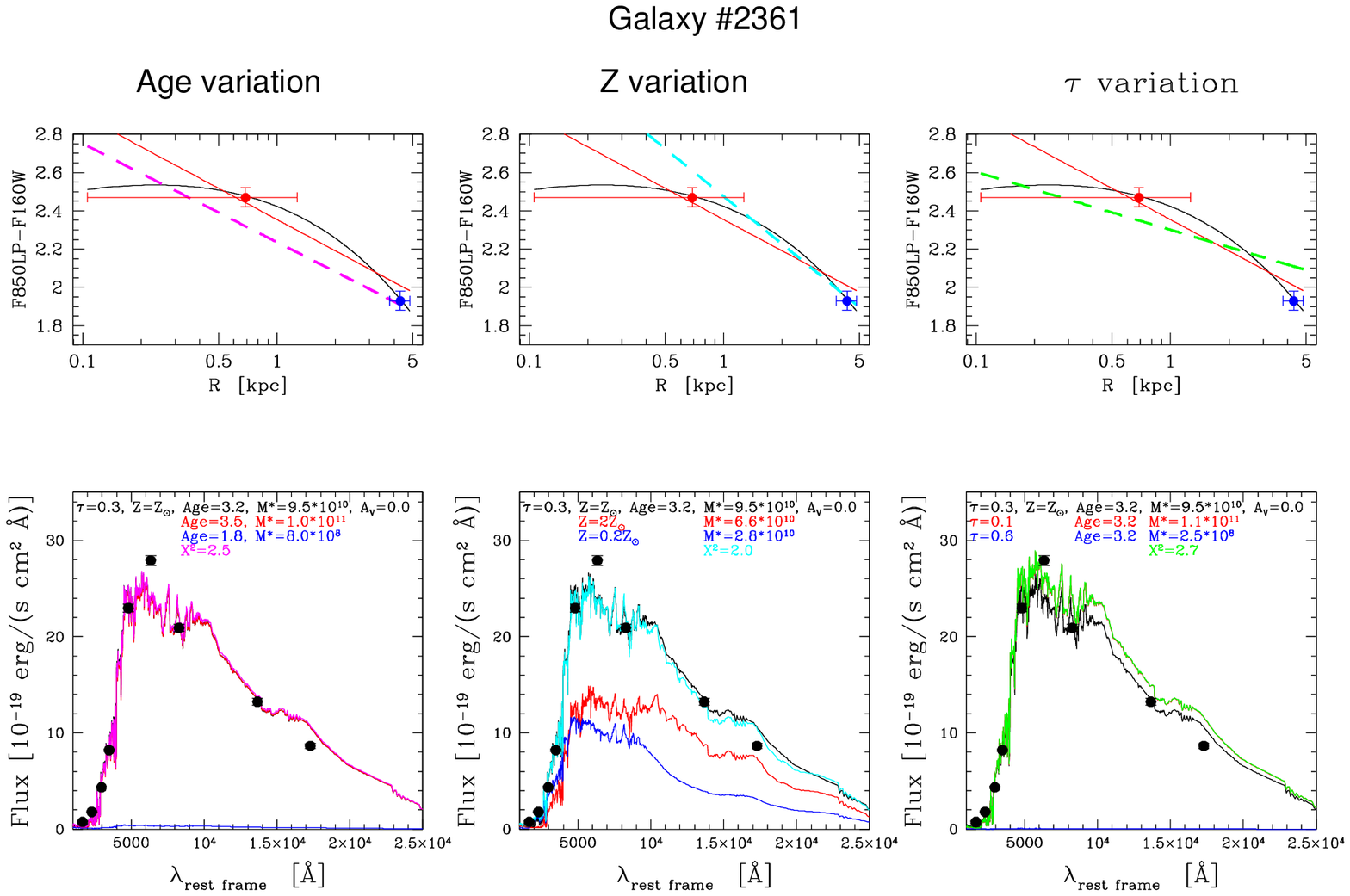}
  \begin{tabular}{c}
   \textbf{Figure 7. (continued)}
  \end{tabular}
\end{figure*}

\begin{landscape} 
\begin{figure}
\begin{center}
   \begin{tabular}{cccc}
   \includegraphics[width=5.0cm,angle=0]{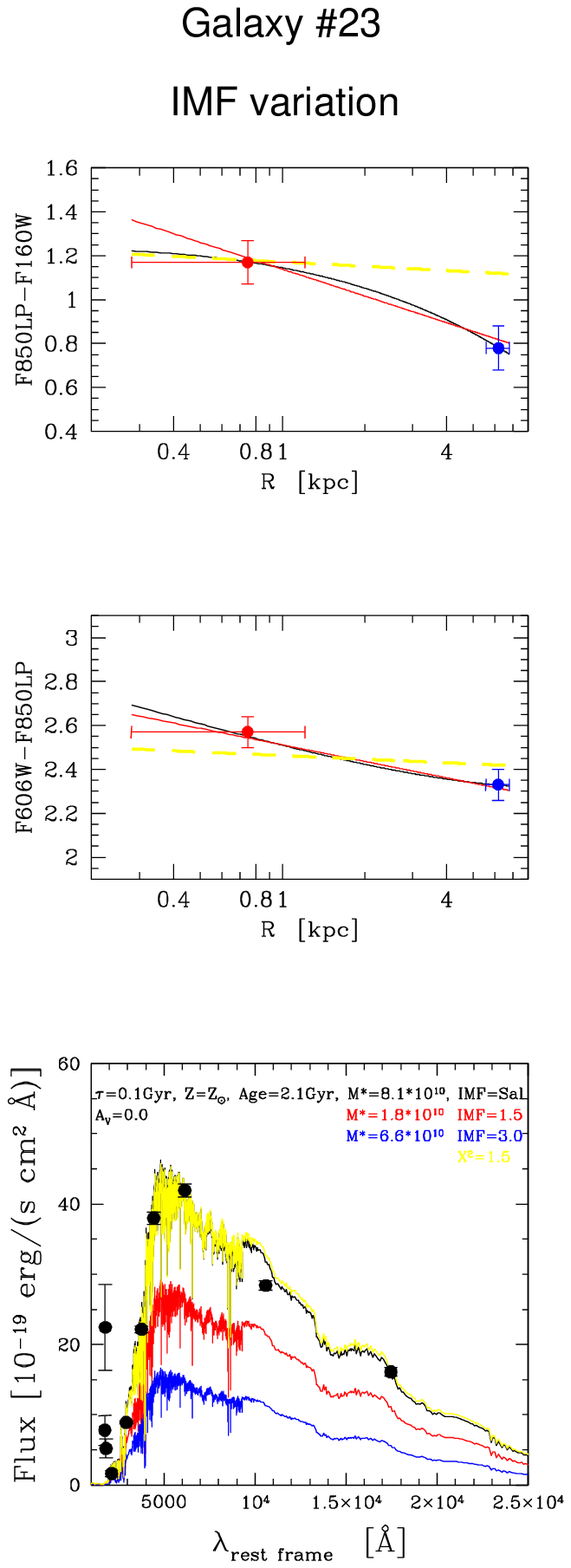} & \includegraphics[width=5.0cm,angle=0]{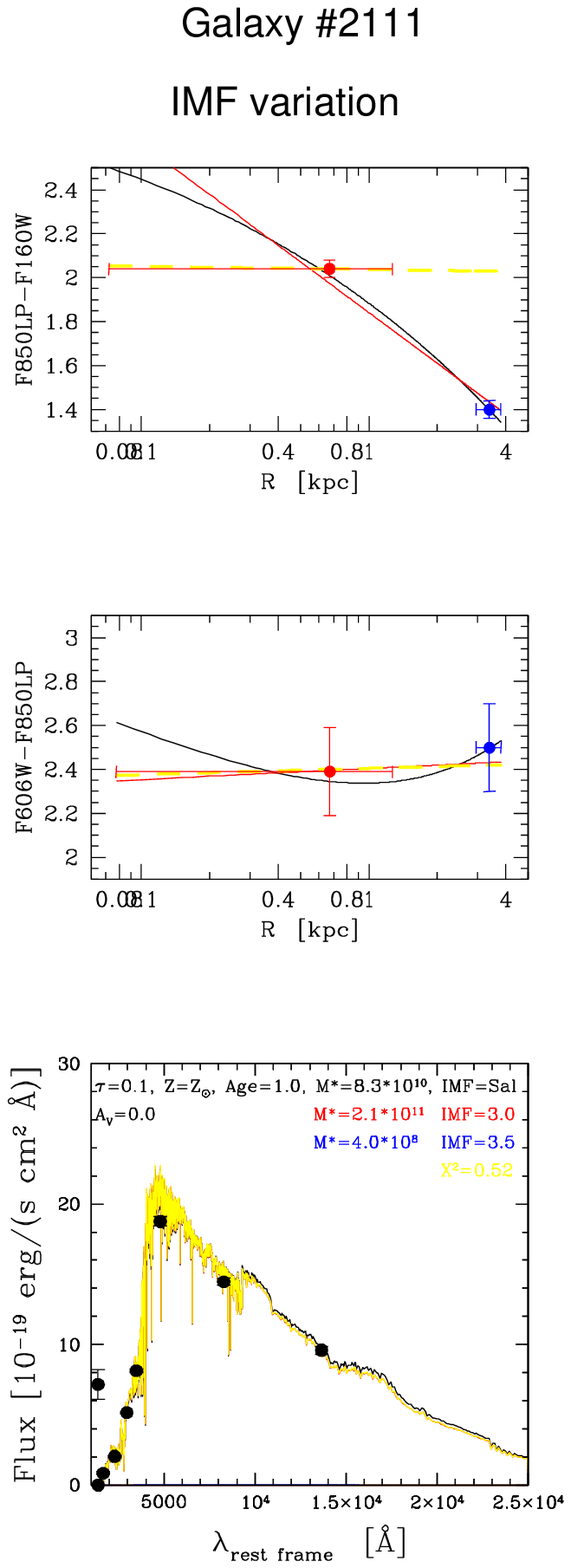} &
   \includegraphics[width=5.0cm,angle=0]{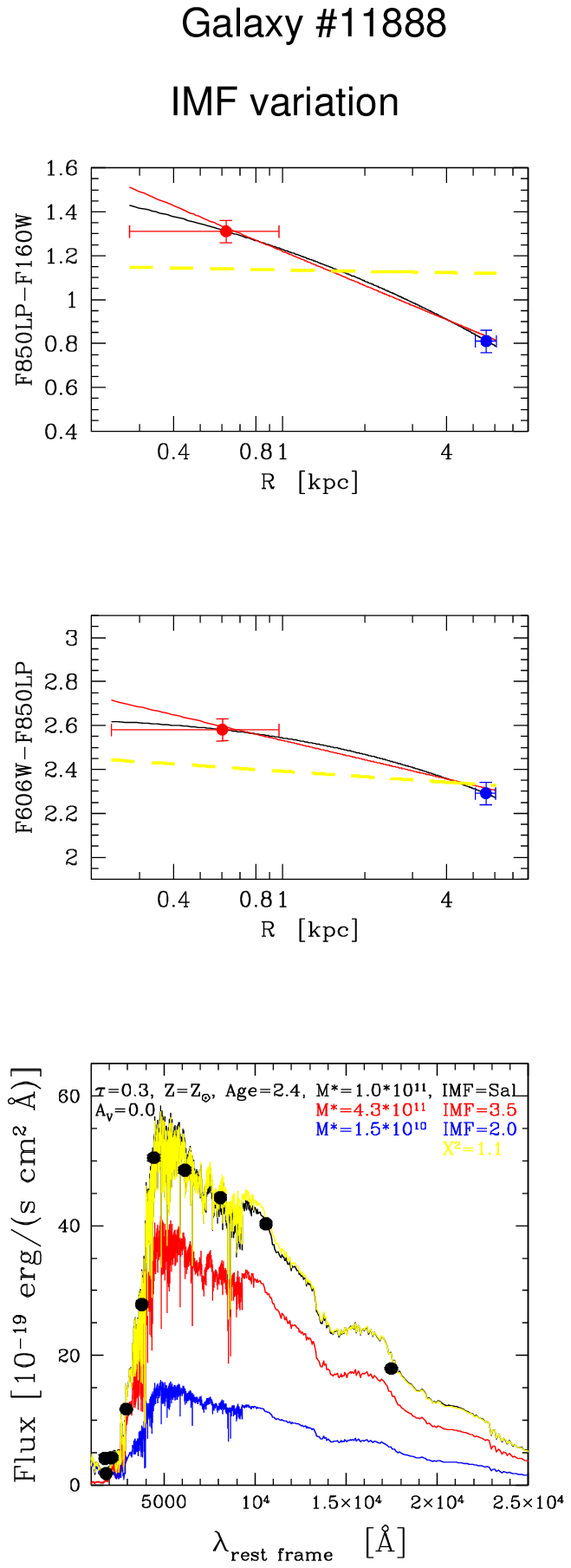} & \includegraphics[width=5.0cm,angle=0]{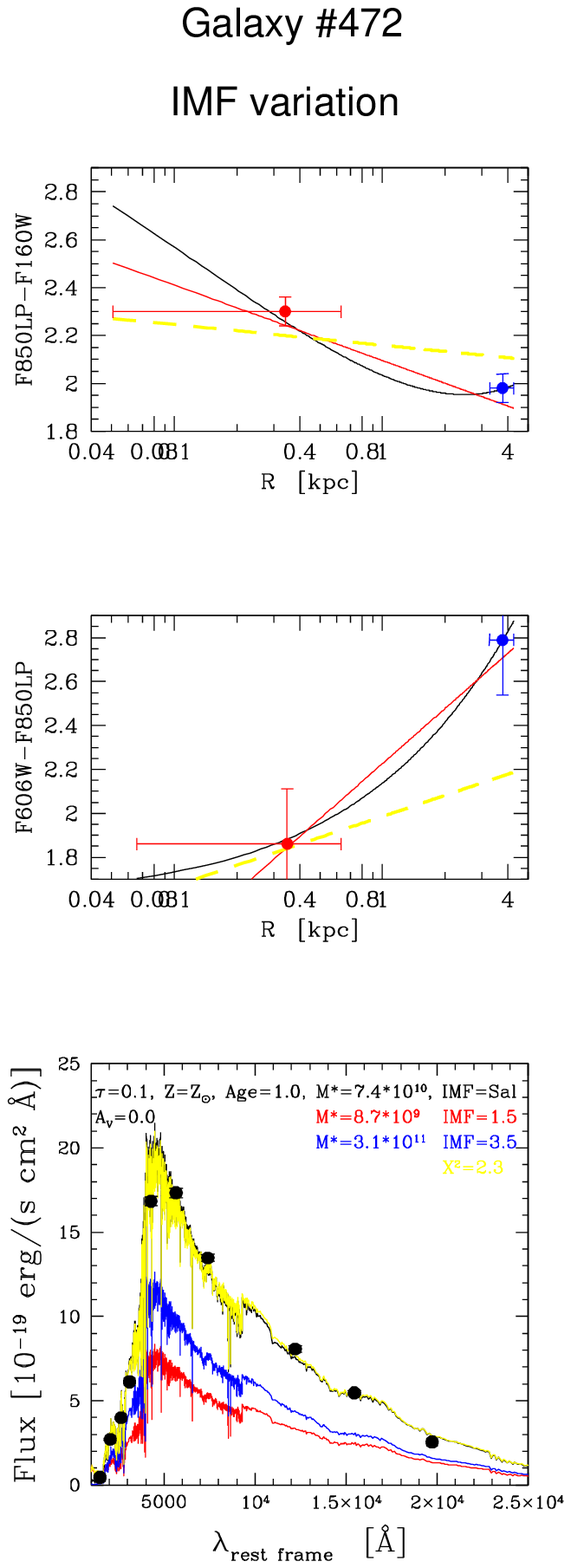} \\
   \end{tabular}
\newline
  \textbf{Figure 10.} Analysis of the radial variation of IMF's slope as possible driver of observed color gradients. 
Top and middle panels show the F850LP-F160W and F606W-F850LP color profiles, respectively (black lines), whit the relative fit (red lines). The bottom panel shows the observed global SED (black dots) as well as the stellar population parameters (black text) derived from its fit (black line) assuming a Salpeter IMF. Among all the stellar populations with varying IMF's slope but age, star-formation time scale, metallicity and dust extinction equal to the value obtained from the fit of the global SED, the red text identifies the one that best simultaneously reproduces the colors observed in the internal region (red dots in the top and middle panel). Similarly, blue text reports the value of the slope of the IMF that best reproduce the external colors (blue dots in the upper panels). The pure linear variation of the IMF's slope from the internal value to the external one, will produce the F850LP-F160W and F606W-F850LP color gradients represented with the dashed yellow lines. The contribution to the total stellar mass of the two populations is fixed in order to best fit the whole SED and are reported in the bottom panel (red and blue lines).
\end{center}
\end{figure}
\end{landscape}

\begin{landscape} 
\begin{figure}
\begin{center}
  \begin{tabular}{cccc}
     \includegraphics[width=4.6cm,angle=0]{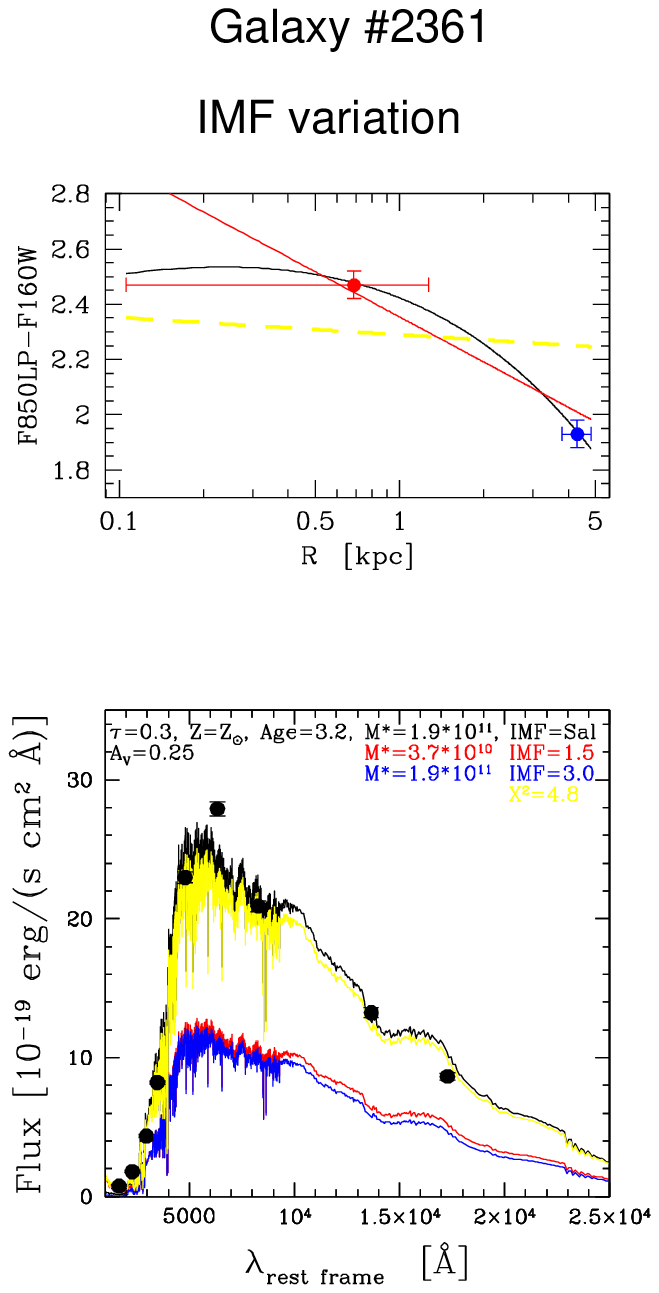} &
     \includegraphics[width=4.6cm,angle=0]{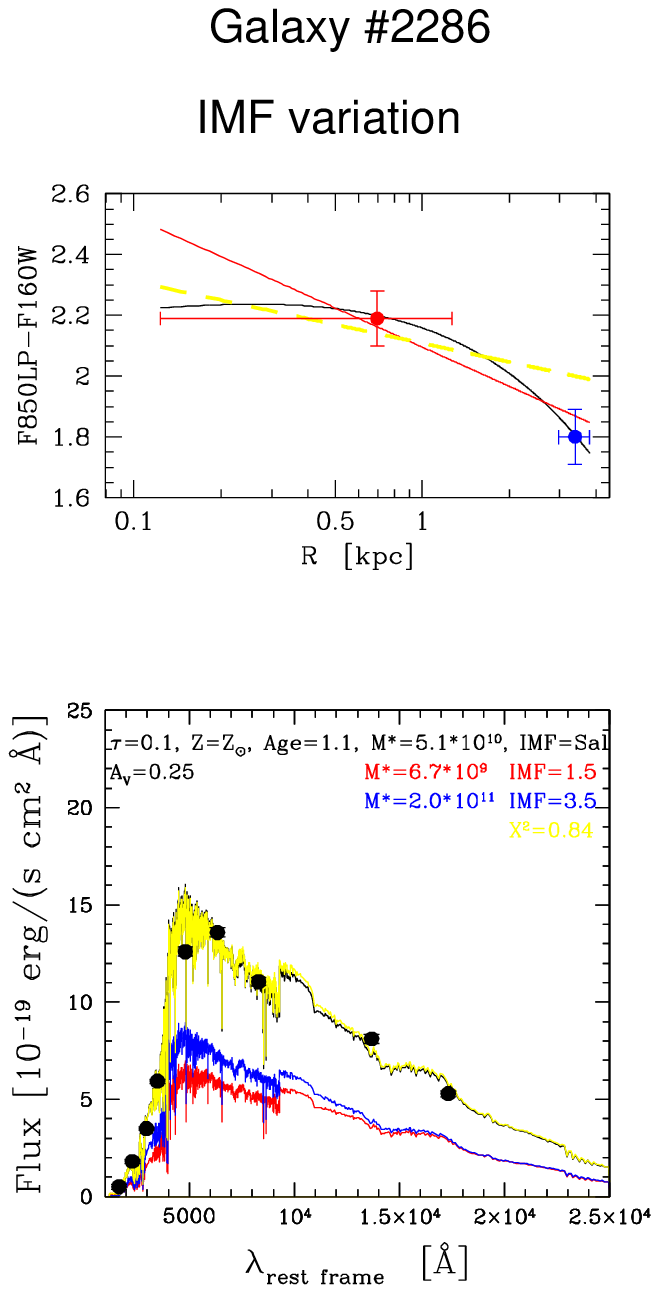} &
     \includegraphics[width=4.6cm,angle=0]{figure13_6.ps} &
     \includegraphics[width=4.6cm,angle=0]{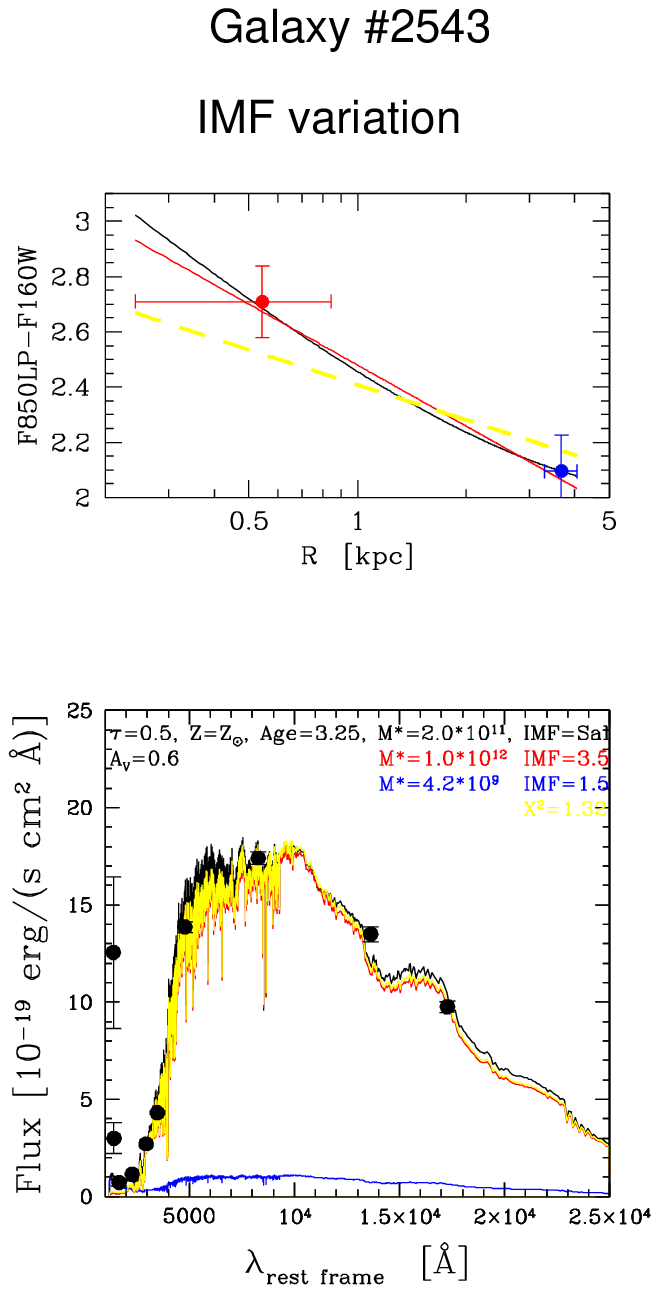}\\
     \includegraphics[width=4.6cm,angle=0]{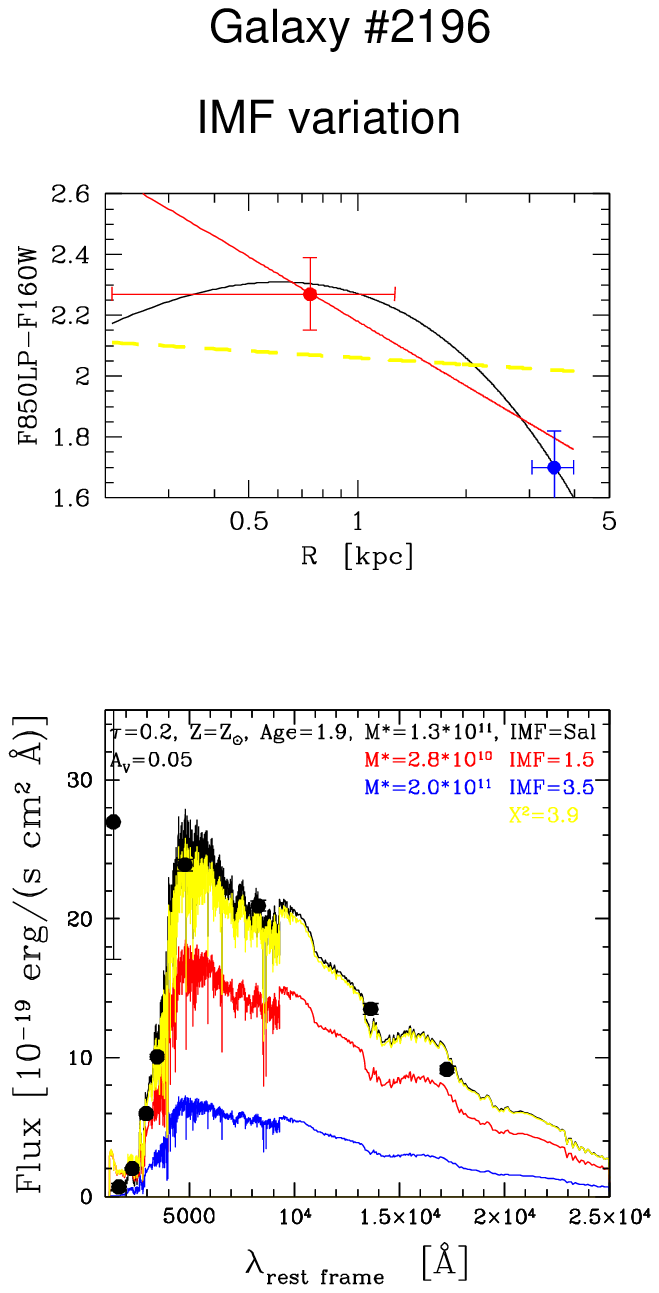} &
     \includegraphics[width=4.6cm,angle=0]{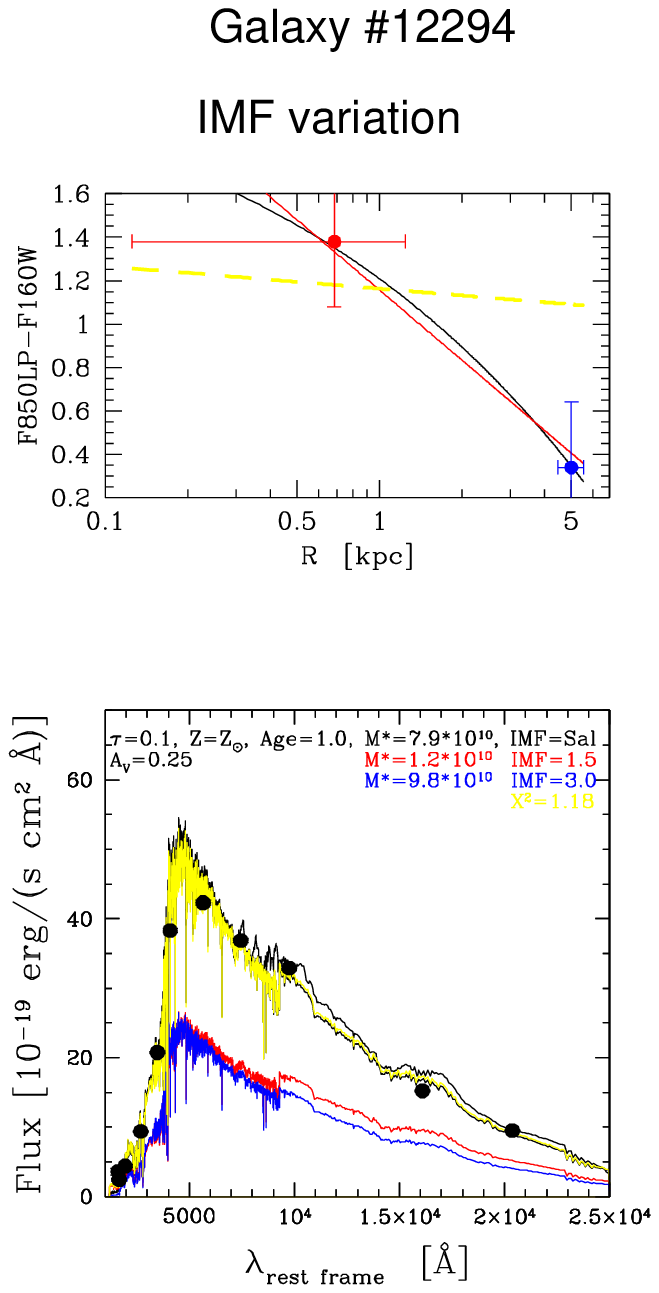} &
     \includegraphics[width=4.6cm,angle=0]{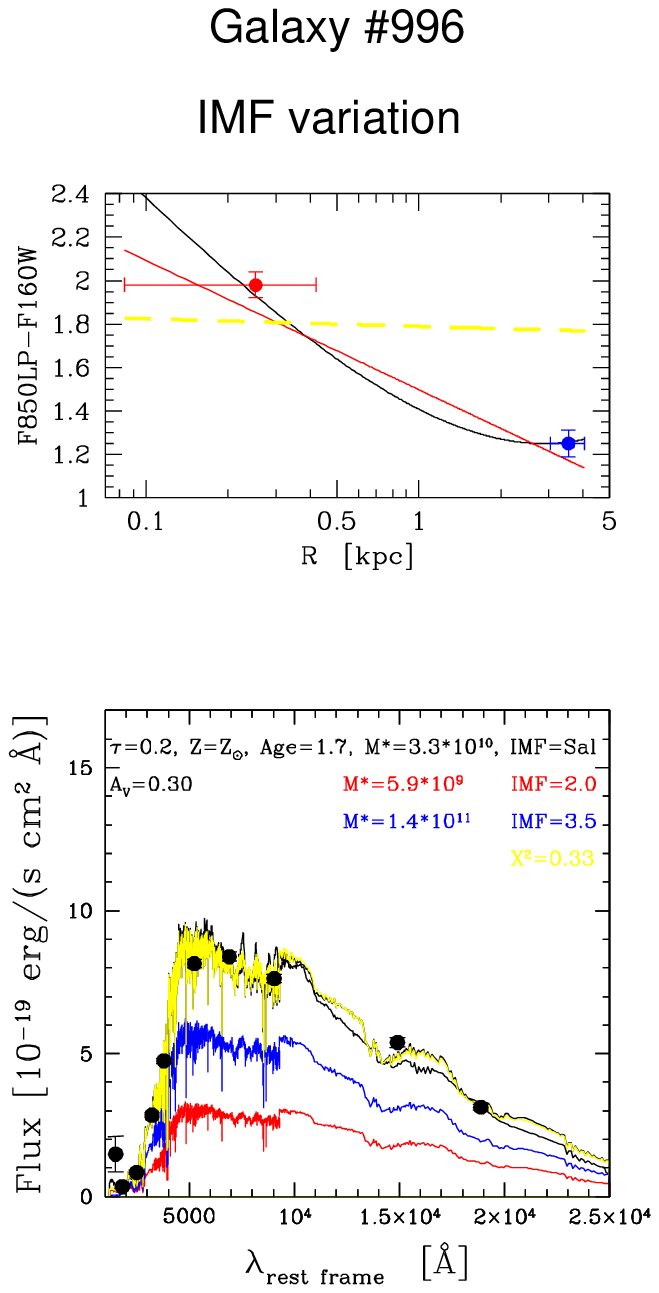} & \\	
    & \textbf{Figure 10.} (continued) & & \\
  \end{tabular}
\end{center}
\end{figure}
\end{landscape}

\end{document}